\documentclass[10pt]{article}
\usepackage[margin=1in]{geometry}
\usepackage{cite}
\usepackage{hyperref}
\usepackage{microtype}
\DisableLigatures[f]{encoding = *, family = * }
\usepackage[utf8]{inputenc}
\usepackage{graphicx}
\usepackage{adjustbox}
\usepackage{dcolumn}
\usepackage{bm}
\usepackage{ulem}
\usepackage{xcolor}
\usepackage{amsmath}
\usepackage{amssymb}
\usepackage{mathtools}
\usepackage{tabularx}
\usepackage{wrapfig}
\usepackage{nameref}
\usepackage{siunitx}
\usepackage{ulem}
\usepackage{lscape}
\usepackage{tocloft}

\newcommand{\der}[2]{\frac{\mathrm{d} #1}{\mathrm{d} #2}}
\newcommand{\secondder}[2]{\frac{\mathrm{d}^2 #1}{\mathrm{d} #2^2}}
\newcommand{\parder}[2]{\frac{\partial #1}{\partial #2}}
\newcommand{\popen}{p_{\text{o}}}

\newcommand{\Vrest}{V_{\text{rest}}}
\newcommand{\Vhalf}{V_{\frac{1}{2}}}
\newcommand{\Vhalfbif}{V^{\text{bif}}_{\frac{1}{2}}}
\newcommand{\tauopen}{\tau_c}
\newcommand{\taurest}{\tau_{\text{rest}}}
\newcommand{\ilocal}{I_c}
\newcommand{\xilocal}{\xi_c}
\newcommand{\iglobal}{I_e}
\newcommand{\tauglobal}{\tau_e}
\newcommand{\xiglobal}{\xi_e}
\newcommand{\itemp}{I_T}
\newcommand{\ivolt}{I_V}
\newcommand{\tautemp}{\tau_T}
\newcommand{\tauvolt}{\tau_V}
\newcommand{\Ntemp}{N_T}
\newcommand{\Nvolt}{N_V}
\newcommand{\ptemp}{p_T}
\newcommand{\pvolt}{p_V}
\newcommand{\Thalf}{T_{\frac{1}{2}}}
\newcommand{\cm}{c_{\text{mem}}}

\newcommand{\noise}{\mathrm{noise}}
\newcommand{\tesc}{t_{\text{AP}}}
\newcommand{\avg}[1]{\left\langle #1 \right\rangle}
\newcommand{\var}[1]{\text{Var} \left(#1\right)}
\newcommand{\fap}{f_{\text{AP}}}

\newcommand{\Xmax}[1]{#1_{\text{max}}}

\newcommand{\itarget}{I_{\text{target}}}
\newcommand{\infoint}{J_{\text{int}}}

\newcommand{\Nm}{N_{\text{m}}}
\newcommand{\Next}{N_{\text{ext}}}
\newcommand{\Neff}{N_{\text{eff}}}
\newcommand{\Vmin}{V_{\text{min}}}
\newcommand{\fisherinfo}{\text{I}}
\newcommand{\fisherinforatevoltage}{\dot{\iota}}
\newcommand{\fisherinforatechannels}{\text{i}_N}
\newcommand{\fisherinforate}{\dot{\text{I}}}
\newcommand{\fisherinfochannel}{\i}
\newcommand{\fisherinforatechannel}{\dot{\i}}
\newcommand{\fisherinfovoltage}{\iota}
\newcommand{\alphaadapt}{\alpha^{\text{a}}}
\newcommand{\tauadapt}{\tau^{\text{a}}}
\newcommand{\vadapt}{\Vhalf^{\text{a}}}
\newcommand{\taus}{\tau_s}
\newcommand{\Vs}{V_s}
\newcommand{\tap}{t_{\text{AP}}}
\newcommand{\tauint}{\tau_{\text{int}}}

\begin{document}

\begin{center}
{\Large
\textbf{A bifurcation integrates information from many noisy ion channels}
}
\bigskip
\\
Isabella R.\ Graf$^{1\star}$,
Benjamin B. Machta$^{1,2\dagger}$
\\
\medskip
\noindent\upshape$^{1}$Department of Physics, Yale University, New Haven, Connecticut 06511, USA
\\
\noindent\upshape$^{2}$Quantitative Biology Institute, Yale University, New Haven, Connecticut 06511, USA
\\
\noindent\upshape$^{\star}$ isabella.graf@yale.edu
\\
\noindent\upshape$^{\dagger}$ benjamin.machta@yale.edu

\end{center}

\vspace{15pt}

\section*{Abstract}
In various biological systems information from many noisy molecular receptors must be integrated into a collective response.
A striking example is the thermal imaging organ of pit vipers.
Single nerve fibers in the organ reliably respond to mK temperature increases, a thousand times more sensitive than their molecular sensors, thermo-TRP ion channels.
Here, we propose a mechanism for the integration of this molecular information.
In our model, amplification arises due to proximity to a dynamical bifurcation, separating a regime with frequent and regular action potentials (APs), from a regime where APs are irregular and infrequent.
Near the transition, AP frequency can have an extremely sharp dependence on temperature, naturally accounting for the thousand-fold amplification.
Furthermore, close to the bifurcation, most of the information about temperature available in the TRP channels’ kinetics can be read out from the timing of APs even in the presence of readout noise.
While proximity to such bifurcation points typically requires fine-tuning of parameters, we propose that having feedback act from the order parameter (AP frequency) onto the control parameter robustly maintains the system in the vicinity of the bifurcation.
This robustness suggests that similar feedback mechanisms might be found in other sensory systems which also need to detect tiny signals in a varying environment.

\vspace{20pt}

\section*{Main text}

\noindent \textbf{Introduction}

Many biochemical pathways serve primarily to amplify a weak signal -- one which is statistically significant only after integrating input from many individually noisy molecular sensors.
A striking example of this is the pit organ, used for thermal imaging by infrared sensing snakes~\cite{bullock_physiology_1952, bullock_properties_1956, goris_infrared_1967, goris_infrared_2011}.
This organ consists of a small opening near the nostril, which forms the aperture of a pinhole camera~\cite{bullock_physiology_1952, sichert_snakes_2006, bakken_imaging_2007}.
Incident radiation heats the pit membrane, a thin ($\approx 10~\mu$m) tissue which lies behind this opening~\cite{goris_infrared_1967}.
A few thousand sensory neurons project primary trigeminal afferents into the pit membrane~\cite{bullock_properties_1956, bakken_imaging_2007}, where resultant temperature changes are detected and transmitted further downstream as changes in the action potential (AP) frequency~\cite{bullock_properties_1956, goris_infrared_1967, goris_infrared_2011}.
Snakes use this organ for nocturnal ambush hunting, and they can detect and localize the tiny radiant heating from warm-blooded prey at distances up to $\approx 1$~m~\cite{ebert_behavioural_2006}.

Careful experiments with single nerve fibers combined with thermal modeling~\cite{bullock_properties_1956, bakken_imaging_2007} suggest that the nerve terminals can reliably detect temperature changes as small as $1$~mK.
At the same time, more recent experimental evidence has demonstrated that the molecular sensor is a TRPA1 thermo-sensitive ion channel~\cite{gracheva_molecular_2010}.
While these channels are among the most sensitive single proteins characterized, they are still three orders of magnitude less sensitive than nerve fibers, opening with increasing temperature according to a sigmoid whose width is about $\Delta T {\approx} 1$~K~\cite{gracheva_molecular_2010}.
In addition, these sensors must opearate in cold-blooded snakes, whose body temperature varies by thousands of milli-Kelvin with ambient changes.

How can the pit organ reliably measure temperature changes with three orders of magnitude higher accuracy than individual channels?
An individual TRPA1 channel provides very little information about mK changes in temperature: It only changes its probability of opening by $0.1\%$.
In principle, it is possible to obtain an arbitrarily more accurate measurement of temperature by averaging over a large number of channels, and/or for a duration much longer than a single channel's opening time $\tauopen$.
Averaging over $N$ channels for a time $\tau_\text{tot}$ could reduce the error in a measurement of temperature by a factor of $\sqrt{N \tau_\text{tot}/\tauopen}$.
These considerations suggest that a measurement of temperature with mK accuracy must somehow integrate over at least a million individual channel events.
These numbers, though impressive, are not necessarily discouraging.
While the density of TRPA1 channels in these nerve terminals has not been measured, it is reasonable to think their number might range in the hundreds of thousands to millions.
With a correlation time of $\tauopen=1-10$~ms~\cite{karashima_bimodal_2007} there is in principle enough information to make mK accurate measurements in $10-100$~ms. 

Still, it remains to be understood how neurons effectively integrate and  amplify the information from these many individual channels into a readout in the form of increased firing of action potentials (APs):
Even if the required accuracy of 1~mK can be achieved by averaging over a million individual channel events, preserving this information requires that the readout sufficiently amplifies the signal above the readout noise.

In this paper we propose a model for how this integration and amplification could be carried out mechanistically.
In our model, amplification arises from the diverging susceptibility at a dynamical bifurcation.
This amplification is driven by positive feedback between individual TRP channels
which are known to be sensitive not only to temperature but also to voltage~\cite{voets_principle_2004, nilius_gating_2005, zheng_molecular_2013, diaz-franulic_allosterism_2016}.
We will argue that a generic model for the voltage dynamics of a single terminal nerve mass that incorporates this positive feedback naturally exhibits a dynamical bifurcation.
In the vicinity of the bifurcation, the rate of action potential firing can have an arbitrarily steep dependence on temperature, and the statistics of the timings of action potentials contain an order-one fraction of all available information distributed amongst individual channels.
Furthermore, we will show how a simple feedback mechanism from the order parameter, the action potential frequency, onto a control parameter of the bifurcation naturally poises the system near this transition, even when background conditions fluctuate dramatically.
While molecular details are likely varied, our proposed feedback mechanism might illustrate a broader design principle for the high sensitivity found in diverse signaling systems made up of many noisy sensors.

\begin{figure*}[t]
\begin{center}
\includegraphics[width=0.6\linewidth]{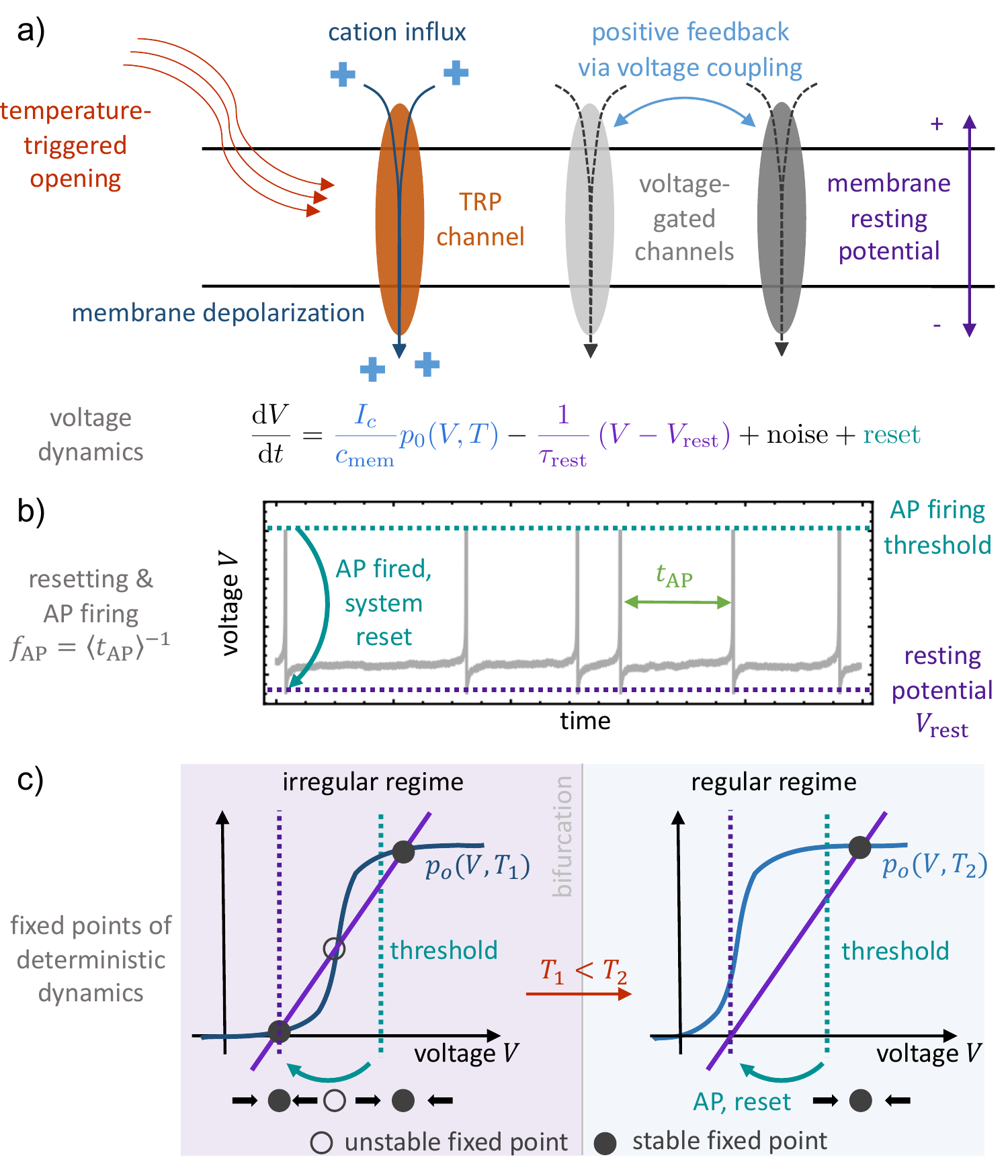}%
\end{center}
\caption{\label{fig:model} a) Model: The membrane voltage increases (the membrane depolarizes) if cations (blue daggers) flow through open channels (blue term in the equation).
The opening probability $p_{\text{o}}$ is as a sigmoid function of voltage, with a temperature-dependent voltage at half maximum (dark vs.\ light blue curve for $T_1<T_2$ in panel c).
An increase in temperature can thus prompt opening and lead to positive feedback: If one channel opens, the voltage across the membrane and thereby the probability for other channels to open increases.
These dynamics can trigger a cascade of opening events, unless the voltage is restored to the membrane resting potential on a faster timescale (purple arrow, purple term in the equation).
b) If the voltage crosses a threshold (turquoise), an action potential is fired and the system is reset back to the membrane resting potential.
c) The deterministic dynamics exhibit two regimes, an ``irregular regime" for low temperatures with two stable (and one unstable) fixed point and a ``regular regime" for high temperatures with a single stable fixed point.
In the irregular regime, after a reset, the voltage crosses the threshold only if noise pushes the dynamical system over the unstable fixed point.
Action potentials are thus rare ($\fap=\avg{\tesc}^{-1}$ is small) and noise-driven.
In contrast, in the regular regime, the voltage increases deterministically after a reset until it hits the threshold; action potentials are frequent ($\fap$ is large) and mostly regular.
The saddle-node bifurcation between the two regimes occurs when the left two fixed points annihilate.
}
\end{figure*}
\vspace{10pt}
\noindent \textbf{Model definition}

We consider a minimal model for the voltage dynamics of a single terminal nerve mass as illustrated in Fig.~\ref{fig:model}.
Voltage increases (the membrane depolarizes) if cations flow through open TRP channels.
We take the opening probability $\popen$ to be a sigmoid function in voltage, 
\begin{align}
    \popen (V,T) = \left(1+e^{-\left(V- \Vhalf(T)\right)/\Delta V} \right)^{-1},
\end{align}
approximating experimental measurements~\cite{voets_principle_2004, nilius_gating_2005, voets_sensing_2005, cordero-morales_cytoplasmic_2011, baez-nieto_thermo-trp_2011, zheng_molecular_2013, diaz-franulic_allosterism_2016},
with a temperature-dependent voltage at half maximum $\Vhalf(T) = \Vhalf (0)  -  (T/\Delta T) \Delta V$, which decreases for increasing temperature $T$ (thus increasing the opening probability); typical values are $\Delta V=30$~mV and $\Delta T= 1$~K.
We assume that when a channel is open, a current $\ilocal$ flows through it~\footnote{For simplicity, we assume that the current $\ilocal$ is voltage-independent but our main conclusions should remain valid even if the current increases (in a non-singular way) with voltage.}.
Together with the voltage-dependent opening probability, this naturally leads to a positive feedback:
A channel opening event leads to an inward current that raises the voltage across the membrane and thereby increases the opening probability of other channels.
An opposing force (``negative feedback") drives the voltage towards the membrane resting potential $\Vrest$ on a timescale $\taurest$, effectively integrating the role of active ion transporters, other channels and passive diffusion along concentration gradients across the membrane~\cite{ermentrout_mathematical_2010}.

These processes lead to the following dynamics:
\begin{align}
	    &\der{V}{t} = \underbrace{ \frac{\ilocal}{\cm} \popen (V,T) - \frac{1}{\taurest} \left(V - \Vrest\right)}_{=:v(V)} + \noise, \label{eq:dynamics}
\end{align}
where $\cm$ is the membrane capacitance per (area of a single) channel and noise is due to stochastic opening of channels, charge and temperature fluctuations, and other sources and is discussed below.
We supplement these continuous voltage dynamics with a simple approximation to action potential dynamics.
In neurons, when voltage reaches a critical value, the voltage goes through a highly stereotyped excursion -- an action potential with nonlinear dynamics driven by voltage sensitive channels~\cite{hodgkin_quantitative_1952}. 
To capture these stereotyped dynamics in an easily tractable model, we augment the voltage dynamics, Eq.~\ref{eq:dynamics}, as follows:
Each time the voltage crosses a threshold, an action potential (AP) is ``fired" and the system is reset back to the resting membrane potential $\Vrest$, see Fig.~\ref{fig:model}.

The relative strength of the positive and negative feedback determines whether action potentials are fired and at what rate. 
As we will show below, when the strength of these opposing forces is matched, the rate of APs can become extremely sensitive to single channel properties and the system achieves maximal amplification.

\vspace{10pt}
\noindent \textbf{Deterministic behavior and amplification}

\begin{figure*}[t]
\includegraphics[width=\linewidth]{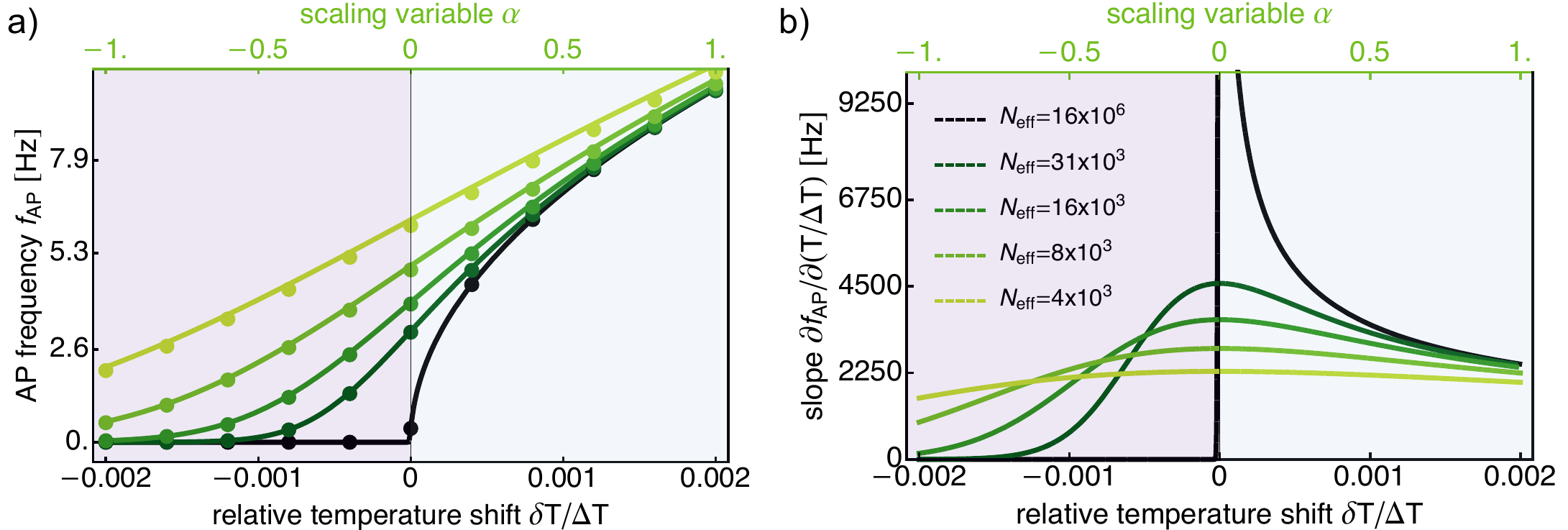}%
\caption{\label{fig:AP_frequency}
a) The frequency of action potentials  $\fap=1/\avg{\tesc}$ increases monotonically from the irregular (purple shaded) to the regular regime (blue shaded).
In the deterministic limit (very small noise, black line), $\fap=0$ below the bifurcation and $\fap \sim (\delta \Vhalf/\Delta V)^{1/2} \sim (\delta T/\Delta T)^{1/2}$ above, leading to non-analytic behavior at the bifurcation point $\delta \Vhalf=0$ (here, $\delta T = 0$).
For larger noise levels, the curve is smooth and the transition region widens (dark to light green).
b) The maximal slope of $\fap$ with respect to $\delta T/\Delta T$ (or $\delta \Vhalf/\Delta V$) (amplification) occurs right at the bifurcation and increases with decreasing noise level (light green to black).
The frequency and its derivative are shown as a function of $\delta T/\Delta T$ (or, equivalently, $\delta \Vhalf/\Delta V$) at the bottom and as a function of the scaling variable $\alpha$, Eq.~\ref{eq:scaling_variable}, for $\Neff = 8 \times 10^3$ at the top.
Lines are analytic results, symbols are from stochastic simulations with parameters $\Delta V=30$~mV, $\Delta T=1$~K, $\taurest=\tauopen=1$~ms, $\ilocal/\cm = 2$~V/s, $\Vrest=-70$~mV, $\Next = \infty$, and $\Neff$ increasing from light green to black as shown in b).
}
\end{figure*}

We first focus on the deterministic dynamics, $\der{V}{t} = v(V)$.
The intersections of the first and second terms in Eq.~\ref{eq:dynamics} correspond to fixed points  where $\der{V}{t}=0$ (see Fig.~\ref{fig:model}).
For high temperatures, there is only a single fixed point at high voltage.
We assume that APs are triggered at a lower voltage than this high-voltage fixed point.
After an AP is fired and the system is reset to the resting potential, voltage again increases monotonically towards the AP threshold.
In this limit, APs occur relatively frequently and regularly (``regular regime"; light blue background in Fig.~\ref{fig:AP_frequency}).

On reducing temperature, two additional fixed points emerge through a saddle node bifurcation, a stable and an unstable one~\footnote{For very small temperatures, the system is again monostable, with only a single low-voltage fixed point.  We do not consider this regime any further but note that the frequency of APs would be vanishingly small in this regime.}.
After a reset the system is now stuck in the lower fixed point~\footnote{We assume throughout that the resting membrane potential is lower than the unstable fixed point and the threshold for firing an AP is higher but lower than the upper fixed point. In this case their exact values do not matter much. Furthermore, we assume $0{<}\Delta V\cm/(\ilocal \taurest){<}1/4$, so that the system can switch between a monostable and bistable regime.}.
In the deterministic limit, the frequency of APs is thus zero in this regime (black curve in Fig.~\ref{fig:AP_frequency}).
Only noise can push the dynamics across the unstable fixed point to trigger an AP, and thus APs are inherently stochastic (``irregular regime"; light purple background in Fig.~\ref{fig:AP_frequency}).

Close to the bifurcation in the regular regime, the AP frequency scales like the square root of the distance to the bifurcation, and so its slope diverges at the bifurcation (see black line in Fig.~\ref{fig:AP_frequency}, cf. also class 1 neurons~\cite{izhikevich2007dynamical}).
Near this bifurcation, tiny changes in temperature are thus strongly amplified into large changes in the AP frequency, naturally accounting for the 1000-fold amplification between single channel properties and the collective neuronal response.

\vspace{10pt}
\noindent \textbf{Noise broadens the bifurcation}

In the absence of noise, our deterministic dynamics predicts an arbitrarily steep relationship between temperature and AP rate near the bifurcation.
However, noise in the voltage dynamics is expected to suppress the singular behavior (diverging slope) in AP frequency and to reduce the maximal possible amplification by broadening the bifurcation region (see Fig.~\ref{fig:AP_frequency}).
To get quantitative insight, we consider two broad categories of noise;  \textit{intrinsic} noise, which is driven by the stochastic nature of the individually noisy molecular sensors, and \textit{extrinsic} noise which includes all other sources.
Since the intrinsic noise arises from the stochastic opening of TRPA1 channels, we take it to be binomial in the number $N$ of these channels, with opening probability $\popen$ and a correlation time of $\tau_c$.
Extrinsic noise is likely dominated by the stochastic opening of other channels, as well as charge and heat fluctuations that provide correlated noise to all parts of the terminal nerve mass.
We model this extrinsic noise phenomenologically via a characteristic correlation time $\tauglobal$ and strength $\iglobal$ and assume that all noise processes are faster than other dynamic features.
In this case, we can approximate the noise processes as being delta-correlated:
\begin{align}
    \noise (t) = \underbrace{\frac{\ilocal}{\cm}\sqrt{\frac{\popen \left( 1{-}\popen\right)}{N}}  \xilocal\left(\frac{t}{\tauopen}\right)}_{\text{intrinsic noise} } {+} \underbrace{\frac{\iglobal}{\cm}\xiglobal \left(\frac{t}{\tauglobal}\right)}_{\text{extrinsic noise}},
\end{align}
where $\xilocal (z)$ and $\xiglobal (z)$ are white noise processes with unit variance.
As expected, with increasing noise, the sharp transition in AP frequency is blurred, see Fig.~\ref{fig:AP_frequency}, with noise making only minimal contributions far away from the transition.

\vspace{10pt}
\noindent \textbf{Scaling analysis of the near-bifurcation region}

In order to quantitatively understand the functional properties of this model and how AP dynamics depend on the strength of noise and the distance to the bifurcation, we focus on the region close to the bifurcation.
There, the timing and frequency of action potentials is dominated by the time it takes the system to cross the region with $v(V) \approx 0$, where the noise can be approximated as:
\begin{align}
        \text{noise near bifurcation} = \frac{\Delta V}{\sqrt{\taurest}} \frac{1}{\sqrt{\Neff}} \xi(t), 
\end{align}
where $\Neff$ is a dimensionless number quantifying the noise strength:
\begin{align}
\Neff^{-1} :=\Nm^{-1}+\Next^{-1},
\hspace{20pt} \text{with} \\
\Nm :=\frac{N \rho \taurest}{\tau_c} \hspace{10pt}\text{and} \hspace{10pt} \Next:=\frac{\Delta V ^2 \cm^2}{I_e^2 \tau_e\taurest}. \nonumber
\end{align}
Here $\Nm$ is the number of measurements of temperature that $N$ independent channels could make per time $\taurest$, and $\Next^{-1}$ parameterizes the extrinsic noise.
Furthermore, we have defined $\rho {:=} \Delta V \cm/\ilocal \taurest$ as a dimensionless measure of channel density; at the bifurcation $\popen(1-\popen)=\rho \leq 1/4$.
With these definitions,  $\Neff$ is the effective number of independent measurements of temperature available in the current in a time $\taurest$.
This number cannot be larger than the number $\Nm$ of measurements in all channels and is limited by extrinsic noise: $\Neff \approx \Next \ll \Nm$ for large extrinsic noise $\Next \ll \Nm$.

To understand the statistics of AP firing times, we approximate the voltage dynamics up to second order around the arg minimum $\Vmin$ of $v(V)$.
By rescaling time and voltage via $s=t/\tau_s$ and $u=(V-\Vmin)/V_s$, with
\begin{align}
        \tau_s &:= \taurest\left( \frac{4 \Neff}{(1-4\rho)} \right)^{1/3}
         \sim N_\text{eff}^{1/3} \taurest \label{eq:rescalings} \\
        V_s &:= \Delta V \left( \frac{4}{(1-4\rho) \Neff^2} \right)^{1/6} \sim N_\text{eff}^{-1/3}\Delta V,  \nonumber
\end{align}
the dynamics can be written simply as
\begin{align}
	\der{u}{s} &=  u^2+\alpha+\xi(s), 
	\label{eq:scaling_eq}
\end{align}
where the noise is of unit strength and $u$ and $s$ are dimensionless quantities.
These rescaled dynamics thus depend on just a single scaling variable, which is linear in the distance to the bifurcation:
\begin{align}
	\alpha :=  \frac{\delta \Vhalf}{V_s} \frac{\tau_s}{\taurest} \sim N_\text{eff}^{2/3}\frac{\delta \Vhalf}{\Delta V}, 
	\label{eq:scaling_variable}
\end{align}
with distance $\delta \Vhalf {:=} \Vhalfbif{-}\Vhalf (T)$ of the voltage $\Vhalf (T)$ at half maximum to its value $\Vhalfbif$ at the bifurcation.
Note that we define $\delta \Vhalf$ so that an increase in temperature corresponds to $\alpha > 0$; in the regular regime we thus have $\alpha > 0$ and in the irregular one $\alpha < 0$.
As one can see from Eq.~\ref{eq:scaling_variable}, noise effectively widens the bifurcation region:
For fixed $\alpha$, the transition region $\delta \Vhalf/\Delta V \sim \Neff^{-2/3}$.

The dimensionless form of the dynamics, Eq.~\ref{eq:scaling_eq}, also allows us to determine a scaling form for the frequency of action potentials (the inverse of the mean time between two action potentials $\avg{\tesc}$).
To this end, we approximate the time between two APs (the interspike time) as the time it takes the dynamical system to go from $u {\rightarrow} {-}\infty$ to $u {\rightarrow} {+}\infty$.
We expect this approximation to hold close to the bifurcation where crossing times are dominated by the region with very small drift,  $v\approx 0$.
As can be seen from the comparison between simulations and theory in Fig.~\ref{fig:AP_frequency}, the agreement is indeed excellent.
In this approximation, the mean interspike time $\avg{\tesc}$ can be written as a non-trivial dimensionless function of $\alpha$, multiplied by the characteristic timescale $\tau_s$:
\begin{align}
	\avg{\tesc} &= \tau_s M(\alpha), 
	\label{eq:mean}
\end{align}
 where $M(\alpha){:=} (2^{1/3} \pi^2) \left(\left(\text{A}_1[{-}2^{2/3} \alpha]\right)^2 + \left(\text{A}_2[{-}2^{2/3} \alpha]\right)^2 \right)$ with $\text{A}$ denoting the Airy function, see SI and Ref.~\cite{hathcock_reaction_2021}.  
In the regime near the bifurcation ($\alpha \approx 0$), the frequency of action potentials $\fap {:=} 1/\avg{\tesc}$ scales as $\fap \sim \tau_s^{-1}\sim N_{\text{eff}}^{-1/3} \taurest^{-1}$.
In contrast, at large positive $\alpha$ we expect the system to be deterministic and $\avg{\tesc}$ to be independent of the noise strength $\Neff^{-1}$.
Using Eqs.~\ref{eq:rescalings} and~\ref{eq:scaling_variable}, it follows that $M(\alpha) \sim \alpha^{-1/2}$ for $\alpha \gg 1$, consistent with the scaling of the deterministic prediction, $\fap \sim \delta \Vhalf^{1/2}$, see also~\cite{pi_critical_2021}.

By taking a derivative of the scaling form for the AP frequency $\fap$, we can determine the susceptibility (slope) $\mathrm{d} \fap/\mathrm{d} \delta \Vhalf$ as
\begin{align}
	\frac{d\fap}{d\delta\Vhalf} &= \frac{1}{\taurest V_s} \left(\frac{-M'(\alpha)}{M^2(\alpha)} \right).
	\label{eq:susceptibility}
\end{align}
As in the deterministic case, the maximal slope is attained for $\alpha=0$, Fig.~\ref{fig:AP_frequency}.
It scales like $\taurest^{-1} V_s^{-1} \sim N_{\text{eff}}^{1/3}$, whereas the log-slope scales as $N_{\text{eff}}^{2/3}$.
As expected, the amplification is thus the larger the smaller the noise $\Neff^{-1}$.

\vspace{10pt}

\noindent \textbf{Information fidelity}

\begin{figure*}[t]
\includegraphics[width=\linewidth]{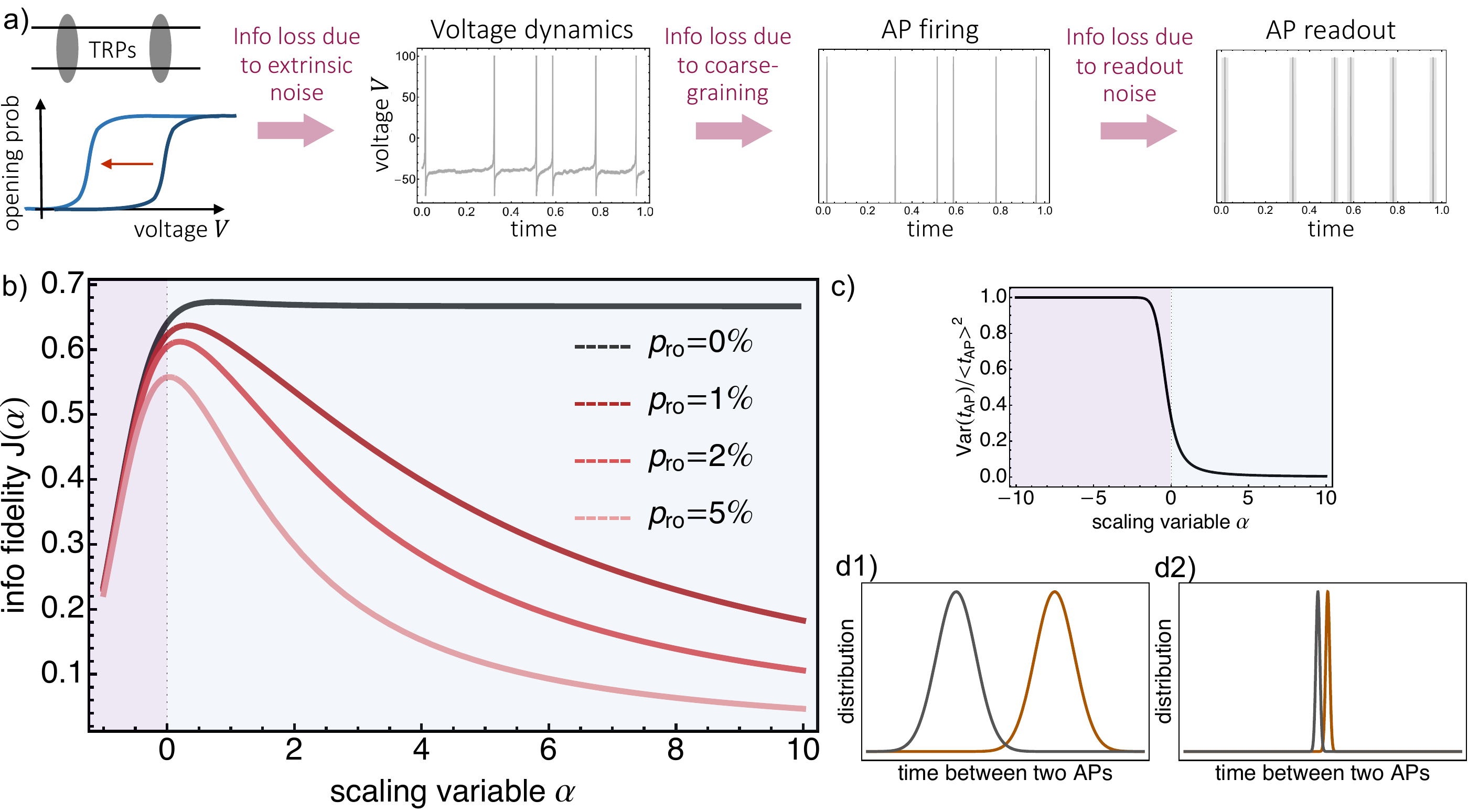}%
\caption{\label{fig:info} 
a) In the model, information about temperature available in the individual channels' opening kinetics is passed down to the voltage dynamics and further to the statistics of AP firing.
The full voltage dynamics $V(t)$ preserves most of the information available in the channels' kinetics and is only corrupted by extrinsic noise.
When going from the voltage dynamics to AP firing, information is lost due to coarse-graining and due to additional noise in the readout of the AP timings.
b) The information fidelity $J(\alpha)$, the fraction of the Fisher information retained in the timing of action potentials (relative to the Fisher information contained in the full voltage dynamics), is maximized close to the bifurcation $\alpha=0$.
If the timing of action potentials can be read out perfectly (black line), the information fidelity saturates in the regular regime $\alpha \gtrsim 1$ at a value of  $J \approx 0.66$.
Deep in the regular regime, the intrinsic coefficient of variation, $\sigma (\tesc)/\avg{\tesc}$, is very small (c) and information is contained in tiny changes in the frequency of action potentials (d2).
If reading out the APs is subject to noise, $\var{\tesc} \rightarrow \var{\tesc} + p_{\text{ro}} \avg{\tesc}^2$, these tiny changes in the frequency cannot be resolved anymore and the information fidelity in the regular regime strongly decreases.
In contrast, close to and to the left of the bifurcation, the intrinsic coefficient of variation is of order 1 (c) and information is contained in strong amplification of the response/a large shift in the mean frequency (d1); readout noise therefore does not suppress the information considerably.
For larger readout noise $p_{\text{ro}}$, the maximum of $J(\alpha)$ is more pronounced and shifts to (slightly) smaller values of $\alpha$.
Similar results are also obtained for constant readout noise, $\var{\tesc} \rightarrow \var{\tesc} + \sigma^2_{\text{ro}}$, see SI.
}
\end{figure*}

In our model, information about temperature contained in the individual channels' opening kinetics is transmitted to the voltage dynamics and then passed down to the statistics of action potential firing, see Fig.~\ref{fig:info}a.
The information contained in a sequence of APs depends on both the mean frequency's dependence on temperature and on the variance in AP timing.
The noisier the timing of APs, the less information can be gained if the frequency changes.
We can determine the variance in AP timing analogously to the mean, Eq.~\ref{eq:mean}, as
\begin{align}
    \var{\tesc} &= \tau_s^2 \ S(\alpha)\sim \taurest^2N_{\text{eff}}^{2/3} ,
	\label{eq:var}
\end{align}
in terms of a scaling function $S(\alpha)$, which we specify in the SI.
Figure~\ref{fig:info}c shows the squared coefficient of variation $\var{\tesc}/\avg{\tesc}^2$, which tends to 1 in the irregular regime and to 0 in the regular regime.

In both cases, the longer the observed trajectory of AP firing, the more information is obtained.
In our model, the timings between subsequent APs are uncorrelated and  information grows linearly in time -- at least for times significantly larger than the typical interspike time $\avg{\tesc}$.
We thus define a \textit{Fisher Information Rate}, $\fisherinforate $, as the Fisher Information of a sequence of spikes, per unit time, to a change in temperature, see also~\cite{toyoizumi_fisher_2006}.
This information rate thus has units of $\text{s}^{-1} \text{K}^{-2}$ and 
is given by
\begin{align}
	\fisherinforate \approx \frac{\left(\partial \avg{\tesc}/\partial T \right)^2}{\avg{\tesc}\var{\tesc}} = \frac{\Neff}{\Nm} \fisherinforatechannels \ J(\alpha)  = \fisherinforatevoltage  \ J(\alpha),  \label{eq:info_rate}
\end{align}
where $\fisherinforatechannels {:=} \rho N/(\tauopen \Delta T^2)$ is the maximal Fisher information rate contained in all $N$ channels if they performed independent measurements (for opening probability evaluated at $\Vmin$) and $\fisherinforatevoltage{=}\fisherinforatechannels \Neff/\Nm$ is the Fisher information rate contained in the full voltage dynamics $V(t)$, see SI.

We call $J(\alpha) {:=} M'(\alpha)^2/(M(\alpha) S(\alpha))$ the ``information fidelity", corresponding to the ratio between the information rate contained in the full voltage dynamics, $\fisherinforatevoltage$, and the information rate available on the level of APs.
$J(\alpha)$ thus quantifies how efficiently information about temperature available in the channels is passed down to the level of AP firing.
The expression for the information rate, Eq.~\ref{eq:info_rate}, provides a lower bound on the actual information rate:
Higher-order terms stemming from changes in the $n{\geq}2$-th cumulants/moments of the interspike time $\tesc$ can only increase the information but are expected to be small.
Including the second order term changes the information rate by ${<}1 \%$, see SI.
Since in our scheme information about the signal (temperature) influences the AP timing only via the membrane voltage, the information rate $\fisherinforatevoltage$ contained in the full voltage dynamics $V(t)$ provides an upper bound on $\dot{I}$, and we indeed numerically find that the information fidelity $J(\alpha) {\leq} 1$; see Fig.~\ref{fig:info}.

That being said, Eq.~\ref{eq:info_rate} shows that information about the signal is transferred efficiently by the proposed signaling system:
The available information $\fisherinforatevoltage$ enters linearly and a substantial fraction of it (roughly $2/3$ if $J(\alpha)=\Xmax{J}$ is maximized over $\alpha$) is captured by the timing of action potentials; see Fig.~\ref{fig:info}.
More generally, since $i_N {\sim} N$, the information rate scales like ${\sim} N/(1+\Nm/\Next)$, and extrinsic noise starts to become relevant for $\Nm \gtrsim \Next$.
This argument provides a rough estimate for the optimal number of channels that is as small as possible but simultaneously maximizes the information that can be obtained about changes in temperature: $N_{\text{opt}} \sim \rho \ilocal^2 \tauopen/(\iglobal^2 \tauglobal)$.

As shown in Fig.~\ref{fig:info}, the information fidelity is small in the irregular regime.
In contrast, in the regular regime it saturates for $\alpha {\gtrsim} 1$ at a value of $2/3$.
In this regime, the variance in the frequency of APs is extremely low, Fig.~\ref{fig:info}c, and information about changes in the temperature is encoded in tiny changes in the frequency, Fig.~\ref{fig:info}d2.
Reading out this information thus requires extremely precise measurements of the APs, and already small noise in the readout of the AP frequencies, strongly suppresses the information for large $\alpha$.
Indeed, if in Eq.~\ref{eq:info_rate} we replace the variance in the interspike time by either $ \var{\tesc} {\rightarrow}  \var{\tesc} + p_{\text{ro}} \avg{\tesc}^2$ for a readout noise that scales with the average time between two APs, Fig.~\ref{fig:info}b, and/or $ \var{\tesc} {\rightarrow}  \var{\tesc} + \sigma^2_{\text{ro}}$ for a constant, extrinsic readout noise,  see SI, the information fidelity features a distinct maximum close to the bifurcation, whose position only slightly depends on the strength of the noise.
Taken together, the accessible information is highest for $\alpha \approx 0$, suggesting that poising the system in the vicinity of the bifurcation is desirable from an information point of view and for being able to sense small changes in the signal.

\vspace{10pt}
\noindent \textbf{Order-parameter feedback}

If changes in the signal always needed to be measured with respect to the same background signal (temperature) and in conditions where the underlying physical properties such as the membrane capacitance or resistance or the channel characteristics are mostly constant, one could imagine that, e.g., evolutionary processes had fine-tuned these properties so as to poise the voltage dynamics at the transition.
However, 
snakes live in deserts, where the ambient background temperature can vary quite considerably on the channels' natural scale of milli-Kelvin.
Membrane channel properties like the conductance are furthermore known to depend on various cellular factors like pip lipids in the membrane or intracellular calcium~\cite{ramsey_introduction_2006, rohacs_regulation_2007,taberner_trp_2015,diaz-franulic_allosterism_2016,hasan_ca2_2018, vangeel_transient_2019}, which are affected by numerous intracellular processes.
It thus seems impossible to achieve proximity to the bifurcation by simply fine-tuning certain channel and membrane properties.

\begin{figure*}[t]
\includegraphics[width=\linewidth]{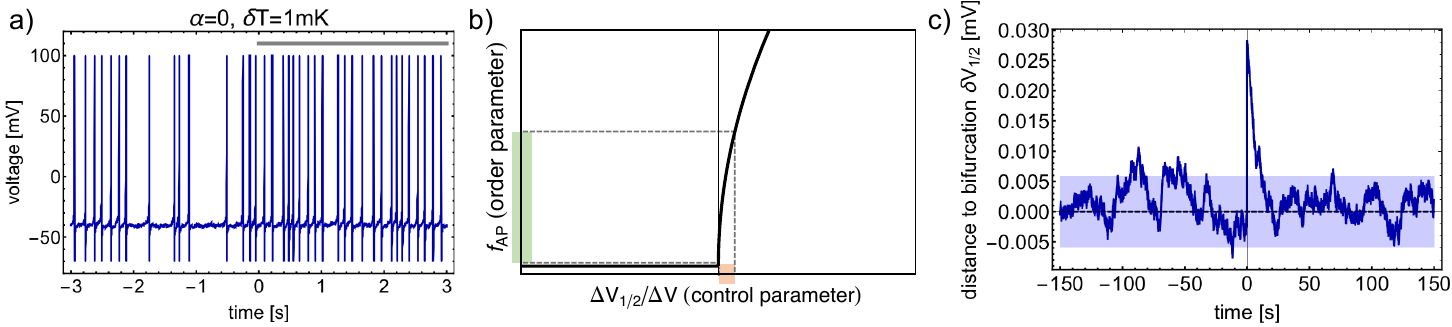}%
\caption{\label{fig:adaptation} Voltage dynamics for a self-adapting system with target frequency $\gamma d^-/d^+=5$~Hz. 
At time $t=0$, a temperature step of $T_0 \rightarrow T_0 +\delta T =T_0 + 1$~mK is applied.
a) As a result of the temperature step, the AP frequency increases transiently (it roughly doubles) and returns to pre-stimulus level after around $10-20$~s (not shown in the time trace on the left).
b) Due to the diverging susceptibility (slope) at the bifurcation point, feedback that maintains the action potential frequency somewhere in the steep part of the curve (green shaded area on the y-axis) tunes the system very close to the bifurcation point (orange shaded area on the x-axis)
c) After a transient increase in the distance $\delta \Vhalf$ to the bifurcation by $\Delta V (1 \text{mK}/\Delta T)=0.03$~mV following the temperature step at time $t=0$, the system adapts back to pre-stimulus level (here $\delta \Vhalf=0$) by increasing its voltage $\Vhalf$ at half maximum by the same amount.
The timescale of the adaptation is chosen so that adaptation happens over roughly 10~s.
The parameters are: $N=2^{19}, \Next = \infty$, $\Vrest=-70$~mV, $\Delta V=30$~mV, $\Delta T=1$~K, $\taurest=\tauopen=1$~ms, $\ilocal/\cm = 2$~V/s, $\gamma^{-1}=1$~ms, $d^-=0.5 \times 10^{-5}$, and $d^+=10^{-3}$, see SI for a more thorough discussion of plausible parameter values.
}
\end{figure*}

Inspired by early work on self-organized criticality~\cite{sornette_critical_1992, sornette_mapping_1995} and recent studies extending these ideas to discontinuous transitions in the framework of so-called self-organized bistability~\cite{di_santo_self-organized_2016, buendia_feedback_2020}, here we propose a specific type of feedback motif, which robustly tunes the dynamical system arbitrarily close to the transition -- without the need for fine-tuning.
The idea is that in systems with a critical or bifurcation point where an order parameter, $\mathcal{O}$, exhibits singular behavior with respect to the control parameter, $\mathcal{C}$, feedback from the order onto the control parameter, which can be as simple as $\mathrm{d} \mathcal{C}/\mathrm{d} t {=} \text{sign} \left( \mathrm{d} \mathcal{O}/\mathrm{d} \mathcal{C} \right) \left( \gamma - \beta \mathcal{O} \right)$, acts to maintain the system in the vicinity of this critical point, Fig.~\ref{fig:adaptation}b.
Due to the diverging susceptibility at the critical or bifurcation point (bif), $\left| \mathrm{d} \mathcal{O}/\mathrm{d} \mathcal{C}\right|_{\text{bif}} {\rightarrow} \infty$, a small change in the control parameter triggers a large change in the order parameter and, conversely, a change $\Delta \mathcal{O}^*$ in the fixed point $\mathcal{O}^* = \gamma/\beta$ of the feedback affects the control parameter only marginally, $\left|\Delta \mathcal{C}^* \right| \sim \left|\Delta \mathcal{O}^* \right| (\mathrm{d} \mathcal{O}/\mathrm{d} \mathcal{C})^{-1}$, thus providing robust tuning to the vicinity of the bifurcation.

In our case, the frequency of APs serves as a collective variable with singular behavior at the bifurcation point, see Fig.~\ref{fig:AP_frequency}.
We implement feedback from the AP frequency onto the control parameter, the voltage at half-maximum $\Vhalf$ or, equivalently, the distance to the bifurcation $\delta \Vhalf$, by decreasing $\Vhalf$ by $d^-$ at rate $\gamma$ in the phases between two APs and deterministically increasing it by $d^+$ when an AP is fired.
Simulating the dynamics with feedback shows the desired adaptation property:
When a temperature step $T_0 {\rightarrow} T_0 {+} \delta T$ is applied after the system has equilibrated to the original temperature, the AP frequency changes transiently but then returns to pre-stimulus level $\gamma d^-/d^+$; see Fig.~\ref{fig:adaptation} and compare with Ref.~\cite{goris_infrared_1967}.
This behavior is equivalent to the signaling protein CheA in \textit{E.\ coli} chemotaxis for which effectively the same type of feedback mechanism (only not in the context of critical behavior) has been suggested as a means to achieve robust adaptation~\cite{barkai_robustness_1997}.
Similar kinds of homeostatic (or control-theoretic) mechanisms also appear to be relevant in other contexts ranging from neural plasticity~\cite{zierenberg_homeostatic_2018, kinouchi_mechanisms_2020, zeraati_self-organization_2021} and regulation of ion channel conductances~\cite{oleary_temperature-robust_2016} to sensitivity in the auditory system~\cite{milewski_homeostatic_2017}.
While we currently do not have a precise idea how an order-parameter feedback in snakes could be implemented physiologically, it seems plausible that it is mediated by intracellular calcium and pip lipids.
In particular, TRP channels are known to be regulated by pip lipids~\cite{ramsey_introduction_2006, rohacs_regulation_2007,diaz-franulic_allosterism_2016,hasan_ca2_2018,vangeel_transient_2019,taberner_trp_2015,harraz_pip2_2020}, which themselves depend on calcium dynamics~\cite{hasan_ca2_2018,vangeel_transient_2019} and thereby plausibly on the firing of action potentials.

\vspace{10pt}
\noindent \textbf{Discussion}

Here we have presented a plausible model for sensitive thermosensing in snake pit organs.
Our model explains how information distributed in many individual ion channels can be read out into an amplified, collective response which can naturally be a thousand times more sensitive than individual channels.
In our model the collective response takes advantage of the diverging susceptibility near a dynamical bifurcation.
The system uses a motif inspired by self-organized criticality~\cite{sornette_critical_1992, sornette_mapping_1995, moreau_balancing_2003} to maintain itself in the narrow, highly amplified region close to the bifurcation -- even while ambient conditions vary over much larger temperature changes, albeit on slower timescales.

Using information theory we trace how information is lost as it is integrated from many channels into AP firing.
We find that proximity to the bifurcation is not needed in principle to preserve information.
The full voltage dynamics $V(t)$ preserve most of the information that is contained in individual channel currents, and an order one fraction of this information is also preserved in the timing of individual spikes, at least in the regime of regular action potentials.
However, except near the bifurcation, reading out this information requires a downstream sensor to be able to distinguish 0.1\% changes in the rate of action potential firing.
Only very near the bifurcation can the signal be arbitrarily amplified with information that is accessible even with an imprecise readout mechanism or in the presence of additional noise in the firing of APs.

Is there a natural way for the voltage dynamics to maintain itself close to the bifurcation where information about temperature is most accessible?
Motivated by work on self-organized criticality~\cite{sornette_critical_1992, sornette_mapping_1995} and control theory~\cite{moreau_balancing_2003}, we suggest that feedback from the order parameter, the AP frequency, onto the control parameter, the voltage at half maximum $\Vhalf (T)$, robustly self-tunes the system to the bifurcation.
In the most simple case, $\Vhalf (T)$, a single-channel property, gets increased (decreased) on average if the AP frequency, a collective property of the system, is above (below) a certain ``target value".
This negative feedback is particularly robust since the diverging susceptibility at the bifurcation point not only leads to high signal amplification but also ensures that the control parameter stays close to its critical value even if the target value varies, see Fig.~\ref{fig:adaptation}b.

Nonetheless, due to the intrinsic noise in the time between two APs (and the ensuing stochasticity in the feedback),  the control parameter exhibits fluctuations around its steady-state value (blue shaded in Fig.~\ref{fig:adaptation}c).
These fluctuations fundamentally limit the maximal sensitivity of the system:
Only temperature changes that (transiently) shift the distance to the bifurcation by considerably more than the typical deviation of the control parameter from the steady-state value can be detected.
As expected, we find a tradeoff between the speed of the adaptation/feedback and the sensitivity:
A more sensitive response requires a larger adaptation time or, conversely, a maximal adaptation time limits the possible sensitivity, see SI.
Furthermore, the control parameter constitutes an unbiased estimator of temperature at steady-state and therefore its precision can be at most the Fisher information (Cramer-Rao bound).
Indeed, as we show in the SI, the typical fluctuations in the order parameter due to intrinsic noise in the feedback scale exactly inversely to the information fidelity and are thus minimal close to the bifurcation if AP firing is subject to additional noise, cf.\ Fig.~\ref{fig:info}.

Feedback from an order onto the control parameter also appears to have relevance in other biological homeostatic mechanisms.
First, when zooming out from the level of a single neuron to networks of neurons, homeostatic plasticity (or spike-rate homeostasis) has been suggested as a means to self-organize the networks close to a critical state with benefits for information processing~\cite{kinouchi_mechanisms_2020, zeraati_self-organization_2021}.
It corresponds to a negative feedback mechanism where the activity of a neuron, a collective property of the neuronal network, suppresses the local synaptic (coupling) strength.
One particular model is the so-called LHG model~\cite{levina_dynamical_2007, kinouchi_mechanisms_2020}, where the synaptic coupling strength (or control parameter) constantly increases in the background while it gets reduced when a presynaptic AP is fired.
It has been shown that this system exhibits self-organized quasi-criticality~\cite{bonachela_self-organization_2010}, sawtooth oscillations that hover around the critical point -- similar to what we see for the control parameter in the case of feedback, see SI.
Similarly, it has been shown that an anti-Hebbian learning rule, where the connectivity matrix evolves according to the neurons' activities, can lead to a self-organized dynamical critical state with infinite susceptibility to external inputs~\cite{magnasco_self-tuned_2009, landmann_self-organized_2021}.
Finally, feedback from the collective output of the signaling system has also been discussed beyond neural systems, for instance as a way to achieve robust adaptation in the context of \textit{E.\ coli} chemosensing~\cite{barkai_robustness_1997} and for self-tuned criticality in the context of hair cells/the auditory system~\cite{camalet_auditory_2000}.

How plausible is such a type of feedback in the context of snakes?
While we are not aware of specific evidence that the opening probability of single channels changes when AP frequency is modulated, it has been suggested that TRP channels act as integrators of calcium signaling~\cite{hasan_ca2_2018}:
TRP channels are activated by membrane phosphatidylinositol-4,5-biphosphate PIP$_2$, which in turn is suppressed by calcium influx, for instance following an AP.
One could thus imagine a scenario in which pip lipids are constantly produced but are rapidly degraded whenever an AP is fired.
In this case, the AP frequency, the order parameter, would directly feed back onto single channel activity, the control parameter, in a negative feedback loop, as desired.
One way to test this hypothesis could be to artificially elicit action potentials in isolated trigeminal nerve fibers of pit vipers, e.g., by injecting currents, and to investigate whether this modification will lead to a (temporary) decrease in AP firing when the injection of currents is paused.
In a similar spirit, one interesting consequence of having order-parameter feedback is that in more complex systems where the order parameter (AP frequency) is affected by various processes and signals, responses to different stimuli get coupled.
In particular, removing a positive, long-lasting stimulus might transiently desensitize the system towards other stimuli.

While the pit organ is ideal for understanding the integration of molecular information since the purpose of the signaling system is very specific and well-defined, the general motif of embedding microscopic sensors in a dynamical system operating close to a bifurcation or critical point may be more common in sensory systems, as also suggested in other contexts~\cite{camalet_auditory_2000, pascual_criticality_2005, Veatch2008, mora_are_2011, machta_critical_2012, shew_functional_2013,  hidalgo_information-based_2014, kimchi_ion_2018, munoz_colloquium_2018, stanoev_organization_2020, autorino_critical_2022, obyrne_how_2022, graf_thermodynamic_2022}
Indeed, other biological systems also amplify weak signals, through mechanistically diverse mechanisms that nonetheless share common features.
For example, \textit{E.\ coli} are able to respond to  1-10\% changes in ligand concentration to climb shallow chemical gradients~\cite{sourjik_receptor_2002}.
This system must also operate over several orders of magnitude in background concentration~\cite{bray_receptor_1998}, a range which is four to five orders of magnitude larger than their sensitivity.
In another example, in the inner ear, faint sounds are mechanically amplified by active processes tuned close to a Hopf bifurcation~\cite{choe_model_1998, eguiluz_essential_2000, camalet_auditory_2000}.
Proximity to the bifurcation enables the ear's enormous dynamic range and frequency discrimination and explains a wide array of observed nonlinearities and the emission of sound in quiet environments~\cite{camalet_auditory_2000,hudspeth_critique_2010,reichenbach_physics_2014}.

One drawback of being near critical points is that there the dynamics suffer from long equilibration and autocorrelation times.
It has been argued that due to this so-called ``critical slowing-down" the rate of information is minimized at criticality even though the susceptibility there is highest~\cite{vennettilli_multicellular_2020}.
While our system exhibits some characteristic features of critical slowing down on the level of the voltage dynamics (the restoring force is zero at the bifurcation), action potentials are either uncorrelated (without adaptation, in which case AP firing corresponds to a Poisson process~\cite{bialek_coding_1990}) or anti-correlated (with adaptation) and the rate of accessible information is highest at the bifurcation, see Fig.~\ref{fig:info}.

One reason why information is most accessible at the bifurcation is due to its relative robustness to noise.
In contrast to the regime of regular firing of APs, the bifurcation regime is inherently noisy and information is contained in highly amplified changes in the AP frequency. 
Additional readout noise thus does not considerably change the sensing capability of the system.
While this robustness against readout noise (and noise in the feedback, see before) is certainly promising, 
there are many other sources of noise that we have not considered here and that might exhibit different robustness properties.
A careful characterization of these would likely require to take into account the spatial distribution of channels within the nerve ending and to determine fundamental limits for temperature sensitivity in the pit membrane~\cite{bialek_physical_1987}.
Furthermore, we here assumed that all noise sources are delta-correlated.
This assumption is likely not true for temperature fluctuations in the pit membrane that plausibly have equilibration timescales much longer than the typical time between two action potentials, see also~\cite{goldwyn_what_2011} for a discussion of noise timescales in the context of AP dynamics.
Finally, while TRP channels seem to have an intrinsic voltage dependence, one could also imagine a system where TRP channels are temperature- but not voltage-gated and instead trigger the opening of other voltage-gated channels.
Preliminary analysis of such a system suggests that the addition of non-TRP channels does not change the bifurcation diagram.
However, additional channels do lower the available information -- the effective noise level roughly doubles if current fluctuations through TRP and non-TRP channels are of the same order of magnitude, see SI. 

Strikingly, the pit organ appears to have evolved independently in snakes and in vampire bats, which use a very similar structure to find blood vessels in the mammals whose blood they feed on~\cite{kurten_thermoperception_1982}.
Vampire bats adapted TRPV1 channels rather than TRPA1, even though both TRPA and TRPV family channels are present in trigeminal neurons of these species~\cite{gracheva_ganglion-specific_2011}.
This observation may suggest that both types of TRP channels, and perhaps all TRP channels, can be coupled together into a bifurcation like the one we propose.
Indeed, the voltage dependence that is crucial for modulating positive feedback in our model is shared by all TRP channels, including both hot- and cold-activating and non-thermally sensitive TRP channels~\cite{nilius_gating_2005}, possibly indicating a common mechanism for amplification among distinct TRP subfamilies.

\vspace{10pt}

\section*{Acknowledgments}
We thank Michael Abbott, Elena Gracheva, Nirag Kadakia, Thomas Shaw, Sarah Veatch, and the members of the Machta group for helpful discussions, and Michael Abbott, Thierry Emonet, Elena Gracheva, and Henry Mattingly for constructive comments on the manuscript. 
This work was supported by NIH R35GM138341 (IRG, BBM), a Simons Investigator award (BBM), and by the Deutsche Forschungsgemeinschaft (DFG, German Research Foundation) – Projektnummer 494077061 (IRG).

\clearpage

\begin{center}
\textbf{\large Supplemental Material to\\ A bifurcation integrates information from many noisy ion channels} \\
\vspace{10pt}
Isabella R.\ Graf$^{1\star}$,
Benjamin B. Machta$^{1,2\dagger}$
\\
\medskip
\noindent\upshape$^{1}$Department of Physics, Yale University, New Haven, Connecticut 06511, USA
\\
\noindent\upshape$^{2}$Quantitative Biology Institute, Yale University, New Haven, Connecticut 06511, USA
\\
\noindent\upshape$^{\star}$ isabella.graf@yale.edu
\\
\noindent\upshape$^{\dagger}$ benjamin.machta@yale.edu
\end{center}
\setcounter{equation}{0}
\setcounter{figure}{0}
\setcounter{table}{0}
\setcounter{section}{0}
\makeatletter
\renewcommand{\theequation}{S\arabic{equation}}
\renewcommand{\thefigure}{S\arabic{figure}}
\renewcommand{\thesection}{S\arabic{section}}

\vspace{10pt}

In this Supplemental Material we derive the analytic expressions for the AP frequency and Fisher information rate as stated in the main text (including a brief discussion of the limiting case $\rho \rightarrow 1/4$) and provide more details about how information is lost when coarse-graining from the channels' kinetics to the voltage dynamics and further to the AP firing statistics.
Furthermore, we comment on a few scenarios for maximizing the amplification and information rate under different constraints and discuss the information fidelity for the case when AP timings are subject to constant readout noise (or a combination of constant and scaled readout noise).
After a brief discussion of plausible parameter values and our choice of parameters for the time trace shown in the main text, we then dedicate one section to a more thorough discussion of the adaptation process (the order-parameter feedback) including the occurrence of sawtooth oscillations and an estimate for the feedback-driven fluctuations in the control parameter.
We end with a section about a modified version of our model with two types of channels that are either voltage- or temperature-gated, respectively.

\clearpage

\renewcommand*{\contentsname}{Table of Contents for SM}
\addtocontents{toc}{\setlength{\cftsecnumwidth}{3em}}
\addtocontents{toc}{\setlength{\cftsubsecnumwidth}{3em}}
\setcounter{tocdepth}{2}
\tableofcontents

\clearpage

\section{Nondimensionalization of the voltage dynamics close to the bifurcation}
\label{SMsec:Nondimensionalize}

In this section, we show how our model for the voltage dynamics in a single nerve fiber can be written in non-dimensionalized form in terms of a single scaling variable close to the bifurcation point.
As explained in the main text, for the voltage $V$ across the membrane we have the following differential equation:
\begin{align}
	    &\der{V}{t} = \underbrace{\frac{1}{\taurest} \left(\Vrest {-}V\right) {+} \frac{\ilocal}{\cm} \popen (V,T)}_{=:v(V)}
	    + \frac{\ilocal}{\cm}\sqrt{\frac{\popen (V,T) \left( 1{-}\popen (V,T)\right)}{N}}  \xilocal\left(\frac{t}{\tauopen}\right) {+} \frac{\iglobal}{\cm}\xiglobal \left(\frac{t}{\tauglobal}\right), \label{SMeq:dynamics}
\end{align}
with the opening probability $\popen (V,T) = \left( 1+e^{-(V-\Vhalf(T))/\Delta V}\right)^{-1}$, deterministic drift $v(V)$ and independent white noise sources $\xilocal, \xiglobal$ with unit variance (due to intrinsic channel noise and extrinsic noise, respectively):
\begin{gather}
    \avg{\xilocal(s) \xilocal(s')} = \delta(s-s') =\avg{\xiglobal(s) \xiglobal(s')}\\
    \avg{\xilocal(s) \xiglobal(s')} =0.
\end{gather}
To understand the scaling behavior of the system close to the bifurcation (where the (local) minimum of the drift term $v$ attains a value of $v(\Vmin) = 0$ and a second stable fixed point appears), we first expand the drift around its minimum:
\begin{align}
    v(V) = \left[\frac{\ilocal}{2 \cm} \left( 1-\sqrt{1-4\rho}\right) - \frac{1}{\taurest} \left( \Vmin-\Vrest\right) \right]+ \frac{1}{2 \taurest \Delta V} \sqrt{1-4\rho}\left( V- \Vmin \right)^2 + \mathcal{O} \left( \left( V-\Vmin \right)^3 \right),
    \label{SMeq:drift}
\end{align}
where 
\begin{align}
    \Vmin = \Vhalf + \Delta V \log \left[ -1+ \frac{1}{2 \rho} \left(1- \sqrt{1-4\rho} \right)\right]
\end{align}
with 
\begin{align}
    \rho = \frac{\cm \Delta V}{\ilocal 
    \taurest}.
\end{align}
Note that we assume throughout that $\rho < 1/4$.
This condition ensures that the maximal slope of $(\ilocal/\cm) \popen (V,T)$ is always larger than $1/\taurest$ and there is a bifurcation where the number of fixed points changes between one and three.
At the bifurcation, the constant term in the drift, Eq.~\ref{SMeq:drift}, is zero and we identify the voltage at half-maximum at the bifurcation to be
\begin{align}
    \Vhalfbif = \frac{\ilocal \taurest}{2 \cm} \left( 1-\sqrt{1-4\rho}\right) +\Vrest  - \Delta V \log \left[ -1+ \frac{1}{2 \rho} \left(1- \sqrt{1-4\rho} \right)\right].
    \label{SMeq:half_voltage_bifurcation}
\end{align}
In terms of the distance $\delta \Vhalf$ to the bifurcation,
\begin{align}
    \delta \Vhalf = \Vhalfbif - \Vhalf,
\end{align}
the drift then simplifies to
\begin{align}
    v(V) = \frac{1}{\taurest} \delta \Vhalf + \frac{1}{2 \taurest \Delta V} \sqrt{1-4\rho}\left( V- \Vmin \right)^2 + \mathcal{O} \left( \left( V-\Vmin \right)^3 \right).
    \label{SMeq:drift_second_order}
\end{align}
Ignoring the higher-order terms, the voltage dynamics is given by
\begin{align}
    \der{V}{t} =  \frac{1}{\taurest} \delta \Vhalf + \frac{1}{2 \taurest \Delta V} \sqrt{1-4\rho}\left( V- \Vmin \right)^2 + \frac{\Delta V}{\sqrt{\taurest}} \nu \ \hat{\xi} \left(t\right),
    \label{SMeq:dynamics_close_to_bif}
\end{align}
with unit-variance white noise $\hat{\xi}$.
Here we have combined the two independent white noise processes into a single one with summed variance and have used that for any white-noise process
\begin{align}
    \xi \left(\frac{t}{\tau} \right) = \sqrt{\tau} \hat{\xi} (t),
\end{align}
which can be seen from
\begin{align}
    \avg{\xi \left( \frac{t}{\tau}\right) \xi \left( \frac{t'}{\tau}\right)} = \delta \left(\frac{t-t'}{\tau}\right) = \tau \delta \left(t-t'\right) = \tau \avg{ \hat{\xi} (t) \hat{\xi} (t')}.
\end{align}
Furthermore, we have defined the effective noise
\begin{align}
    \nu = \sqrt{\frac{\tauopen}{N \taurest \rho} + \frac{\iglobal^2 \tauglobal \taurest}{\left( \Delta V \right)^2 \cm^2}} = \sqrt{\Nm^{-1} + \Next^{-1}} = \sqrt{\Neff^{-1}},
    \label{SMeq:effective_noise}
\end{align}
which scales like the inverse square root of the effective number $\Neff$ of independent measurements that are available during the membrane relaxation time $\taurest$.
For zero extrinsic noise $\Next=\infty$, this effective number equals the number $\Nm$ of measurements that $N$ independent channels can make during that time.
Note that here we made the simplifying assumption that the noise term is independent of voltage, by taking $\popen (V,T)$ to be constant and setting it to its value at the arg minimum $\Vmin$ of the drift, $ \popen (V,T)= \popen (\Vmin, T)$.
As a result, $\popen (V,T) (1-\popen (V,T)) = \rho$.
Since our analytic results agree well with results from stochastic simulations, see Fig.~2 in the main text, this assumption seems to be justified a posteriori.

Equation~\ref{SMeq:dynamics_close_to_bif} can be recast in non-dimensionalized form by rescaling time and voltage,
\begin{align}
    s=\sigma t = t/\taus \hspace{50pt} u=\omega \left( V-\Vmin\right) = \left( V-\Vmin\right)/\Vs.
    \label{SMeq:rescale_time_volt}
\end{align}
In terms of these non-dimensionalized variables, we find
\begin{align}
    \der{u}{s} = \frac{\omega}{\sigma} \left[ \frac{1}{\taurest} \delta \Vhalf + \frac{1}{2 \taurest \Delta V} \sqrt{1-4\rho} \frac{1}{\omega^2} u^2 + \frac{\Delta V}{\sqrt{\taurest}} \nu \ \sqrt{\sigma} \tilde{\xi} \left(s\right)\right],
\end{align}
where we again used that $\hat{\xi}\left(\frac{s}{\sigma} \right) = \sqrt{\sigma} \tilde{\xi} \left( s\right)$.
We can now choose $\omega$ and $\sigma$ to set the coefficients in front of $u^2$ and $\tilde{\xi} (s)$ to zero:
\begin{align}
    \frac{1}{2 \taurest \Delta V} \sqrt{1-4\rho} \frac{1}{\omega \sigma} &\overset{!}{=} 1 \\
    \frac{\Delta V}{\sqrt{\taurest}} \nu \frac{\omega}{\sqrt{\sigma}} &\overset{!}{=} 1.
\end{align}
This choice gives
\begin{align}
    \omega &= 
    \left( \frac{1-4\rho}{4 \nu^4}\right)^{1/6} \frac{1}{\Delta V} = \frac{1}{\Vs} \\
    \sigma &= \left( \frac{\nu^2 \left( 1-4\rho \right)}{4 }\right)^{1/3} \frac{1}{\taurest}= \frac{1}{\taus}.
    \label{SMeq:rescale_parameters}
\end{align}
The rescaled dynamics is then
\begin{align}
    \der{u}{s} = \alpha + u^2 + \tilde{\xi} (s) =: f(u) + \tilde{\xi} (s),
    \label{SMeq:dynamics_rescaled}
\end{align}
with the scaling variable
\begin{align}
    \alpha = \frac{\omega}{\sigma} \frac{1}{\taurest} \delta \Vhalf = \frac{\taus}{\taurest} \frac{\delta \Vhalf}{\Vs}  = \left( \frac{4}{1-4\rho} \right)^{1/6} \frac{1}{\nu^{4/3}}\frac{\delta \Vhalf}{\Delta V},
    \label{SMeq:scaling_variable_alpha}
\end{align}
which is linear in the distance to the bifurcation $\delta \Vhalf$ (relative to the width of the sigmoid $\Delta V$) and scales like $\nu^{-4/3}$.
Lower noise thus makes the transition sharper: $\delta \Vhalf/\Delta V\sim \nu^{4/3}$ for constant $\alpha$.

\section{Action potential frequency and variance in action potential timing}
\label{SMsec:AP_frequency}

In our model, an AP is fired if the voltage crosses a threshold, which we assume to be close to but to the left of the high-voltage fixed point~\footnote{We assume throughout that the temperature is not so low that there is only a low-voltage fixed point. That is, we focus on the monostable case with a high-voltage fixed point and the bistable case with a low- and a high-voltage fixed point.}, and then the system is reset to the membrane resting potential.
Close to the bifurcation, the minimal drift between the resting potential and the threshold is close to zero.
We therefore expect that the time it takes the voltage to reach the threshold after a reset (the time $\tap$ between two action potentials) is dominated by the region around the minimum.
Correspondingly, we approximate $\tap$ by the time it takes the system to go from $V=-\infty$ to $V=\infty$.
To this end, following Ref.~\cite{hathcock_reaction_2021}, we rewrite the non-dimensionalized dynamics, Eq.~\ref{SMeq:dynamics_rescaled}, in terms of the corresponding backwards Fokker-Planck equation that can then be used to calculate the (moments of the) first-passage time $\tap$:
\begin{align}
    \partial_s p \left(\left.u,s\right| u_0 \right) = f(u_0) \partial_{u_0} p \left(\left.u,s\right| u_0 \right) + \frac{1}{2} \partial_{u_0}^2 p \left(\left.u,s\right|  u_0 \right),
    \label{SMeq:backwards_FPE}
\end{align}
where $p \left(\left.u,s\right| u_0 \right)=p \left(\left.u(s)=u\right| u(0) = u_0 \right)$ denotes the probability density that $u(s) = u$ given that the process starts at $u_0$ at $s=0$. 
Note that the derivatives on the right-hand side are with respect to the initial condition (rather than the current value of $u$).

In general, the first passage time $S_{u_f|u_0}$ is defined as the time it takes a process starting at $u_0$ at $s=0$ to first cross the value $u_f > u_0$.
Its probability density $P_{S,u_f|u_0} (s)$ can be written in terms of the survival probability,
\begin{align}
    \Pi(s; u_f|u_0) = \int_{-\infty}^{u_f} \mathrm{d} u' p(u',s|u_0),
    \label{SMeq:survival_prob}
\end{align}
as
\begin{align}
    P_{S,u_f|u_0} (s) = - \der{\Pi(s;u_f|u_0)}{s}.
\end{align}
Using the backwards Fokker-Planck equation, Eq.~\ref{SMeq:backwards_FPE}, gives
\begin{align}
    -1 &= \int_{-\infty}^{u_f} \mathrm{d} u' \left[ p(u',\infty|u_0)- p(u',0|u_0)\right] = \int_{-\infty}^{u_f} \mathrm{d} u' \ \int_0^{\infty} \mathrm{d} s \ \partial_s p(u',s|u_0) = \\
    &=\int_{-\infty}^{u_f} \mathrm{d} u'  \ \int_0^{\infty} \mathrm{d} s \ \left[ f(u_0) \partial_{u_0} p \left(\left.u',s\right| u_0 \right) + \frac{1}{2} \partial_{u_0}^2 p \left(\left.u',s\right|  u_0 \right)\right] =\\
    &= \int_0^{\infty}\mathrm{d} s \ \left[ f(u_0) \partial_{u_0} \Pi \left(s;u_f| u_0 \right) + \frac{1}{2} \partial_{u_0}^2 \Pi \left(s;u_f| u_0 \right)\right] = \\
    &= f(u_0) \partial_{u_0} \avg{S_{u_f|u_0}} + \frac{1}{2} \partial_{u_0}^2 \avg{S_{u_f|u_0}}. \label{SMeq:first_passage_mean_diff_eq}
\end{align}
The first equality is true since $p(u',\infty|u_0) = 0$ for all $u' <u_f$ (the system eventually crosses the threshold $u_f$) and $\int_{-\infty}^{u_f} \mathrm{d} u' p(u',0|u_0)=1$ (the system starts to the left of $u_f$ at $u_0<u_f$).
Furthermore, in the last step, we made use of the following equality,
\begin{align}
    \avg{S_{u_f|u_0}} &= \int_0^{\infty} \mathrm{d} s \ s P_{S,u_f|u_0} (s) = -\int_0^{\infty} \mathrm{d} s \ s \der{\Pi(s;u_f|u_0)}{s} = -\left[s \Pi(s;u_f|u_0)\right]_{0}^{\infty}+\int_0^{\infty} \mathrm{d} s \ \Pi(s;u_f|u_0) = \\
    &=\int_0^{\infty} \mathrm{d} s \ \Pi(s;u_f|u_0).
\end{align}
This equation also allows for the derivation of a differential equation for the mean squared first passage time.
More specifically,
\begin{align}
    -\avg{S_{u_f|u_0}}  &= \int_0^{\infty} \mathrm{d} s \ s \der{\Pi(s;u_f|u_0)}{s} = \int_{-\infty}^{u_f} \mathrm{d} u' \int_0^{\infty} \mathrm{d} s \ s \ \partial_s p(u',s|u_0) = \\
    &=\int_{-\infty}^{u_f} \mathrm{d} u' \int_0^{\infty} \mathrm{d} s \ s \left[f(u_0) \partial_{u_0} p \left(\left.u',s\right| u_0 \right) + \frac{1}{2} \partial_{u_0}^2 p \left(\left.u',s\right|  u_0 \right) \right] =\\
    &= \left[f(u_0) \partial_{u_0}  + \frac{1}{2} \partial_{u_0}^2\right] \int_{-\infty}^{u_f} \mathrm{d} u' \int_0^{\infty} \mathrm{d} s \ s \ p \left(\left.u',s\right|  u_0 \right) = \frac{1}{2}\left[f(u_0) \partial_{u_0}  + \frac{1}{2} \partial_{u_0}^2\right] \avg{S^2_{u_f|u_0}},
\label{SMeq:first_passage_squared_diff_eq}
\end{align}
where we plugged in the definition of the survival probability, Eq.~\ref{SMeq:survival_prob}, and the backwards Fokker-Planck equation, Eq.~\ref{SMeq:backwards_FPE}.
Furthermore, we used that
\begin{align}
    \int_{-\infty}^{u_f} \mathrm{d} u' \int_0^{\infty} \mathrm{d} s \ s \ p \left(\left.u',s\right|  u_0 \right) &= \int_0^{\infty} \mathrm{d} s \ s \ \Pi(s; u_f|u_0) = \left[ \frac{1}{2} s^2 \Pi(s; u_f|u_0) \right]_0^{\infty} - \int_0^{\infty} \mathrm{d} s \ \frac{1}{2} s^2  \ \der{\Pi(s; u_f|u_0)}{s} = \\
    &=\int_0^{\infty} \mathrm{d} s \ \frac{1}{2} s^2 P_{S, u_f|u_0} (s) = \frac{1}{2} \avg{S^2_{u_f|u_0}}.
\end{align}

The differential equations for the mean first-passage time, Eq.~\ref{SMeq:first_passage_mean_diff_eq}, and the mean squared first-passage time, Eq.~\ref{SMeq:first_passage_squared_diff_eq}, are of the form 
\begin{align}
    -H(u_0) = f(u_0) Q'(u_0) + \frac{1}{2} Q''(u_0)
    \label{SMeq:first_passage_general_diff}
\end{align}
and can be solved with the ansatz
\begin{align}
    Q'(u_0) = c(u_0) e^{2 V(u_0)},
    \label{SMeq:ansatz}
\end{align}
where $V$ is the potential corresponding to the drift, $f(u_0) = -V'(u_0)$, and $Q'(u_0) = c e^{2 V(u_0)}$ solves the homogeneous equation $-V'(u_0) Q'(u_0) = \frac{1}{2} Q''(u_0)$.
Plugging in this ansatz, Eq.~\ref{SMeq:ansatz}, into the differential equation, Eq.~\ref{SMeq:first_passage_general_diff}, yields
\begin{align}
    c'(u_0) = -2 H(u_0) e^{-2 V(u_0)}.
\end{align}
Both the mean first-passage time and the mean squared first-passage time satisfy $Q'(-\infty) = 0$ and $Q(u_f)=0$, and so by integrating we find
\begin{align}
    Q(u_0) = \int_{u_0}^{u_f} \mathrm{d} u'' \ \int_{-\infty}^{u''} \mathrm{d} u' \ 2 H(u') e^{-2(V(u')-V(u''))}.
\end{align}

For the mean first-passage time, $H(u_0) = 1$, and for the mean squared first-passage time, $Q(u_0) = 2 \avg{S_{u_f|u_0}}$, so we find
\begin{align}
    \avg{S_{u_f|u_0}} &= 2 \int_{u_0}^{u_f} \mathrm{d} u'' \ \int_{-\infty}^{u''} \mathrm{d} u' \ e^{-2(V(u')-V(u''))}\\
    \avg{S^2_{u_f|u_0}} &= 4 \int_{u_0}^{u_f} \mathrm{d} u'' \ \int_{-\infty}^{u''} \mathrm{d} u' \ \avg{S_{u_f|u'}} e^{-2(V(u')-V(u''))} = \\
    &=8 \int_{u_0}^{u_f} \mathrm{d} u'' \ \int_{-\infty}^{u''} \mathrm{d} u' \
    \int_{u'}^{u_f} \mathrm{d} u'''' \
    \int_{-\infty}^{u''''} \mathrm{d} u''' \
    e^{-2(V(u')-V(u'')+V(u''')-V(u''''))},
    \label{SMeq:first_passage_integral_expressions}
\end{align}
where $V(u) = - \alpha u - u^3/3$.

As mentioned before, the drift term for $u \rightarrow \pm \infty$ is large and we will approximate the first passage time by the time it takes the system to go from $u_0 = -\infty$ to $u_f = \infty$.
In this case, we find an explicit solution for the mean first-passage time and express the variance of the first-passage time in terms of a two-fold integral:
\begin{align}
    \avg{S_{\infty|-\infty}} &= 2 \int_{-\infty}^{\infty} \mathrm{d} u'' \ \int_{-\infty}^{u''} \mathrm{d} u' \ e^{-2(V(u')-V(u''))} = \int_0^{\infty} \mathrm{d} y \ \int_{-\infty}^{\infty} \mathrm{d} z \ e^{-2\left(V\left(\frac{1}{2} \left(z-y \right)\right)-V\left(\frac{1}{2} \left(z+y \right)\right)\right)} =\\
    &= 2^{1/3} \pi^2 \left( \text{A}_1^2 \left[-2^{2/3} \alpha \right]+\text{A}_2^2 \left[-2^{2/3} \alpha \right]\right) =: M(\alpha),
    \label{SMeq:first_passage_mean_analytic}
\end{align}
with the Airy functions $\text{A}_{1/2}$ of the first and second kind, respectively; see also Ref.~\cite{hathcock_reaction_2021}.

The variance of the first-passage time is
\begin{align}
    &\var{S_{\infty|-\infty}} = \avg{S^2_{\infty|-\infty}} - \avg{S_{\infty|-\infty}}^2 = \\
    &=8 \int_{-\infty}^{\infty} \mathrm{d} u'' \ \int_{-\infty}^{u''} \mathrm{d} u' \
    \int_{u'}^{\infty} \mathrm{d} u'''' \
    \int_{-\infty}^{u''''} \mathrm{d} u''' \
    e^{-2(V(u')-V(u'')+V(u''')-V(u''''))} - \\
    &-4 \int_{-\infty}^{\infty} \mathrm{d} u'' \ \int_{-\infty}^{u''} \mathrm{d} u' \
    \int_{-\infty}^{\infty} \mathrm{d} u'''' \
    \int_{-\infty}^{u''''} \mathrm{d} u''' \
    e^{-2(V(u')-V(u'')+V(u''')-V(u''''))} =\\
    &= \avg{S_{\infty|-\infty}}^2 - 8 \int_{-\infty}^{\infty} \mathrm{d} u'' \ \int_{-\infty}^{u''} \mathrm{d} u' \
    \int_{-\infty}^{u'} \mathrm{d} u'''' \
    \int_{-\infty}^{u''''} \mathrm{d} u''' \
    e^{-2(V(u')-V(u'')+V(u''')-V(u''''))}.
\end{align}
We did not manage to solve the latter (4-fold) integral analytically, but it is possible to reduce it to the following 2-fold integral that can be evaluated numerically:
\begin{align}
    \var{S_{\infty|-\infty}} = \avg{S_{\infty|-\infty}}^2 - 8 \pi \int_0^{\infty} \mathrm{d} r \ \int_0^{\pi/2} \mathrm{d} \varphi \ r \text{erfc} \left[\frac{r^3}{\sqrt{2}} \sin \varphi \cos \varphi \right] e^{-\frac{1}{48} r^6 (5+3 \cos (4 \varphi)) - 2 \alpha r^2} =: S(\alpha),
    \label{SMeq:first_passage_var_analytic}
\end{align}
with the complementary error function $\text{erfc}$.

Going back to real units, we find for the mean and variance of the timing $\tap$ between two action potentials,
\begin{align}
    \avg{\tap} &= \taus \avg{S_{\infty|-\infty}} = \taurest \left( \frac{4}{\nu^2 (1-4\rho)}\right)^{1/3} M(\alpha) = \frac{\taurest}{\nu^{2/3}} \left(\frac{4}{1-4\rho} \right)^{1/3} \ M(\alpha),\label{SMeq:AP_time}\\
    \var{\tap} &= \taus^2 \var{S_{\infty|-\infty}} = \frac{\taurest^2}{\nu^{4/3}} \left(\frac{4}{1-4\rho} \right)^{2/3} \ S(\alpha) \label{SMeq:AP_time_var}.
\end{align}

\section{Fisher information rate}
\label{SMsec:FIM_rate}

How much information about temperature is contained in the timings between action potentials?
To quantify this information, we first look at two limiting distributions, one where the AP timings are Gaussian distributed and one where they are exponentially distributed, in each case with a control parameter (temperature) dependent mean $\mu=\mu(c)$.

To quantify how much information is contained in an infinitesimal change in the mean, we look at the Fisher information of the respective probability distributions with respect to a change in the control parameter $c$.
The Fisher information (metric), $\fisherinfo$, corresponds to the Hessian matrix of the Kullback-Leibler divergence or relative entropy (with respect to the control parameter), which itself is a measure of the difference between two probability distributions.
In the case of a single control parameter $c$ defining the probability distributions $p(x|c)$, the (scalar) Fisher information can be written as 
\begin{align}
    \fisherinfo = -\avg{\partial_c^2 \log p(x|c)},
\end{align}
where the average is over the probability space, i.e., $x$.

If the probability distribution is Gaussian with mean $\mu=\mu (c)$ and variance $\sigma$, 
\begin{align}
    p(x|c) = \frac{1}{\sqrt{2 \pi \sigma^2} } e^{-\frac{(x-\mu(c))^2}{2 \sigma^2}},
\end{align}
the Fisher information is
\begin{align}
    \fisherinfo = \frac{1}{2 \sigma^2}\avg{\partial_c^2 \left(x-\mu (c)\right)^2} = \frac{1}{\sigma^2}\avg{\left(\mu'(c)\right)^2 - \mu''(c) \left( x-\mu(c)\right)}= \left(\frac{\mu'(c)}{\sigma}\right)^2.
\end{align}
Here, we used that $\avg{x} = \mu(c)$.

Similarly, for an exponential distribution with mean $\mu(c)$,
\begin{align}
    p(x|c) =\frac{1}{\mu (c)} e^{-x/\mu(c)},
\end{align}
we find
\begin{align}
    \fisherinfo = \avg{\partial_c^2 \frac{x}{\mu(c)}} + \partial_c^2 \log \left(\mu(c) \right) = \left(-\frac{\mu''}{\mu^2}+\frac{2 (\mu')^2}{\mu^3} \right) \avg{x} + \left( \frac{\mu''}{\mu} - \frac{(\mu')^2}{\mu^2}\right) = \left(\frac{\mu'(c)}{\sigma}\right)^2,
\end{align}
since $\avg{x}=\mu = \sigma$.

In both cases, the Fisher information is thus determined by the ratio of the change in mean as compared to the standard deviation.
This expression reflects the intuitive expectation: The larger the change in mean as compared to the width of the distribution, the larger the information.

\subsection{Fisher information for exponential family}
\label{SMsubsec:FIM_rate_exp_fam}

In the general case (including ours), however, it is not only the mean which depends on the control parameter (temperature), but also the higher-order moments and cumulants carry information about temperature.
The information about a temperature change contained in a change in the distribution of AP timings is thus generally higher than the information contained in just the change in the mean.
In order to gauge the magnitude of the higher-order terms, we imagine that we have knowledge about the first $k$ moments $\mu_i (c) = \avg{x^i}_{p(x|c)}, i=1,\ldots,k$ and know how they depend on the control parameter $c$, but do not know anything about the higher-order ones.
The corresponding unbiased, maximum-entropy distribution, which contains the least additional information but is consistent with the known moments, is the exponential family
\begin{align}
    p(x|c) = p\left(x|\vec{\lambda}(c) \right)= \frac{1}{Z\left(\vec{\lambda} (c)\right)} h(x) e^{-\sum_{i=1}^k \lambda_i (c) x^i}, 
\end{align}
where $\vec{\lambda}^T(c) = \left\{ \lambda_i (c) \right\}_{i=1,\ldots,k}$ are the Lagrange multipliers and $Z\left(\vec{\lambda} (c)\right)$ is the partition function
\begin{align}
    Z\left(\vec{\lambda} (c)\right) = \int \mathrm{d} x \ h(x) e^{-\sum_{i=1}^k \lambda_i (c) x^i}.
\end{align}

The Fisher information is then given by
\begin{align}
    \fisherinfo &= -\avg{\partial_c^2 \log \left( p(x|c) \right)} = - \avg{\partial_c \left[ \sum_{i=1}^k \left(\partial_{\lambda_i} \log \left( p(x|c) \right)\right) \der{\lambda_i}{c} \right]} =\\
    &=- \avg{\left[ \sum_{i=1}^k \left(\partial_{\lambda_i} \log \left( p(x|c) \right)\right) \secondder{\lambda_i}{c} \right]}- \avg{\left[ \sum_{i,j=1}^k \der{\lambda_i}{c} \left(\partial_{\lambda_i} \partial_{\lambda_j} \log \left( p(x|c) \right)\right) \der{\lambda_j}{c} \right]} = \\
    &= \sum_{i,j=1}^k \der{\lambda_i}{c} A_{ij} \der{\lambda_j}{c}
    \label{SMeq:FI_lagrange_mult}
\end{align}
where we used that
\begin{align}
    \avg{\partial_{\lambda_i} \log \left( p(x|c) \right)} = -\avg{x^i} - \partial_{\lambda_i} \log Z\left(\vec{\lambda}\right) = 0 \ \forall i=1,\ldots,k,
\end{align}
and have defined the covariance matrix
\begin{align}
    A_{ij} &= - \avg{\partial_{\lambda_i} \partial_{\lambda_j} \log \left( p(x|c) \right)} = \avg{\partial_{\lambda_i} \partial_{\lambda_j} \left[\log Z\left(\vec{\lambda}\right) +  \sum_{l=1}^k \lambda_l x^l\right]} = \\
    &=\partial_{\lambda_i} \partial_{\lambda_j} \log Z\left(\vec{\lambda}\right) 
    = \frac{1}{Z} \partial_{\lambda_i} \partial_{\lambda_j} Z  - \frac{1}{Z^2} \left( \partial_{\lambda_i} Z  \right) \left( \partial_{\lambda_j} Z  \right) = \\
    &=\avg{x^{i+j}} - \avg{x^i} \avg{x^j}.
\end{align}
The expression for the Fisher information as in Eq.~\ref{SMeq:FI_lagrange_mult} still depends on the Lagrange multipliers.
To rewrite it in terms of the moments, we realize that 
\begin{align}
    \parder{\mu_i}{\lambda_j} = - \partial_{\lambda_i} \partial_{\lambda_j} \log Z =-A_{ij},
\end{align}
since $\mu_i = \avg{x^i} = -\partial_{\lambda_i} \log Z$.
As a result, in matrix notation,
\begin{align}
    \Delta \vec{\mu} &= - A \Delta \vec{\lambda} \hspace{30pt} \text{or} \\
    \Delta \vec{\lambda} &= - A^{-1} \Delta \vec{\mu},
\end{align}
and we find
\begin{align}
    \fisherinfo = \sum_{n,i,j,m=1}^k \mu_n'(c) \left(A^{-1}\right)_{in} A_{ij} \left(A^{-1}\right)_{jm} \mu_m'(c) = \sum_{n,m=1}^k \mu_n'(c) \left(A^{-1}\right)_{nm} \mu_m'(c) = \vec{\mu}'(c)^T A^{-1} \vec{\mu}'(c)
    \label{SMeq:FI_moments}
\end{align}
for the Fisher information in terms of the changes $\mu_i'(c)$ of the moments with respect to the control parameter and the covariance matrix $A$.

\subsection{First- and second-order approximation}
\label{SMsubsec:FIM_rate_approx}

If only information about the first moment (mean) is available, $k=1$, the expression for the Fisher information, Eq.~\ref{SMeq:FI_moments}, reduces to
\begin{align}
    \fisherinfo_1 = \frac{(\mu_1')^2}{A_{11}}  = \frac{(\mu_1')^2}{\avg{x^2} - \avg{x}^2} =  \frac{(\mu_1')^2}{\mu_2 -\mu_1^2} =\frac{(\mu_1')^2}{\sigma^2},
    \label{SMeq:FIM_1st_order}
\end{align}
agreeing with what we have seen before for the examples of the Gaussian and exponential distribution.

If both the first and second moment are considered, we find the following expression for the Fisher information in terms of the cumulants $\kappa_i = \kappa_i (c), i=1,\ldots,4$:
\begin{align}
    \fisherinfo_2 = \frac{\kappa_2 \left( \kappa_2'\right)^2 - 2 \kappa_3 \kappa_1' \kappa_2'+ \left( 2 \kappa_2^2+\kappa_4\right) \left( \kappa_1' \right)^2}{2 \kappa_2^3 - \kappa_3^2 + \kappa_2 \kappa_4}.\label{SMeq:FIM_2nd_order}
\end{align}

\begin{figure*}[t]
\centering
\includegraphics[width=0.5\linewidth]{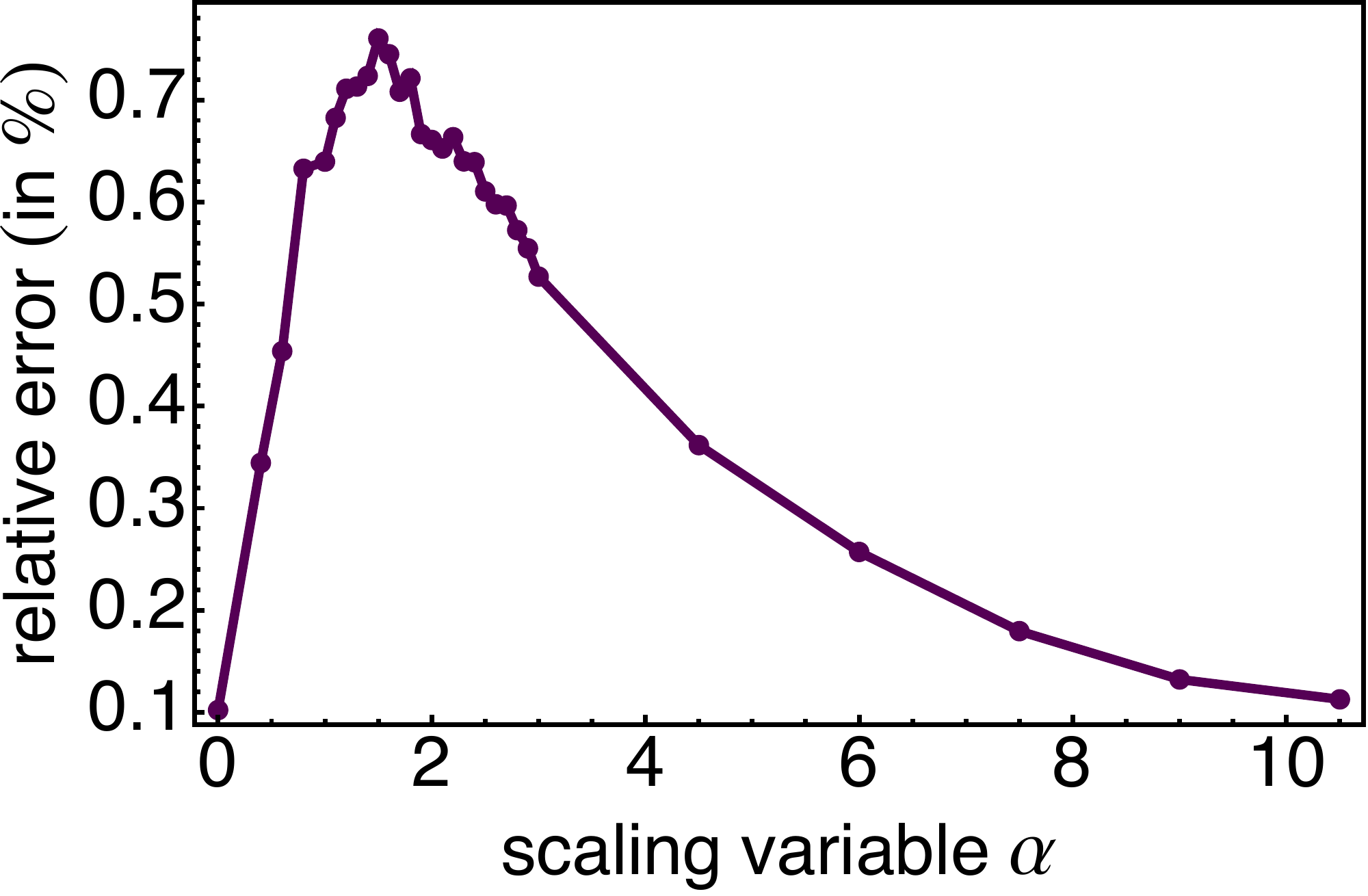}%
\caption{\label{SMfig:rel_error_FI}  Relative error $(\fisherinfo_2 - \fisherinfo_1)/\fisherinfo_1$ of the second-order approximation $\fisherinfo_2$ of the Fisher information, Eq.~\ref{SMeq:FIM_2nd_order}, with respect to the lowest-order approximation $\fisherinfo_1$, Eq.~\ref{SMeq:FIM_1st_order}.
}
\end{figure*}

How does this second-order approximation $\fisherinfo_2$ compare to the first-order approximation $\fisherinfo_1$?
Since the third and fourth cumulant require calculation of a 6- and 8-fold integral, respectively, (analogous to the 2- and 4-fold integral for the mean and variance, respectively, see Eq.~\ref{SMeq:first_passage_integral_expressions}), we cannot reliably determine them numerically.
Instead, we determine the third and fourth cumulant from stochastic simulations and use our explicit expressions for the first and second cumulants, Eqs.~\ref{SMeq:first_passage_mean_analytic},~\ref{SMeq:first_passage_var_analytic}, and their derivatives, to calculate the relative error 
\begin{align}
    \frac{\fisherinfo_2 - \fisherinfo_1}{\fisherinfo_1}.
    \label{SMeq:FIM_rel_error}
\end{align}
The result is shown in Fig.~\ref{SMfig:rel_error_FI}.
The error is largest for $\alpha \approx 1-3$ but is consistently below $1\%$, indicating that the first-order approximation $\fisherinfo_1$ might be a reasonable estimate.
Furthermore, as expected, the error is always positive since the additional knowledge of the second moment can only increase the information, $\fisherinfo_2 \geq \fisherinfo_1$.
The information (rate) as given in the main text thus always tends to underestimate the information contained in the full distribution of the times between action potentials.
At the same time, we know that the information fidelity is bounded by $1$, thus providing a (weak) upper bound.

\subsection{Fisher information rate in the AP dynamics}

The Fisher information $\fisherinfo$ discussed so far is for a single observation, i.e., a single measurement of the time between two consecutive action potentials.
Since in our model (at least without order-parameter feedback), the timings between subsequent action potentials are independent, information grows linearly with the number of observations (and therefore with time).
On average, one such observation requires a time of $\avg{\tap}$ and we can define the information rate (in the AP dynamics) as
\begin{align}
    \fisherinforate^{(T)} &= \frac{\fisherinfo}{\avg{\tap}} \approx \frac{\fisherinfo_1}{\avg{\tap}} = \frac{1}{\taus M(\alpha)} \frac{\left( \der{M(\alpha (T))}{T}\right)^2}{S(\alpha)} = \frac{1}{\taus M(\alpha)} \frac{\left( M'(\alpha) \der{\alpha}{T}\right)^2}{S(\alpha)} = \\
    &= \frac{1}{\taus} \left(\frac{\taus}{\taurest} \frac{\Delta V}{\Vs}\right)^{2} \left(\frac{\der{\Vhalf}{T}}{\Delta V}\right)^2 J(\alpha)   =\frac{1}{\taus} \left(  \frac{4 }{\nu^8 (1-4\rho)} \right)^{1/3} \left(\frac{1}{\Delta T}\right)^2 J(\alpha) = \\
    &=\frac{1}{\taurest} \frac{1}{\nu^2 (\Delta T)^2} J(\alpha) = \frac{\rho N}{\tauopen (\Delta T)^2} \frac{\Neff}{\Nm} J(\alpha)
    \label{SMeq:Fisher_info_rate_temp}
\end{align}
in the lowest-order approximation of the Fisher information, where we used that $\der{\Vhalf}{T} = - \Delta V/\Delta T$ and defined
\begin{align}
    J(\alpha) =\frac{\left( M'(\alpha)\right)^2}{M(\alpha) S(\alpha)},
    \label{SMeq:info_fidelity}
\end{align}
which we call the ``information fidelity" (see below).
The Fisher information rate for a small change in temperature $\delta T$ is then
\begin{align}
    \fisherinforate^{(T)} \left( \delta T\right)^2 = \frac{\rho N}{\tauopen} \frac{\Neff}{\Nm} J(\alpha) \left( \frac{\delta T}{\Delta T}\right)^2,
\end{align}
where $\Delta T$ denotes the width of the sigmoid opening probability with respect to temperature.

Note that the Fisher information rate as given in Equation~\ref{SMeq:Fisher_info_rate_temp} is with respect to an infinitesimal change in temperature rather than distance to the bifurcation.
The latter is 
\begin{align}
    \fisherinforate^{(V)} = \fisherinforate^{(T)} \left(\frac{\Delta T}{\Delta V} \right)^2 = \frac{\rho N}{\tauopen (\Delta V)^2} \frac{\Neff}{\Nm} J(\alpha).
    \label{SMeq:Fisher_info_rate_voltage}
\end{align}

\subsection{Fisher information in the channels' kinetics}

How does the information rate in the AP dynamics, Eq.~\ref{SMeq:Fisher_info_rate_temp}, relate to the information rate that is available in all the individual channels' kinetics?

A single channel just follows a Bernoulli distribution with (opening) probability $\popen (T,V)$.
The Fisher information for a single channel $\fisherinfochannel_1^{(T)}$ (with respect to temperature) is thus
\begin{align}
    \fisherinfochannel_1^{(T)} &= -\sum_{x=0,1} p(x|\popen) \partial_T^2 \log p(x|\popen) = - (1-\popen) \partial_T^2 \log (1-\popen) - \popen \partial_T^2 \log \popen = \\
    &= \frac{\left(\der{\popen}{T}\right)^2}{\popen (1-\popen)} = \popen (1-\popen) \frac{1}{\left(\Delta T\right)^2},
\end{align}
where we used that
\begin{align}
    \der{\popen}{T} &= \der{ }{T} \left( 1+ e^{-(V-\Vhalf(T))/\Delta V}\right)^{-1} = -\popen^2 e^{-(V-\Vhalf(T))/\Delta V} \frac{1}{\Delta V} \der{\Vhalf(T)}{T} = \\
    &=\popen^2 e^{-(V-\Vhalf(T))/\Delta V} \frac{1}{\Delta T} = \popen (1-\popen) \frac{1}{\Delta T}.
\end{align}
Evaluating the opening probability at the arg minimum of the drift term, as we have done above when we introduce the rescaled variables, Eq.~\ref{SMeq:rescale_parameters}, gives 
\begin{align}
    \fisherinfochannel_1^{(T)} = \frac{\rho}{\left(\Delta T\right)^2}
\end{align}
for the information in a single channel.
Each channel can make one independent measurement per channel opening time $\tauopen$.
Therefore, the maximal information rate with respect to temperature in all $N$ channels (if they performed independent measurements), $\fisherinforatechannel_N^{(T)}$, is given by
\begin{align}
    \fisherinforatechannel_N^{(T)} = \frac{N}{\tauopen} \fisherinfochannel_1^{(T)} = \frac{N \rho}{\tauopen \left( \Delta T \right)^2}. \label{SMeq:Fisher_info_rate_all_channels}
\end{align}
Similarly, the one with respect to a change in the distance to the bifurcation $\delta \Vhalf$ is 
\begin{align}
    \fisherinforatechannel_N^{(V)} = \frac{N}{\tauopen} \fisherinfochannel_1^{(V)} = \frac{N \rho}{\tauopen \left( \Delta V \right)^2}. \label{SMeq:Fisher_info_rate_voltage_all_channels}
\end{align}

\subsection{Information fidelity}

Combining Equations~\ref{SMeq:Fisher_info_rate_temp},~\ref{SMeq:Fisher_info_rate_all_channels}, the Fisher information rate in the AP dynamics (with respect to temperature) can be expressed in terms of the Fisher information in all channels as
\begin{align}
    \fisherinforate^{(T)} = \fisherinforatechannel_N^{(T)} \frac{\Neff}{\Nm} J(\alpha),
\end{align}
or, analogously for the distance to the bifurcation/voltage,
\begin{align}
    \fisherinforate^{(V)} = \fisherinforatechannel_N^{(V)} \frac{\Neff}{\Nm} J(\alpha).
\end{align}
The ``information fidelity" $J(\alpha)$ thus quantifies what fraction of the information contained in all the channels is still available on the level of the AP dynamics.
More precisely, as we will explain in more detail in the next section, $\fisherinforatechannel_N \Neff/\Nm$ is the information rate one would obtain if one were to measure the entire voltage dynamics.
Thus, $J(\alpha)$ quantifies how much information is lost when instead of the entire voltage dynamics only the timings of action potentials are recorded.
Since information can only be lost during the coarse-graining step from the voltage dynamics to the AP dynamics, $J(\alpha)$ always has to be $\leq 1$, as we observe when numerically evaluating the analytic integral-expression for $J(\alpha)$ or determining it from simulations.

In the next section, we discuss in more detail how information is lost in the different coarse-graining steps (from channels to voltage to AP dynamics).

\section{Information transmission}
\label{SMsec:info_transmission}

To understand why only some of the information contained in the channels is available on the level of action potentials and where the remaining information is lost, we first consider the information contained in the full voltage dynamics.
We will see that the voltage dynamics contains the same information as the channels (except for corruption by extrinsic noise).
We will then show that the timing of action potentials is, however, not a sufficient statistic of the voltage dynamics, so that information is lost when only the timing of action potentials is recorded instead of the full voltage dynamics.

\subsection{Information in full voltage dynamics}
\label{SMsubsec:info_full_voltage}

Suppose if, instead of measuring just the timings of action potentials, the whole voltage dynamics were recorded, how much information would one obtain about temperature?
To answer this question, we consider the rescaled dynamics, Eq.~\ref{SMeq:dynamics_rescaled},
\begin{align}
    \der{u}{s} = \alpha + u^2 + \tilde{\xi}(s),
\end{align}
and consider the joint probability density
\begin{align}
    P \left(\vec{u} | \alpha(T)\right) = P\left(\{u_0, u_1, \ldots, u_M \}| \alpha(T)\right)
\end{align}
for the dynamics to attain values $u(s_i)=u_i$ at times $s_i = i \Delta s$. 
The condition makes explicit that the non-dimensionalized parameter/the scaling variable $\alpha = \alpha (T)$, Eq.~\ref{SMeq:scaling_variable_alpha}, depends on temperature $T$.
If the (rescaled) voltage is measured at all times $s_i, i=0,\ldots, M$, the Fisher information (with respect to temperature) is
\begin{align}
    \fisherinfovoltage_{M,\Delta s}^{(T)} = -\int \mathrm{d}^{M+1} \vec{u} \ P \left(\vec{u} | \alpha(T)\right) \partial_T^2 \log P \left(\vec{u} | \alpha(T)\right).
\end{align}
Since the voltage dynamics is a Markov process, the joint probability is given by
\begin{align}
    P \left(\vec{u} | \alpha(T)\right) = P \left(u_0\right)  \prod_{i=1}^M \Theta \left(u_{i-1}\left. \rightarrow u_i \right| \alpha(T) \right),
\end{align}
with the transition probabilities $\Theta \left(u_{i-1}\left. \rightarrow u_i \right| \alpha(T) \right)$ from state $u_{i-1}$ at time $s_{i-1}$ to $u_i$ at time $s_i$, and the initial condition $P \left(u_0\right)$.
The Fisher information then becomes
\begin{align}
    \fisherinfovoltage_{M,\Delta s}^{(T)} = -\sum_{i=1}^M \int \mathrm{d}^{M+1} \vec{u} \ P \left(\vec{u} | \alpha(T)\right) \partial_T^2 \log \Theta \left(u_{i-1}\left. \rightarrow u_i \right| \alpha(T) \right).
\end{align}
Integrating the stochastic differential equation, Eq.~\ref{SMeq:dynamics_rescaled}, over an infinitesimal $\Delta s$ yields
\begin{align}
    u_i-u_{i-1} = \Delta s \left( \alpha + u_{i-1}^2\right) + \sqrt{\Delta s} Y,
\end{align}
where $Y\sim \mathcal{N} (0,1)$ is a standard Normal distribution.
In the limit $\Delta s \rightarrow 0$, $M \rightarrow \infty$ (with $M\Delta s = s_{\mathrm{tot}}$ fixed), the transition probability is thus given by
\begin{align}
    \Theta \left(u_{i-1}\left. \rightarrow u_i \right| \alpha(T) \right) = \frac{1}{\sqrt{2\pi \Delta s}} e^{-\left(u_i - u_{i-1} -\Delta s \left(\alpha + u_{i-1}^2 \right) \right)^2/ \left(2 \Delta s\right)},
\end{align}
and the Fisher information is rewritten as 
\begin{align}
    \fisherinfovoltage_{M \rightarrow \infty, M\Delta s = s_{\mathrm{tot}}}^{(T)} = \sum_{i=1}^M \int \mathrm{d}^{M+1} \vec{u} \ P\left( \vec{u} | \alpha (T) \right) \Delta s \left( \der{\alpha}{T} \right)^2 = s_{\mathrm{tot}} \left( \der{\alpha}{T} \right)^2,
    \label{SMeq:fisher_info_voltage_tot_der_alpha}
\end{align}
where we used that $\alpha$ is linear in $T$ and that the probability is normalized.
Performing analogous steps as in Eq.~\ref{SMeq:Fisher_info_rate_temp}, we get
\begin{align}
    \fisherinfovoltage_{M \rightarrow \infty, M\Delta s = s_{\mathrm{tot}}}^{(T)} =  s_{\mathrm{tot}} \taus \frac{\rho N}{\tauopen \left( \Delta T\right)^2} \frac{\Neff}{\Nm}.
\end{align}
Finally, transforming back to real time via $\sigma$, Eq.~\ref{SMeq:rescale_parameters}, the Fisher information rate in the entire voltage dynamics is
\begin{align}
    \fisherinforatevoltage = \frac{\fisherinfovoltage}{s_{\mathrm{tot}}} \sigma= \frac{\fisherinfovoltage}{s_{\mathrm{tot}}} \frac{1}{\taus}  =  \frac{\rho N}{\tauopen \left( \Delta T\right)^2} \frac{\Neff}{\Nm}.
    \label{SMeq:Fisher_info_rate_entire_voltage_dynamics}
\end{align}
The voltage dynamics thus preserves all of the information available in the channels if there is no extrinsic noise ($\Next=\infty$ or $\Neff=\Nm$).
Non-zero extrinsic noise ($\Next<\infty$ or $\Neff<\Nm$) causes fluctuations in the voltage dynamics (in addition to those due to intrinsic channel noise) and thereby reduces the available information about temperature.

\subsection{Information loss when coarse-graining to action potential timing}

As shown before, Eq.~\ref{SMeq:Fisher_info_rate_temp}, the information rate in the timings of action potentials is given by 
\begin{align}
    \fisherinforate = \fisherinforatevoltage \ J(\alpha),
\end{align}
where we used Eq.~\ref{SMeq:Fisher_info_rate_entire_voltage_dynamics}.
As shown in the main text $J(\alpha) <1$ in both the regular and irregular regime.
So information is lost when instead of the voltage dynamics only the timing of the action potentials is recorded.
This information loss is intuitive in the irregular regime, where the voltage dynamics very rarely elicits action potentials and mostly hovers around the lower stable fixed point.
But why is information lost deep in the regular regime where the voltage increases (almost) deterministically from the resting potential to the threshold for AP firing?
In particular, it is known that the total traveled distance is a sufficient statistic for a drift-diffusion process with position-independent drift.
In contrast, as we will see in the next section, the timing of action potentials is not a sufficient statistic of the voltage dynamics -- the drift is not constant along the trajectory and by averaging over the regions with different drift, information is lost.

\subsubsection{AP timing is not a sufficient statistic}

Suppose we have a drift-diffusion process with position-independent drift,
\begin{align}
    \der{\tilde{u}}{s} = \tilde{\alpha} + \tilde{\xi} (s),
    \label{SMeq:const_drift_diff}
\end{align}
then the probability density for finding the system at $\{ \tilde{u} (s_0) = \tilde{u}_0, \ldots, \tilde{u}(s_M) = \tilde{u}_M\}$ with $s_i < s_j \ \forall i<j$ is given by
\begin{align}
    P(\{ \tilde{u}_0, \ldots, \tilde{u}_M\}|\tilde{\alpha}) &= P(\tilde{u}_0) \prod_{i=1}^M \frac{1}{\sqrt{2\pi (s_i-s_{i-1})}} e^{-\left(\tilde{u}_i - \tilde{u}_{i-1}-\tilde{\alpha} \left( s_i - s_{i-1}\right) \right)^2/\left(2 \left( s_i - s_{i-1}\right) \right)} =\\
    &=\underbrace{P(\tilde{u}_0) \left[\prod_{i=1}^M \frac{1}{\sqrt{2\pi (s_i-s_{i-1})}} e^{-\left(\tilde{u}_i - \tilde{u}_{i-1} \right)^2/\left(2 \left( s_i - s_{i-1}\right) \right)} \right]}_{=: \tilde{q}\left(\{ \tilde{u}_0, \ldots, \tilde{u}_M\}\right)} \underbrace{e^{-\frac{\tilde{\alpha}^2}{2} (s_M-s_0)} e^{\tilde{\alpha}(\tilde{u}_M - \tilde{u}_0)}}_{=: \tilde{Q}\left(\tilde{u}_M-\tilde{u}_0 | \tilde{\alpha} \right)},
\end{align}
factorizing as in the Fisher–Neyman factorization theorem.
The total traveled distance, $\tilde{u}_M-\tilde{u}_0$, between $s_0$ and $s_M$ is thus a sufficient statistic for the trajectory during that interval.
As a result, recording just the first-passage time $\tilde{S}_{\tilde{u}_f|\tilde{u}_0}$ instead of the whole trajectory between $\tilde{u}_0$ and $\tilde{u}_f > \tilde{u}_0$ contains the same information in the deterministic limit, i.e., for $\tilde{\alpha}>0$:
For  $\tilde{\alpha}>0$, the first-passage time for the constant-drift-diffusion process, Eq.~\ref{SMeq:const_drift_diff}, is distributed according to 
\begin{align}
    P_{\tilde{S},\tilde{u}_f|\tilde{u}_0} (s|\tilde{\alpha})= \frac{\tilde{u}_f-\tilde{u}_0}{\sqrt{2\pi s^3}} e^{-\left(\tilde{u}_f - \tilde{u}_0 - \tilde{\alpha} s \right)^2/\left(2s \right)}.
\end{align}
Correspondingly, the Fisher information in the first-passage time (with respect to the drift $\tilde{\alpha}$) is
\begin{align}
    \tilde{\fisherinfo} = -\int_0^{\infty} \mathrm{d} s \ P_{\tilde{S},\tilde{u}_f|\tilde{u}_0} (s|\tilde{\alpha}) \partial_{\tilde{\alpha}}^2 \log P_{\tilde{S},\tilde{u}_f|\tilde{u}_0} (s|\tilde{\alpha}) = \int_0^{\infty} \mathrm{d} s \ P_{\tilde{S},\tilde{u}_f|\tilde{u}_0} (s|\tilde{\alpha}) s = \frac{\tilde{u}_f-\tilde{u}_0}{\tilde{\alpha}}.
\end{align}
Performing analogous steps as in Section~\ref{SMsubsec:info_full_voltage}, the information (about $\tilde{\alpha}$) in the full trajectory $\{\tilde{u}(s)\}_{s\in[0,\avg{\tilde{S}_{\tilde{u}_f|\tilde{u}_0}}]}$ (up to the mean first-passage time $\avg{\tilde{S}_{\tilde{u}_f|\tilde{u}_0}}$) is
\begin{align}
    \tilde{\fisherinfovoltage} = \avg{\tilde{S}_{\tilde{u}_f|\tilde{u}_0}} = \frac{\tilde{u}_f-\tilde{u}_0}{\tilde{\alpha}}.
\end{align}
In the deterministic limit, where the first-passage time becomes deterministic, the information in the voltage dynamics thus equals the information in the first-passage time.
For a drift-diffusion process with position-independent drift, the information fidelity (as defined above) would thus equal 1 in the regular regime.

In our case, the drift is, however, not independent of position (voltage), and the total traveled distance is not a sufficient statistic:
\begin{align}
    P(\{ u_0, \ldots, u_M\}|\alpha) &= P(u_0) \prod_{i=1}^M \frac{1}{\sqrt{2\pi (s_i-s_{i-1})}} e^{-\left(u_i - u_{i-1}-\left( \alpha + u_{i-1}^2\right) \left( s_i - s_{i-1}\right) \right)^2/\left(2 \left( s_i - s_{i-1}\right) \right)} =\\
    &=q\left(\{ u_0, \ldots, u_M\}\right) Q_1\left(u_M-u_0 | \alpha \right) Q_2\left( \left.\sum_{i=1}^M (s_i-s_{i-1}) u_{i-1}^2 \right| \alpha \right),
\end{align}
where
\begin{align}
    q\left(\{ u_0, \ldots, u_M\}\right)=P(u_0) \left[\prod_{i=1}^M \frac{1}{\sqrt{2\pi (s_i-s_{i-1})}} e^{-\left(u_i - u_{i-1}-u_{i-1}^2 (s_i-s_{i-1}) \right)^2/\left(2 \left( s_i - s_{i-1}\right) \right)} \right]
\end{align}
is independent of $\alpha$ and there are two additional factors depending on $\alpha$:
\begin{align}
    Q_1\left(u_M-u_0 | \alpha \right) &= e^{-\frac{\alpha^2}{2} (s_M-s_0)} e^{\tilde{\alpha}(u_M - u_0)},
\end{align}
which depends on the total traveled distance, $u(s_M) - u(0)$, and
\begin{align}
    Q_2\left( \left.\sum_{i=1}^M (s_i-s_{i-1}) u_{i-1}^2 \right| \alpha \right) &= e^{-\alpha \sum_{i=1}^M\left(s_i-s_{i-1} \right) u_{i-1}^2},
\end{align}
which depends on the total squared distance
\begin{align}
    \sum_{i=1}^M (s_i-s_{i-1}) u_{i-1}^2 \approx \int_0^{s_M} \mathrm{d} s \ u(s)^2.
\end{align}
A sufficient statistic is thus given by a combination of the total traveled distance and the total squared traveled distance.
Accordingly, information is lost when only the total traveled distance (the timing of APs) is considered.

\section{Limiting case $\rho \rightarrow 1/4$: bifurcation right at inflection point}

In the limit $\rho \rightarrow \left. \frac{1}{4} \right.^-$, the second order term of the drift, Eq.~\ref{SMeq:drift_second_order}, is zero, and the lowest non-constant term is the third order term.
We thus approximate the drift as
\begin{align}
    v(V) = \frac{1}{\taurest}  \left(2 \Delta V - \Vhalf + \Vrest \right) - \frac{1}{\taurest} \frac{1}{12 \Delta V^2} (V-\Vhalf)^3 + \mathcal{O} \left((V-\Vhalf)^4 \right)
\end{align}
around the minimum $\Vmin=\Vhalf$.
The bifurcation is thus 
\begin{align}
    \Vhalfbif= 2 \Delta V + \Vrest
\end{align}
and in terms of the distance to the bifurcation $\delta \Vhalf = \Vhalfbif-\Vhalf$ we have
\begin{align}
    v(V) = \frac{1}{\taurest} \delta \Vhalf - \frac{1}{\taurest} \frac{1}{12 \Delta V^2} (V-\Vhalf)^3.
\end{align}

Close to the bifurcation, we can thus approximate the dynamics as 
\begin{align}
    \der{V}{t} =  \frac{1}{\taurest} \delta \Vhalf - \frac{1}{\taurest} \frac{1}{12 \Delta V^2} (V-\Vmin)^3 + \frac{\Delta V}{\sqrt{\taurest}} \nu \ \hat{\xi} \left(t\right),    \label{SMeq:dynamics_close_to_bif_rho_1_4}
\end{align}
with the previous definition of the noise level $\nu$.
As before, we now rescale time and voltage \begin{align}
    s=\tilde{\sigma} t \hspace{50pt} u=\tilde{\omega} \left( V-\Vmin\right).
\end{align}
In terms of these non-dimensionalized variables, we find
\begin{align}
    \der{u}{s} = \frac{\tilde{\omega}}{\tilde{\sigma}} \left[ \frac{1}{\taurest} \delta \Vhalf - \frac{1}{\taurest} \frac{1}{12 \Delta V^2} \frac{u^3}{\tilde{\omega}^3} + \frac{\Delta V}{\sqrt{\taurest}} \nu \ \sqrt{\tilde{\sigma}} \hat{\xi}  \left(s\right) \right].
\end{align}
Choosing $\tilde{\omega}$ and $\tilde{\sigma}$ so that the coefficients in front of $u^3$ and $\hat{\xi} (s)$ are equal to -1 and 1, respectively,
\begin{align}
    \tilde{\sigma} &= \frac{\nu}{12^{1/2} \taurest} \\
    \tilde{\omega} &= \frac{1}{12^{1/4}\Delta V \nu^{1/2}},
\end{align}
we get the following non-dimensionalized dynamics
\begin{align}
    \der{u}{s} = \tilde{\alpha} - u^3 + \hat{\xi} (s),
\end{align}
with scaling variable
\begin{align}
    \tilde{\alpha} = \frac{\tilde{\omega}}{\tilde{\sigma}} \frac{1}{\taurest} \delta \Vhalf = \frac{12^{1/4}}{\nu^{3/2}} \frac{\delta \Vhalf}{\Delta V},
\end{align}
which is still linear in the distance to the bifurcation but exhibits a different scaling with respect to the noise level $\nu$ (here: $\alpha \sim \nu^{-3/2}$ instead of $\alpha \sim \nu^{-4/3}$ previously).

The mean time between action potentials, i.e., the mean time it takes the system to go from $V=\Vrest \approx -\infty$ to $V=V_{\text{th}}$~\footnote{The threshold voltage $V_{\text{th}}$ cannot be set to $\infty$ here since the drift gets negative for large enough $V$.}, and the corresponding variance are then given by
\begin{align}
    \avg{\tap} &= \frac{\sqrt{12} \taurest}{\nu} \tilde{M}(\tilde{\alpha}, u_{\text{th}}), \\
    \var{\tap} &= \frac{12 \taurest^2}{\nu^2} \tilde{S}(\tilde{\alpha}, u_{\text{th}}),
\end{align}
which now not only depend on the scaling parameter but also on the rescaled threshold value
\begin{align}
    u_{\text{th}} = \frac{1}{12^{1/4} \nu^{1/2}}\frac{V_{\text{th}}}{\Delta V}.
\end{align}

Putting everything together we find for the information rate in the AP dynamics
\begin{align}
    \fisherinfovoltage = \frac{N \rho}{\tauopen \left(\Delta V\right)^2} \frac{\Neff}{\Nm} \tilde{J}(\tilde{\alpha}) = \fisherinforatechannel_N^{(V)}\frac{\Neff}{\Nm} \tilde{J}(\tilde{\alpha}),
\end{align}
where 
\begin{align}
    \tilde{J}(\tilde{\alpha}) = \frac{\left(\partial_{\tilde{\alpha}}\tilde{M}(\tilde{\alpha},u_{\text{th}}) \right)^2}{\tilde{M}(\tilde{\alpha},u_{\text{th}}) \tilde{S}(\tilde{\alpha},u_{\text{th}})}.
\end{align}
The information rate in the AP dynamics thus scales linearly with the information rate in all the channels, as before.

\section{Maximal amplification and information rate under different constraints}
\label{SMsec:max_amp_info_constraints}

In this section, we briefly discuss a few scenarios for maximizing the (log-)slope and information (rate) under different constraints.
The list of scenarios presented here is definitely non-exhaustive and one could imagine various other reasonable optimization schemes.
We will focus on optimization with respect to either $\rho$ and/or the number of channels $N$, which appear to be more accessible for control than, e.g., the overall noise level $\nu$. 

\subsection{Maximizing the (log-)slope}

\subsubsection*{Maximize the slope with respect to $\rho$ (and $\alpha$)}
The slope of the action potential frequency with respect to the distance to the bifurcation (or, equivalently, the temperature) is maximized for $\rho \rightarrow 0$ (assuming that we simultaneously keep the noise level/the number of measurements $\nu$ constant and that $\alpha$ can be chosen independently via, e.g., $\delta \Vhalf$):
\begin{align}
    \der{\fap}{\delta \Vhalf} &= \der{}{\delta \Vhalf} \left[ \frac{1}{\taus M(\alpha)} \right] = -\frac{1}{\taus} \frac{M'(\alpha)}{\left( M(\alpha)\right)^2} \left(\frac{\taus}{\Vs \taurest}\right) = \\
    &= -\frac{1}{\taurest} \frac{M'(\alpha)}{\left( M(\alpha)\right)^2} \left(\frac{(1-4\rho)}{4 \nu^4 }\right)^{1/6} \frac{1}{\Delta V},
\end{align}
where we used Eq.~\ref{SMeq:AP_time} together with the definition of the scaling variable $\alpha$, Eq.~\ref{SMeq:scaling_variable_alpha}, and of the scaling parameter $\Vs$, Eq.~\ref{SMeq:rescale_parameters}.
Assuming that we can choose $\alpha$ independently of $\rho$, e.g., via $\delta \Vhalf$ and keep $\nu$ fixed, the slope is maximized for $\rho \rightarrow 0$.
Note, however, that the limit $\rho \rightarrow 0$  requires that $N\rightarrow \infty$ if $\nu$ is to be kept constant.

Taking the derivative of $M'(\alpha)/M(\alpha)^2$ with respect to $\alpha$, we furthermore see that the slope is maximized for $\alpha=0$, as mentioned in the main text:
\begin{align}
    \left. \der{}{\alpha} \frac{M'(\alpha)}{M(\alpha)^2}\right|_{\alpha=0}=0 \\
     -\left. \der{^2}{\alpha^2} \frac{M'(\alpha)}{M(\alpha)^2}\right|_{\alpha=0}<0.
\end{align}

\subsubsection*{Maximize the log-slope with respect to $\rho$}

Maximizing the log-slope yields
\begin{align}
    \der{\log \fap}{\delta\Vhalf} = -\der{\log M(\alpha)}{\delta \Vhalf} = -\frac{M'(\alpha)}{M(\alpha)} \der{\alpha}{\delta\Vhalf} = -\frac{M'(\alpha)}{M(\alpha)} \left(\frac{4}{\nu^8 (1-4\rho)}\right)^{1/6} \frac{1}{\Delta V},
\end{align}
where we used the expression for the average time between action potentials, Eq.~\ref{SMeq:AP_time}, and the definition of the scaling variables $\alpha$ and $\taus$.
For fixed $\alpha$ and $\nu$, the log-slope is thus the larger, the larger $\rho$ and diverges in the limit $\rho \rightarrow (1/4)^-$.

In contrast to the slope, the log-slope is not maximized for $\alpha=0$ but $-M'(\alpha)/M(\alpha)$ increases towards $\alpha \rightarrow -\infty$. 
The log-slope is thus maximal deep in the irregular regime.
However, since the time between two action potentials increases exponentially with $|\alpha|$ in this regime~\cite{hathcock_reaction_2021}, it would take the system exceedingly long to realize a change in AP frequency.

\subsection{Maximizing the information}

Depending on the presence and specifics of, for instance, energetic constraints, the pit organ might try to maximize information relative to different quantities.
Here, we briefly touch upon maximizing the information rate per current through the channels, the information per action potential, the information rate per channel and the bare information rate (with respect to $\rho$ and the number of channels $N$, respectively).

\subsubsection*{Maximize the information rate per current through the channels with respect to $\rho$}

The current through the TRP channels is given
\begin{align}
    \dot{q} = N \ilocal \popen.
\end{align}
Approximating the (time-dependent) opening probability by its value at the minimum of the drift term as done in Section~\ref{SMsec:Nondimensionalize}, we can express $\rho$ in terms of the current as
\begin{align}
    \rho = \popen (1-\popen) = \frac{\dot{q}}{N \ilocal} \left( 1-\frac{\dot{q}}{N \ilocal}\right),
\end{align}
giving for the information rate per current through the channels,
\begin{align}
    \frac{\fisherinforate^{(T)}}{\dot{q}} = \frac{1}{\tauopen \ilocal (\Delta T)^2} \frac{\Neff}{\Nm} J(\alpha) \left( 1-\frac{\dot{q}}{N \ilocal}\right),
\end{align}
where we used Eq.~\ref{SMeq:Fisher_info_rate_temp}.
The information rate per current is thus the larger the smaller the current $\dot{q}$ or, equivalently, the smaller $\rho<1/4$ (for fixed $\Nm$, $\Neff$, and $\alpha$).

\subsubsection*{Maximize the information per action potential with respect to $\rho$}

The information per action potential is given by
\begin{align}
    \fisherinforate^{(T)} \avg{\tap} = \left(\frac{4}{\nu^8 \left( 1-4\rho\right)} \right)^{1/3} \frac{J(\alpha) M(\alpha)}{(\Delta T)^2},
\end{align}
according to Eqs.~\ref{SMeq:AP_time} and ~\ref{SMeq:Fisher_info_rate_temp}. 
For fixed $\alpha$, $\nu$, the information per action potential thus increases with increasing $\rho$ and diverges in the limit $\rho \rightarrow (1/4)^-$, that is, in the opposite limit as the information rate per current.

\subsubsection*{Maximize the information rate per channel with respect to the number of channels $N$}

The information rate per channel is given by 
\begin{align}
    \frac{\fisherinforate^{(T)}}{N}=\frac{\rho J(\alpha)}{\tauopen (\Delta T)^2} \frac{\Neff}{\Nm}=\frac{\rho J(\alpha)}{\tauopen (\Delta T)^2} \frac{1}{1+\Nm/\Next},
\end{align}
using Eq.~\ref{SMeq:Fisher_info_rate_temp}.
Due to the extrinsic noise term and since $\Nm \sim N$, the information rate per channel thus decreases if the number of channels increases, and it is maximized for $N=1$.

\subsubsection*{Maximize the information rate with respect to the number of channels $N$}

The total information rate is
\begin{align}
    \frac{\rho J(\alpha)}{\tauopen (\Delta T)^2} \frac{N}{1+\Nm/\Next},
\end{align}
according to Eq.~\ref{SMeq:Fisher_info_rate_temp}.
Although $\Nm \sim N$, it thus increases if the number of channels increases and is maximized in the limit $N \rightarrow \infty$.
For $\Nm\gg \Next$ (or, equivalently, $N \gg (\ilocal^2 \tauopen \rho)/(\iglobal^2 \tauglobal)$), however, $N/(1+\Nm/\Next)$ saturates  and a further increase of $N$ only has a marginal effect.

\section{Constant readout noise}
\label{SMsec:const_readout_noise}

For the discussion of the information (rate) in the SM so far, we have taken into account the intrinsic and extrinsic noise in the voltage dynamics/channels but have not considered noise in reading out the timing of action potentials. 
This noise will be particularly relevant in the regular regime, where information is contained in tiny changes in the very regular firing of action potentials.
In the main text, we have shown that readout noise proportional to the squared mean time between two action potentials, $ \var{\tesc} {\rightarrow}  \var{\tesc} + p_{\text{ro}} \avg{\tesc}^2$ (or, equivalently, $ S(\alpha) {\rightarrow}  S(\alpha) + p_{\text{ro}} M(\alpha)^2$), strongly suppresses the information in the regular regime and leads to a pronounced maximum close to the bifurcation.

\begin{figure*}[t]
\centering
\includegraphics[height=0.25\linewidth]{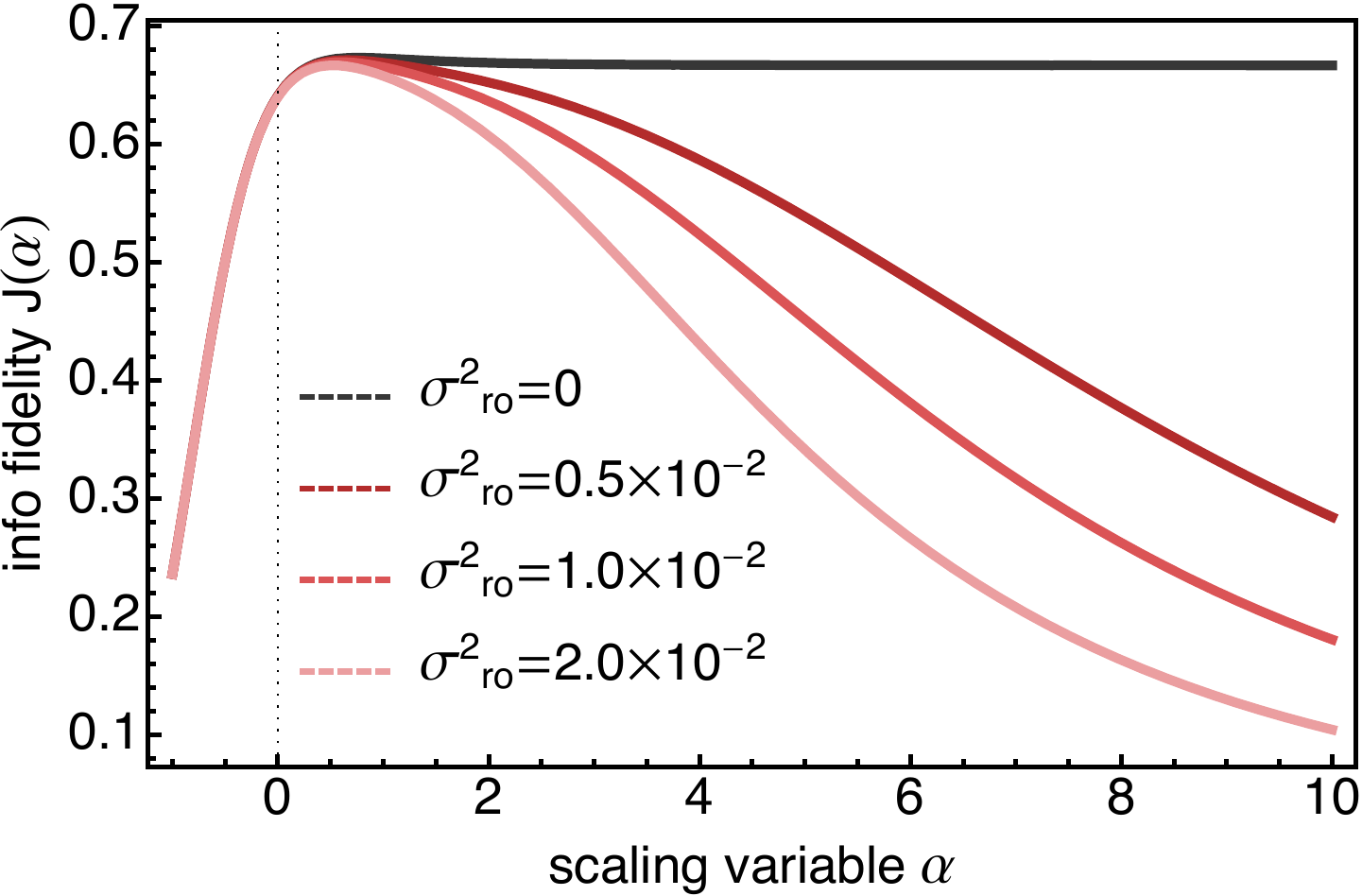}%
\hfill
\includegraphics[height=0.25\linewidth]{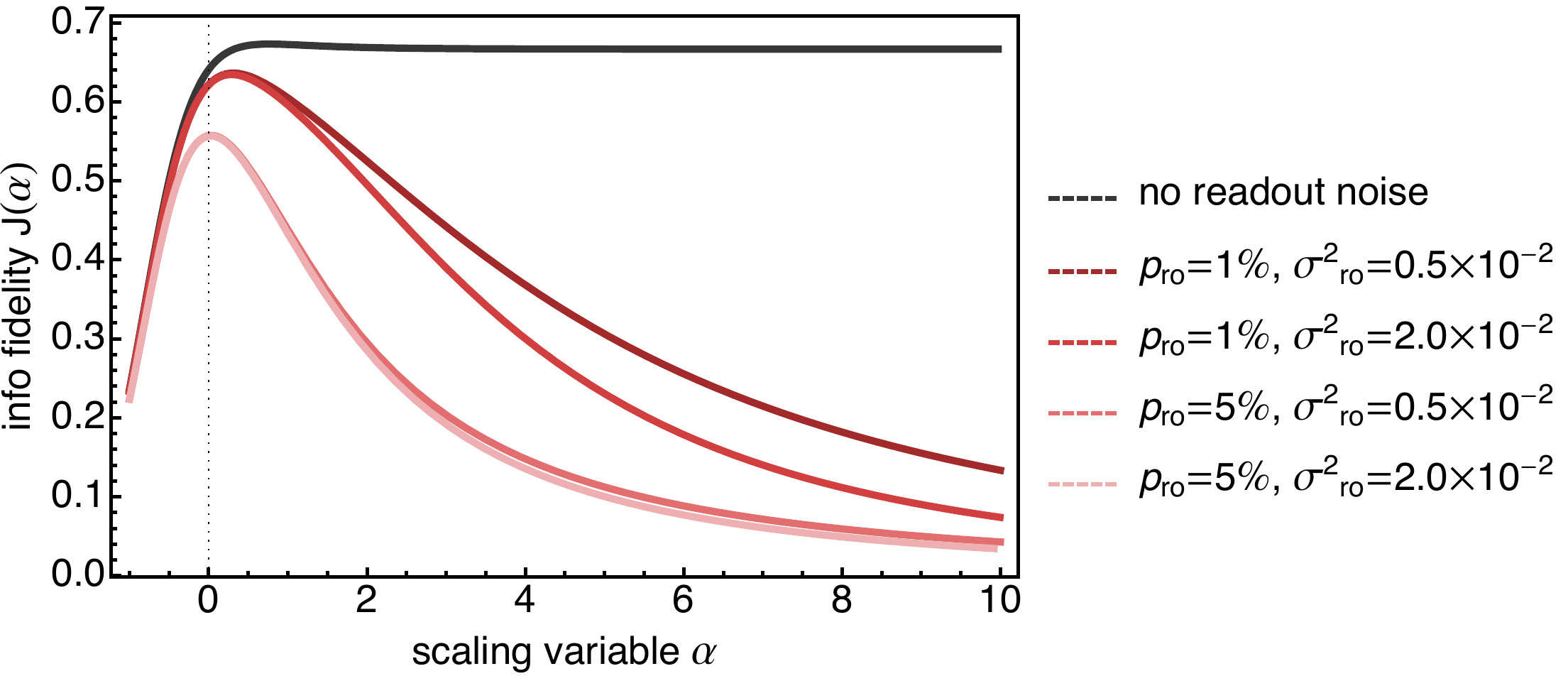}
\caption{\label{SMfig:constant_readout_noise}  Information fidelity $J(\alpha)$ if the time between two APs cannot be perfectly read out but is subject to a constant noise $\sigma^2_{\text{ro}}$, i.e., $ S(\alpha) {\rightarrow}  S(\alpha) + \sigma^2_{\text{ro}}$ (left), or a combination of constant and scaled readout noise $ S(\alpha) {\rightarrow}  S(\alpha) +p_{\text{ro}} M(\alpha)^2 + \sigma^2_{\text{ro}}$ (right).
Readout noise suppresses the information fidelity most strongly in the regular regime so that it develops a maximum close to the bifurcation.
For large enough noise, the maximal value of the information fidelity deviates considerably from the saturation value $2/3$, as expected.
}
\end{figure*}

The same is true for a constant readout noise, $ S(\alpha){\rightarrow}  S(\alpha) + \sigma^2_{\text{ro}}$ (or a combination of scaled and constant readout noise), see Fig.~\ref{SMfig:constant_readout_noise}.

\section{Plugging in numbers: Estimation of parameter values}

In the main text, we argue that, in principle, arbitrarily high amplification is possible if the system is poised at the bifurcation.
Here we show that for a $1$~mK temperature change macroscopic (two-fold) amplification and $\mathcal{O}(1)$ bits of information can indeed be achieved for plausible parameter values and in a reasonable time.
More specifically, we will demonstrate that the following constraints can be satisfied simultaneously for a membrane relaxation timescale $\taurest \approx 1$~ms and with a large enough number of channels $N$:
\begin{itemize}
    \item $\mathcal{O}(1)$ bits of information about a temperature change of $\delta T = 1$~mK can be obtained for an integration time $ \tauint<0.5$~s.
    \item The resting frequency (in the absence of stimulus) is of the order $f_0 \approx 5$~Hz.
    \item The frequency roughly doubles when temperature is increased by $\delta T = 1$~mK.
\end{itemize}
To this end, we will assume that the system is poised right at the bifurcation point, $\alpha=0$, where the amplification (the information fidelity) is maximal (close to maximal).

For the frequency of a system poised at the bifurcation (``resting frequency") we have
\begin{align}
    f_0 = \frac{1}{\taus} \frac{1}{M(0)} =: \frac{\phi^{1/3}}{\taurest} \frac{1}{M(0)},
    \label{SMeq:resting_frequency}
\end{align}
where we defined
\begin{align}
    \phi = \frac{\nu^2 (1-4\rho)}{4}.\label{SMeq:def_phi}
\end{align}
Furthermore, the Fisher information for an ``infinitesimal" temperature change of $\delta T = 1$~mK (i.e., very small compared to $\Delta T = 1$~K) gained over an integration time $\tauint$ is
\begin{align}
    \infoint=\fisherinforate (\delta T)^2 \tauint = \frac{ J(0)}{\nu^2} \frac{\tauint}{\taurest} \left(\frac{\delta T}{\Delta T}\right)^2.
    \label{SMeq:info_integration_time}
\end{align}
Here, we again assumed that the system is poised at the bifurcation.
Solving Eq.~\ref{SMeq:resting_frequency} for $\phi$ and plugging the solution into Eq.~\ref{SMeq:info_integration_time} yields
\begin{align}
    \infoint = \frac{1-4\rho}{4 \phi} J(0) \frac{\tauint}{\taurest} \left(\frac{\delta T}{\Delta T}\right)^2 = \frac{1-4\rho}{4 (f_0 \taurest)^3}  \frac{J(0)}{M(0)^3}\frac{\tauint}{\taurest} \left(\frac{\delta T}{\Delta T}\right)^2,
    \label{SMeq:info_integration_time_rho}
\end{align}
where we used Eq.~\ref{SMeq:def_phi}.
Similarly, we find for the relative amplification of the frequency
\begin{align}
    a=\left. \parder{\log \fap}{T} \right|_{f_0} \delta T &= -\frac{M'(0)}{M(0)}\der{\alpha}{T} \delta T = -\frac{M'(0)}{M(0)}\frac{\taus \Delta V}{\Vs \taurest \Delta T} \delta T = \\
    &=-\frac{M'(0)}{M(0)}\frac{\sqrt{1-4\rho}}{2 \phi^{2/3}} \frac{\delta T}{\Delta T}= -\frac{M'(0)}{M(0)^3}\frac{\sqrt{1-4\rho}}{2 (f_0 \taurest)^2} \frac{\delta T}{\Delta T}.
    \label{SMeq:amplification_rho}
\end{align}
We observe that, for fixed $f_0$, the integrated Fisher information, Eq.~\ref{SMeq:info_integration_time_rho}, and the amplification, Eq.~\ref{SMeq:amplification_rho}, are the larger the smaller $\rho$.
Furthermore, the smaller $\rho$, the smaller the integration time $\tauint$ can be for fixed $f_0$ and target information $\itarget = \fisherinforate (\delta T)^2 \tauint$.
Conversely, for fixed $f_0$ (fixed $\phi$), a smaller $\rho$ requires a smaller noise level $\nu = \sqrt{4 \phi/(1-4\rho)}$, according to Eq.~\ref{SMeq:def_phi}.

If noise is not a limitation (large enough number of channels $N$ and small extrinsic noise) and so $\rho << 1/4$, we can approximate the information and amplification as 
\begin{align}
    \infoint &= \frac{1}{4 (f_0 \taurest)^3}  \frac{J(0)}{M(0)^3}\frac{\tauint}{\taurest} \left(\frac{\delta T}{\Delta T}\right)^2 \\
    a&=-\frac{M'(0)}{M(0)^3}\frac{1}{2 (f_0 \taurest)^2} \frac{\delta T}{\Delta T}.
\end{align}
Solving $\infoint = 4$, for which $\log_2 (\sqrt{\infoint}) \sim 1$~(bit), and $a=1$ (the frequency doubles) for $\delta T = 10^{-3} \Delta T$ and $f_0 = 5$~Hz, yields
\begin{align}
    \taurest \approx 1 \text{ms} \\
    \tauint \approx 400 \text{ms}.
\end{align}
$\mathcal{O}(1)$ bits of information and a doubling of frequency can thus be achieved if the membrane relaxation time is roughly 1~ms and the system integrates over 400~ms.

\begin{figure*}[t]
\centering
\includegraphics[width=0.5\linewidth]{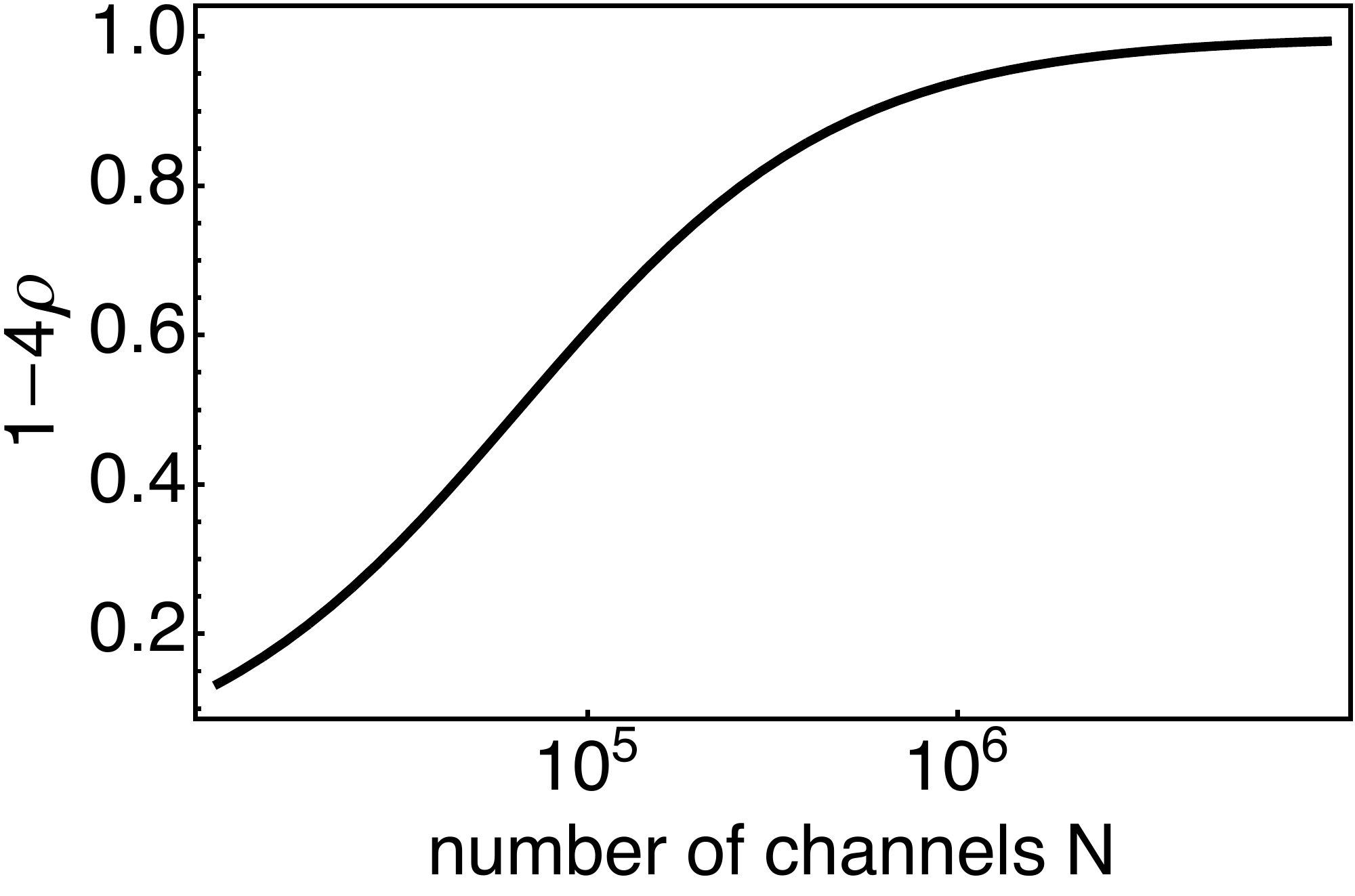}%
\caption{\label{SMfig:factor1-4rho_number_channels} 
Increasing the number of channels and thereby decreasing $\rho$ increases the sensitivity and information via the factor $\sqrt{1-4\rho}$ and $1-4\rho$, respectively.
Here we plot $1-4\rho$ as a function of the number of channels $N$, according to Eq.~\ref{SMeq:rho_N_phi}, with the value of $\phi$ chosen as in Eq.~\ref{SMeq:choose_phi}.
}
\end{figure*}

Of course, $\rho=0$ cannot be achieved in practice, and the amplification and information will be lower in any realistic system (or, alternatively, to achieve the same amplification and information, $\taurest$ needs to be decreased and/or $\tauint$ increased). 
To estimate the effect of the number of channels $N$ and the extrinsic noise on $\rho$, we assume that for a given extrinsic noise level the number of channels is optimized so that the intrinsic and extrinsic noise are of the same order (increasing the number of channels further does not reduce the overall noise substantially), i.e., $\Nm\approx \Next$.
In this case,
\begin{align}
    \nu^2 = \frac{2}{\Nm} = \frac{2 \tauopen}{N \rho \taurest} \approx \frac{2}{N \rho},
\end{align}
according to Eq.~\ref{SMeq:effective_noise}, where we also took into account that typical channel opening times $\tauopen$ of TRPA1 seem to be of the order of $1$~ms.
As a result, according to Eq.~\ref{SMeq:def_phi},
\begin{align}
    \rho = \frac{2}{N \nu^2} = \frac{2}{N} \frac{1-4\rho}{4\phi} \hspace{20pt} \text{or} \hspace{20pt} \rho = \frac{1}{4+2N\phi}. \label{SMeq:rho_N_phi}
\end{align}
For the parameters as calculated before ($f_0=5$~Hz, $\taurest=1$~ms), we have
\begin{align}
    \phi = \left(f_0 \taurest M(0) \right)^3 \approx 3 \times 10^{-5}, \label{SMeq:choose_phi}
\end{align}
for which $1-4\rho>0.5$ for roughly $N \gtrsim 10^5$, see Fig.~\ref{SMfig:factor1-4rho_number_channels}.

\subsection{Parameter values for the time series in the main text}
\label{SMsubsec:param_values}

To generate the voltage time series in the main text for a temperature step $T_0 \rightarrow T_0 + 1$~mK at time $t=0$, we chose the parameter values $N=2^{19}, \Next = \infty$ (or, equivalently, $\Nm=\Next$ and $N=2^{20}\approx 10^6$), $\Vrest=-70$~mV, $\Delta V=30$~mV, $\Delta T=1$~K, $\taurest=\tauopen=1$~ms, $\ilocal/\cm = 2$~V/s, and $\Vrest=-70$~mV. Furthermore, the voltage at half maximum is decreased by $d^-=0.5 \times 10^{-5}$~mV at constant background rate $\gamma^{-1}=1$~ms and is increased by $d^+=10^{-3}$~mV each time an AP is fired.
For these values, we have $\rho=1.5 \times 10^{-2}$, $\nu\approx 1.1 \times 10^{-2}$, $\phi\approx 3 \times 10^{-5}$, and $\alpha \approx 5 \times 10^2 \delta \Vhalf/\Delta V$.
At the same time, the voltage $\Vhalf$ at half maximum self-adapts to a value where the AP frequency is $f_0 = \gamma d^-/d^+=5$~Hz.
After the system has adapted, we thus have
\begin{align}
    \frac{\taurest}{\phi^{1/3}} M(\alpha)=200~\text{ms},
\end{align}
according to Eq.~\ref{SMeq:AP_time}.
Plugging in the parameter values gives 
\begin{align}
    \alpha \approx 0.009
\end{align}
or
\begin{align}
    \delta \Vhalf \approx 2 \times 10^{-5} \Delta V = 6 \times 10^{-4}~\text{mV} \approx 0.
\end{align}
The system is thus maintained basically right at the bifurcation by tuning the voltage at half maximum to its value at the bifurcation
\begin{align}
    \Vhalfbif = \Vrest+\Delta V \left[ \frac{1}{2\rho} \left( 1-\sqrt{1-4\rho}\right)-\log \left( -1+\frac{1}{2\rho} \left( 1-\sqrt{1-4\rho}\right)\right)\right] \approx 85.5342~\text{mV},
\end{align}
according to Eq.~\ref{SMeq:half_voltage_bifurcation}.

\section{Self-organized dynamics of the control parameter, the voltage at half-maximum}

\subsection{Sawtooth oscillations}

\begin{figure*}[t]
\centering
\includegraphics[width=0.5\linewidth]{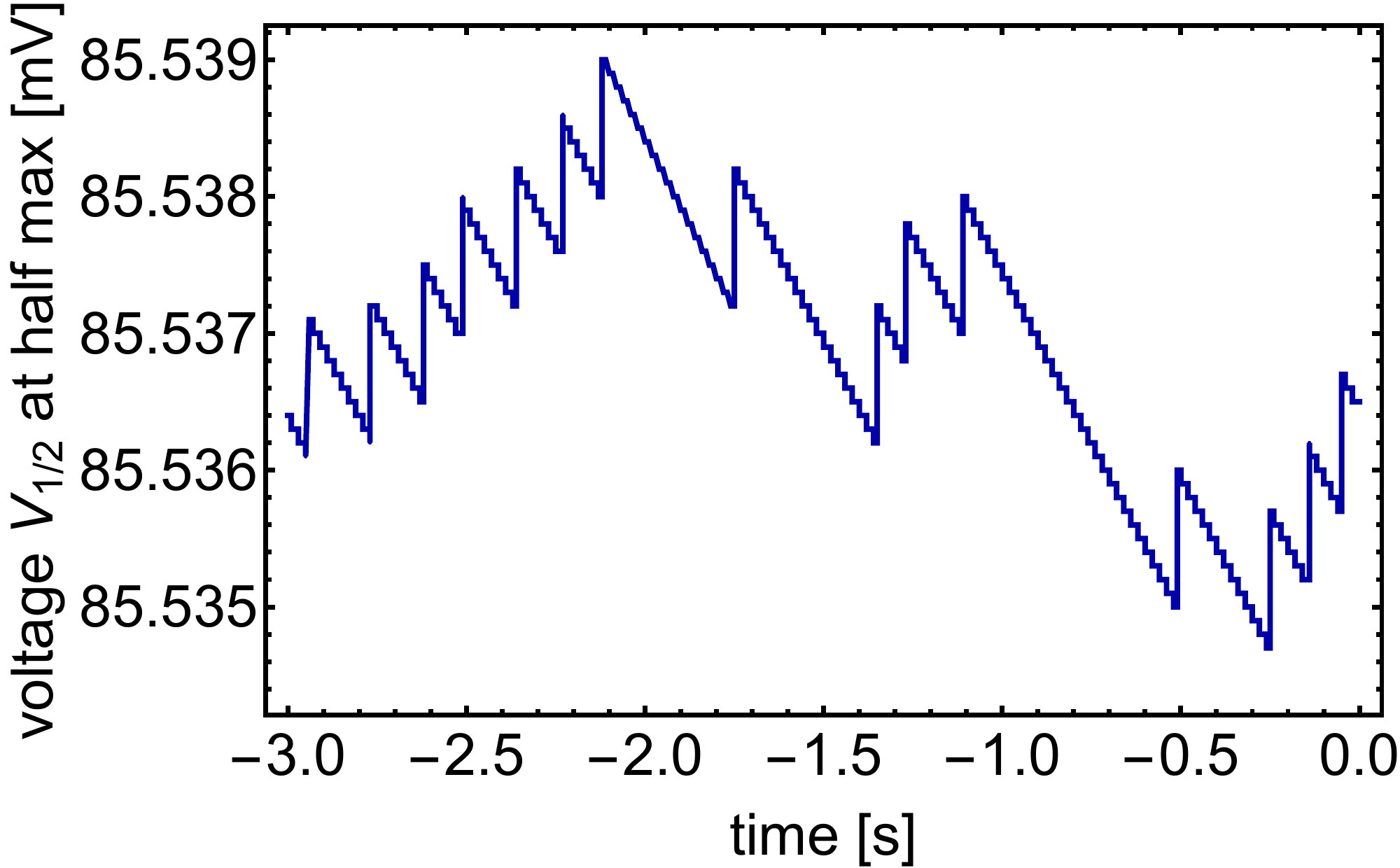}%
\caption{\label{SMfig:half_voltage_sawtooth} 
The voltage at half maximum, the control parameter, shows the characteristic sawtooth-shaped time trace.
Data are from the same simulation as for the time traces shown in the main text, Fig.4.
}
\end{figure*}

As mentioned in the main text, the way we have implemented the order-parameter feedback is inspired by previous work in the context of self-organized criticality.
In particular, we use a constant (deterministic) background rate $\gamma$ at which the voltage $\Vhalf (0)$ at half-maximum is decreased by $d^-$ and an opposing order-parameter dependent increase $d^+$ each time an action potential is fired.
Between two action potentials, the voltage at half-maximum thus exhibits a sawtooth shape, Fig.~\ref{SMfig:half_voltage_sawtooth}, as discussed in other related models, see e.g.~\cite{bonachela_self-organization_2010}.
Since the order parameter, the AP frequency, is intrinsically stochastic, the net change in the control parameter from one AP to the next one fluctuates around zero even if the system has already equilibrated to the current background temperature (we will call these fluctuations ``equilibrium fluctuations" or ``adaptation noise", in each case in slight abuse of notation).
These equilibrium fluctuations limit the temperature sensitivity of the system:
If they are of the same order of magnitude as the change in the voltage at half maximum due to a temperature change, this temperature change cannot be detected. 

\subsection{Estimate for the temporal ``equilibrium" fluctuations, the ``adaptation noise"}

To estimate the magnitude of the equilibrium fluctuations and to understand how the magnitude depends on different noise sources in the adaptation feedback, we consider a slightly more general version of the feedback than in the simulation.
In particular, we assume that the voltage $\Vhalf (0)$ at half-maximum is decreased by $d^-$ at a (stochastic) background rate $\gamma$ and is increased by $d^+$ (+ noise) each time an action potential is fired.
Overall, this setup incorporates two sources of noise:
\begin{itemize}
    \item noise due to the intrinsic randomness of the order parameter, the timing between two APs
    \item extrinsic noise in the feedback.
\end{itemize}

In the limit where the feedback is slow compared to the AP dynamics, the change $\Delta \Vhalf$ in the voltage at half maximum between one AP and the next one does not considerably change the expected time between APs.
In this limit, we can thus assume that the value of $\Vhalf$ at the beginning of the AP is sufficient to determine the time until the next AP is fired.
$\Delta \Vhalf$ is then given by
\begin{align}
    \Delta \Vhalf = - d^- \mathcal{P} + d^+ + \xi,
\end{align}
where $\mathcal{P}=\mathcal{P}\left(\gamma \tap (\Vhalf)\right)$ is Poisson distributed with mean $\gamma \tap (\Vhalf)$ (corresponding to the mean number of background increases of $\Vhalf$ during two APs) and $\xi$ is Gaussian white noise with standard deviation $\sigma$.
The mean change is then
\begin{align}
    \avg{\Delta \Vhalf} = - d^- \gamma \avg{\tap (\Vhalf)} + d^+,
    \label{SMeq:half_voltage_drift}
\end{align}
reflecting the fact that the voltage at half maximum increases (decreases) if the time between two APs is smaller (larger) than the target interspike time $d^+/(\gamma d^-)$.
Since increasing (decreasing) $\Vhalf$ decreases (increases) the opening probability of channels and thereby increases (decreases) the time between two APs, $\Vhalf$ self-tunes to a value where the frequency between APs matches the target frequency $\gamma d^-/d^+$.

Noise in the feedback leads to fluctuations of $\Vhalf$ around its mean value, see Fig.~\ref{SMfig:half_voltage_sawtooth} where the starting point of each sawtooth shape performs a random walk.
To get an estimate for the typical range of this random walk, we first determine the variance $\var{\Delta \Vhalf}$ in the change $\Delta \Vhalf$ between two APs and then map the dynamics of $\Vhalf$ onto an approximate Ornstein-Uhlenbeck process, for which the variance is known.

A Poisson process $X \sim \text{Poisson} (\gamma T)$ with constant rate $\gamma$ evaluated at a random time $T$ (with probability density $p$) has a probability distribution
\begin{align}
    \text{Prob} (X=n) = \int_0^{\infty} \mathrm{d} t \ p(T=t) \frac{1}{n!} \left( \gamma t\right)^n e^{-\gamma t}.
\end{align}
Its variance is thus given by
\begin{align}
    \var{X} &= \int_0^{\infty} \mathrm{d} t \ p(T=t) \sum_{n=0}^{\infty} n^2 \frac{1}{n!} \left( \gamma t\right)^n e^{-\gamma t} - \left[ \int_0^{\infty} \mathrm{d} t \ p(T=t) \sum_{n=0}^{\infty} n \frac{1}{n!} \left( \gamma t\right)^n e^{-\gamma t} \right]^2 = \\
    &=\int_0^{\infty} \mathrm{d} t \ p(T=t) \left( \gamma t + (\gamma t)^2 \right) - \left[ \int_0^{\infty} \mathrm{d} t \ p(T=t) \gamma t \right]^2 = \\
    &= \gamma \avg{T} + \gamma^2 \avg{T^2} - \gamma^2 \avg{T}^2 = \\
    &=\gamma \avg{T} + \gamma^2 \var{T}.
\end{align}
As a result, the variance in the increments of $\Vhalf$ between one AP and the next one is
\begin{align}
    \varsigma^2 (\Vhalf) = \var{\Delta \Vhalf} = \left(d^-\right)^2 \left(\gamma \avg{\tap (\Vhalf)} + \gamma^2 \var{\tap (\Vhalf)}\right) + \sigma^2.
    \label{SMeq:half_voltage_diffusion}
\end{align}
Combining this expression with the expression for the mean change in $\Delta \Vhalf$, Eq.~\ref{SMeq:half_voltage_drift}, we can write the (coarse-grained~\footnote{not resolving the sawtooth-like shape of the trajectory but only the starting/mean position}) dynamics of $\Vhalf$ as 
\begin{align}
    \der{\Vhalf}{t} = - d^- \gamma \avg{\tap (\Vhalf)} + d^+ + \varsigma(\Vhalf) \eta (t),
\end{align}
where time is in units of AP firing and $\eta (t)$ is unit white noise.
Expanding the drift term up to linear order around zero (that is, for the ``equilibrium" case where the system has already converged to the fixed point value, $\Vhalf = \vadapt$), we find
\begin{align}
    \der{\Vhalf}{t} \approx  \underbrace{d^- \gamma \frac{\taurest}{\nu^2}  \frac{2}{\sqrt{1-4\rho}}\frac{1}{\Delta V} \left. M'(\alpha) \right|_{\Vhalf=\vadapt}}_{=:-\theta (\vadapt)} \left(\Vhalf - \vadapt \right)+ \varsigma(\vadapt) \eta (t),
    \label{SMeq:half_voltage_OU}
\end{align}
where we have used that $\avg{\tap (\Vhalf)} = \taurest \nu^{-2/3} 4^{1/3} (1-4\rho)^{-1/3} \ M(\alpha) $, Eq.~\ref{SMeq:AP_time}, and that $\alpha = 4^{1/6} (1-4\rho)^{-1/6} \nu^{-4/3} \delta \Vhalf/\Delta V$, Eq.~\ref{SMeq:scaling_variable_alpha}, and $\Vhalf =\Vhalfbif - \delta \Vhalf$.
The fixed point value $\vadapt$ is determined as the solution to $\avg{\tap (\vadapt)} = d^+/(d^- \gamma)$.
This approximated dynamics of the voltage at half maximum corresponds to an Ornstein-Uhlenbeck process whose variance $W$ is given by
\begin{align}
    W = \frac{\varsigma^2}{2 \theta}.
    \label{SMeq:variance_OU}
\end{align}
Furthermore, the relaxation/adaptation time of the (linearized) dynamics is given by
\begin{align}
    \tauadapt = \frac{\avg{\tap}}{\theta},
\end{align}
where we have transformed back to real time (instead of time in units of AP firing).
For fixed adaptation time $\tauadapt$, how does the variance $W$ of the equilibrium fluctuations (the strength of the adaptation noise) scale with the distance to the bifurcation?
To answer this question, we fix $\tauadapt$ and express $d^- \gamma$, $d^+$, and $\theta$ in terms of $\tauadapt$, the fixed point value $\alpha = \alphaadapt$, and the other system's parameters:
\begin{align}
    \theta &= \frac{\avg{\tap}}{\tauadapt} = \frac{\taurest}{\tauadapt} \nu^{-2/3} 4^{1/3} (1-4\rho)^{-1/3} M(\alphaadapt) \\
    d^- \gamma &= -\theta  \frac{\nu^2 \Delta V \sqrt{1-4\rho}}{2\taurest M'(\alphaadapt)} = - \frac{1}{\tauadapt}  \nu^{4/3} \Delta V \left(\frac{1-4\rho}{4 } \right)^{1/6}\frac{M(\alphaadapt)}{M'(\alphaadapt)} \\
    d^+&= (d^- \gamma) \avg{\tap} = - \frac{\taurest}{\tauadapt}  \nu^{2/3} \Delta V \left(\frac{4}{1-4\rho} \right)^{1/6}\frac{\left(M(\alphaadapt)\right)^2}{M'(\alphaadapt)},
\end{align}
using the explicit formula for the mean time between APs, Eqs.~\ref{SMeq:AP_time}.
Plugging these expressions, together with Eq.~\ref{SMeq:half_voltage_diffusion}, into the formula for the variance, Eq.~\ref{SMeq:variance_OU}, we find
\begin{align}
    W &= \frac{\taurest}{2\tauadapt}  \nu^2 (\Delta V)^2 \frac{M(\alphaadapt) S (\alphaadapt)}{(M'(\alphaadapt))^2} &\hspace{10pt} \text{\textit{(intrinsic  noise in AP firing)}} \label{SMeq:contribution_adaptation_noise_intrinsic} \\
    &+ \frac{1}{2\gamma \tauadapt}  \nu^{8/3} (\Delta V)^2 \left(\frac{1-4\rho}{4} \right)^{1/3}\frac{(M(\alphaadapt) )^2}{(M'(\alphaadapt))^2} &\hspace{10pt} \text{\textit{(extrinsic noise in background process)}} \label{SMeq:contribution_adaptation_noise_extrinsic_background}\\
    &+ \sigma^2 \frac{\tauadapt}{2\taurest}  \nu^{2/3} \left(\frac{1-4\rho}{4}\right)^{1/3} \frac{1}{M(\alphaadapt)} &\hspace{10pt} \text{\textit{(extrinsic noise in AP-dependent process)}} \label{SMeq:contribution_adaptation_noise_extrinsic_AP}
\end{align}
where we used Eq.~\ref{SMeq:AP_time_var}.

\begin{figure*}[t]
\centering
\includegraphics[width=0.8\linewidth]{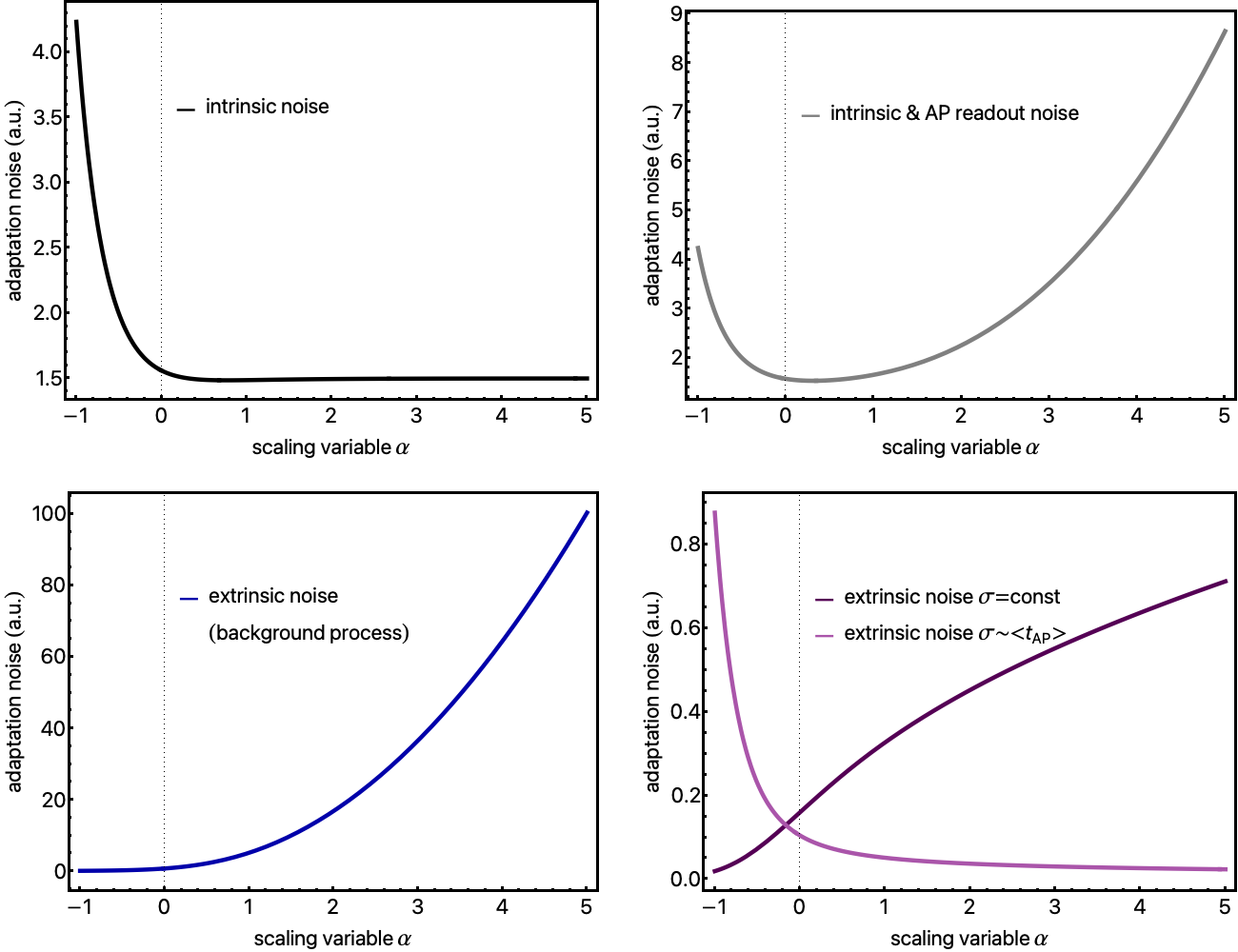}%
\caption{\label{SMfig:adaptation_noise} 
Scaling of the adaptation noise (the expected variance in the voltage at half maximum due to the feedback) as a function of the scaling variable $\alpha$.
The different curves (in arbitrary units) correspond to different contributions to the adaptation noise.
Upper panel shows the contribution of intrinsic noise due to the variance in the timings between APs (left: no readout noise, right: with readout noise: $S(\alphaadapt) \rightarrow S(\alphaadapt) + 0.1$).
Lower panel shows the contribution of extrinsic noise (left: due to a stochastic background rate, right: due to noise added each time an AP is fired).
}
\end{figure*}

Interestingly, the contribution from the intrinsic noise in the firing of APs, Eq.~\ref{SMeq:contribution_adaptation_noise_intrinsic}, scales exactly inversely (with respect to $\alpha$) compared to the information fidelity, Eq.~\ref{SMeq:info_fidelity}.
This relationship reflects that in the absence of extrinsic noise (and in the limit where adaptation is much slower than firing of APs, which we consider throughout) the voltage $\Vhalf$ at half maximum contains the same information about temperature as the firing of APs:
The timing between APs can be inferred from the dynamics of $\Vhalf$ and vice versa.
The intrinsic noise contribution to the adaptation noise thus saturates at a low value for all $\alpha \gtrsim 0$, see Fig.~\ref{SMfig:adaptation_noise} upper left.
Furthermore, if firing of APs is not only subject to intrinsic noise but also has an extrinsic component (cf.\ ``readout noise" in the discussion of information fidelity), i.e., $S(\alphaadapt) \rightarrow S(\alphaadapt) + \text{const}$, the strength of the adaption noise $W$ is minimal close to the bifurcation, see Fig.~\ref{SMfig:adaptation_noise} upper right.
As we can see from Eq.~\ref{SMeq:contribution_adaptation_noise_intrinsic}, there is a tradeoff between the speed of adaptation and the magnitude of the intrinsic adaptation noise: The faster the system needs to adapt (smaller $\tauadapt$), the larger the noise $W$ will be, thereby limiting the possible temperature sensitivity.
Specifically, for the parameter values used for the time trace in the main text and as stated in section~\ref{SMsubsec:param_values}, the standard deviation of the equilibrium fluctuations of the voltage $\Vhalf$ at half maximum is given by $\sqrt{W}/\Delta V \approx 0.01/\sqrt{\tauadapt[\text{ms}]} \approx 10^{-4}$ 
with the adaptation time $\tauadapt [\text{ms}]\approx 10^4$ measured in ms.
For this adaptation time of 10~s, the standard deviation is thus an order of magnitude lower than the change $\left| \delta \delta \Vhalf \right|= \left|\delta T/\Delta T\right| \Delta V = 10^{-3} \Delta V$ following a temperature step of $\delta T = 10^{-3} \Delta T =1$~mK.
Consequently, the response in $\Vhalf$ to a temperature step, as shown in the time trace in the main text, clearly stands out against the equilibrium (or steady-state) noise in $\Vhalf$ (the shaded area corresponds to $\pm 2 \sqrt{W}$).
For much faster adaptation times, this would not be true any more.

Extrinsic noise in the adaptation process, Eqs.~\ref{SMeq:contribution_adaptation_noise_extrinsic_background} and~\ref{SMeq:contribution_adaptation_noise_extrinsic_AP}, leads to adaptation noise whose scaling differs from the intrinsic noise scaling and depends on details of the underlying extrinsic noise.
For instance, noise that scales with the time between action potentials (e.g.\ if a Poisson process underlies the background decrease in $\Vhalf$), has a large contribution deep in the regular regime where the time between APs is small, see Fig.~\ref{SMfig:adaptation_noise} lower left.
Similarly, noise added each time an AP is fired only has a small contribution in the regular regime if its strength scales with $\sigma \sim \avg{\tap}$ but increases towards $\alpha \rightarrow \infty$ if $\sigma = \text{const}$ is independent of the time between APs, see Fig.~\ref{SMfig:adaptation_noise} lower right.

\section{System with independent voltage- and temperature-gated channels}

In the version of the model presented in the main text, TRP channels fulfill a double role: 
They sense temperature and amplify the signal by being temperature- and voltage-gated, respectively.
One might wonder whether signal amplification and high information fidelity can also be achieved in a system where temperature- and voltage-gating occur in two distinct types of channels.
As we will argue here, this decoupling does not eliminate the existence of a bifurcation point with high signal amplification but adds an additional contribution to the noise term.

To see this, we consider the following voltage dynamics:
\begin{align}
    \der{V}{t} &= \underbrace{- \frac{1}{\taurest} \left(V - \Vrest \right) + \frac{\itemp}{\cm} \ptemp (T) + \frac{\ivolt}{\cm} \frac{\Nvolt}{\Ntemp} \pvolt (V)}_{=:v(V)} + \\
    &+ \frac{\itemp}{\cm} \sqrt{\frac{\tautemp}{\Ntemp} \ptemp (T) \left( 1-\ptemp (T)\right)}  \xi_T (t) + \frac{\ivolt}{\cm} \frac{\Nvolt}{\Ntemp} \sqrt{\frac{\tauvolt}{\Nvolt} \pvolt (V) \left( 1-\pvolt (V)\right)} \xi_V (t) + \frac{\iglobal}{\cm}  \sqrt{\tauglobal} \xi_e (t),
\end{align}
where $\cm$ is the membrane capacitance per TRP channel, $\Ntemp$ and $\Nvolt$ are the number of TRP and non-TRP (voltage-gated) channels, respectively, and $\itemp$ and $\ivolt$ ($\tautemp$ and $\tauvolt$) the corresponding currents through an open channel (channel correlation times).
The opening probabilities $\ptemp$ and $\pvolt$ are taken to be sigmoids in temperature and voltage, respectively:
\begin{align}
    \ptemp (T) = \left( 1+ e^{-\frac{T-\Thalf}{\Delta T}}\right)^{-1} \\
    \pvolt (V) = \left( 1+ e^{-\frac{V-\Vhalf}{\Delta V}}\right)^{-1},
\end{align}
with the voltage $\Vhalf$ and temperature $\Thalf$ at half-maximum and the characteristic widths $\Delta T$ and $\Delta V$.
Note that, in contrast to before, $\Vhalf$ is now constant and does not depend on temperature $T$.
Similar to before, we take into account intrinsic (binomial) channel noise, now for TRP and non-TRP channels, and extrinsic noise that scales like the system size.
All noises processes $\xi_T$, $\xi_V$, and $\xi_e$ are assumed to be white with unit variance.

To get some quantitative insight, we perform similar steps as before. 
We first expand the deterministic part of the dynamics, the drift $v(V)$, up to second order around its minimum and then non-dimensionalize the approximated dynamics to find the scaling variable $\alpha$.

Expanding the drift term $v(V)$ up to second order around its arg minimum $\Vmin$ yields
\begin{align}
    v(V)= \frac{1}{\taurest} \delta \Vhalf + \frac{1}{2 \Delta V \taurest} \sqrt{1-4\rho} (V-\Vmin)^2,
\end{align}
where
\begin{align}
    \rho &= \frac{\cm \Delta V}{\ivolt \taurest} \frac{\Ntemp}{\Nvolt} \\
    \Vmin &= \Vhalf + \Delta V \log \left( -1 + \frac{1}{2\rho} \left( 1-\sqrt{1-4\rho}\right)\right) \\
    \delta \Vhalf &= \Vhalfbif - \Vhalf \\
    \Vhalfbif &= \Vrest + \Delta V \left[ \frac{\itemp \Ntemp \ptemp}{\ivolt \Nvolt \rho} + \frac{1}{2\rho} \left(1-\sqrt{1-4\rho} \right) - \log \left( -1+\frac{1}{2\rho} \left(1-\sqrt{1-4\rho}\right)\right)\right]. \label{SMeq:volt_half_max_bif_indep_channels}
\end{align}
The drift thus takes on the same functional form as before but with $\rho$, $\Vmin$, and $\Vhalfbif$ now denoting slightly different quantities.
In particular, note that here $\Vhalfbif$ is a function of temperature (via $\ptemp$) whereas $\Vhalf$ is not, and that -- similar to before -- we will assume $0<\rho<1/4$ throughout.

We can thus approximate the voltage dynamics as 
\begin{align}
    \der{V}{t} &= \frac{1}{\taurest} \delta \Vhalf + \frac{1}{2 \Delta V \taurest} \sqrt{1-4\rho} (V-\Vmin)^2 + \\
    &+ \sqrt{\frac{\itemp^2 \ptemp (1-\ptemp) \tautemp}{\cm^2 \Ntemp} + \frac{\ivolt^2 \pvolt (1-\pvolt) \tauvolt \Nvolt}{\cm^2 \Ntemp^2} + \frac{\iglobal^2 \tauglobal}{\cm^2}} \xi(t).
\end{align}
Rescaling time and voltage as before, Eq.~\ref{SMeq:rescale_time_volt},
\begin{align}
    s=\sigma t \hspace{50pt} u=\omega \left( V-\Vmin\right),
\end{align}
and approximating the noise by its value at the minimum of the drift term (where $\left. \pvolt (1-\pvolt) \right|_{V=\Vmin}= \rho$)
we can set the coefficients in front of $u^2$ and $\xi(s)$ to be 1, i.e.\ $\der{u}{s} = \alpha + u^2 + \xi(s)$, by choosing
\begin{align}
    \sigma &= \frac{1}{\taurest} \left(\frac{\nu^2 (1-4\rho)}{4} \right)^{1/3}\\
    \omega &= \frac{1}{\Delta V} \left(\frac{(1-4\rho)}{4 \nu^4} \right)^{1/6}.
    \label{SMeq:scaling_param_indep_channels}
\end{align}
Here, the noise term $\nu$ is given by 
\begin{align}
    \nu^2 &= \frac{\tauvolt}{\Nvolt \rho \taurest} \left( 1+\frac{\itemp^2 \Ntemp \tautemp \varrho}{\ivolt^2 \Nvolt \tauvolt \rho} + \frac{\iglobal^2 \Ntemp^2 \tauglobal}{\ivolt^2 \Nvolt \tauvolt\rho}\right) =\\
    &=\frac{\itemp^2 \Ntemp \tautemp \varrho}{\ivolt^2 \Nvolt^2 \taurest \rho^2} \left( 1+\frac{\ivolt^2 \Nvolt \tauvolt \rho}{\itemp^2 \Ntemp \tautemp \varrho} + \frac{\Ntemp\iglobal^2 \tauglobal}{\itemp^2 \tautemp \varrho}\right),
    \label{SMeq:noise_indep_channels}
\end{align}
where we have defined $\varrho:=\ptemp(1-\ptemp)$.
Furthermore, the scaling variable is
\begin{align}
    \alpha = \frac{\omega}{\sigma} \frac{\delta \Vhalf}{\taurest},
\end{align}
as before.
For the Fisher information rate in the AP dynamics, we thus find
\begin{align}
    \fisherinforate^{(T)} = \frac{1}{\avg{\tap}} \frac{\left(\parder{\avg{\tap}}{T}\right)^2}{\var{\tap}} = \sigma \left( \der{\alpha}{T} \right)^2 J(\alpha) = \frac{1}{\taurest^2} \frac{\omega^2}{\sigma} \left( \der{\delta \Vhalf}{T}\right)^2 J(\alpha),
    \label{SMeq:Fisher_info_rate_AP_indep_channels}
\end{align}
where we have assumed that the noise is (approximately) independent of the temperature, i.e.\ $\ptemp (1-\ptemp) = \varrho \approx \text{const}$, in analogy to the case with simultaneously voltage- and temperature-gated TRP channels.
Plugging in the expressions for the scaling parameters, Eq.~\ref{SMeq:scaling_param_indep_channels}, and using that 
\begin{align}
    \der{\delta \Vhalf}{T} = \der{\Vhalfbif}{T} = \Delta V \frac{\itemp \Ntemp}{\ivolt \Nvolt \rho} \der{\ptemp}{T} = \frac{\Delta V}{\Delta T} \frac{\itemp \Ntemp}{\ivolt \Nvolt \rho} \ptemp (1-\ptemp) = \frac{\Delta V}{\Delta T} \frac{\itemp \Ntemp \varrho}{\ivolt \Nvolt \rho},
\end{align}
according to Eq.~\ref{SMeq:volt_half_max_bif_indep_channels} yields
\begin{align}
      \fisherinforate^{(T)} = \frac{\left(\Delta V\right)^2}{\taurest^2} \frac{\omega^2}{\sigma} \left( \frac{\itemp \Ntemp \varrho}{\ivolt \Nvolt \rho}\right)^2 \frac{1}{\left( \Delta T\right)^2}J(\alpha) = \frac{1}{\left( \Delta T\right)^2} \frac{\Ntemp \varrho}{\tautemp}  \left(1+\frac{\ivolt^2 \Nvolt \tauvolt \rho}{\itemp^2 \Ntemp \tautemp \varrho} + \frac{\Ntemp\iglobal^2 \tauglobal}{\itemp^2 \tautemp \varrho} \right)^{-1} J(\alpha),
\end{align}
where we used Eqs.~\ref{SMeq:scaling_param_indep_channels} and \ref{SMeq:noise_indep_channels}.
Since the Fisher information rate contained in all $\Ntemp$ temperature-gated TRP channels is given by
\begin{align}
    \fisherinforatechannel_{\Ntemp}^{(T)} = \frac{\Ntemp \varrho}{\tautemp \left( \Delta T \right)^2},
\end{align}
we find
\begin{align}
      \fisherinforate^{(T)} = \fisherinforatechannel_{\Ntemp}^{(T)}  \left(1+\frac{\ivolt^2 \Nvolt \tauvolt \rho}{\itemp^2 \Ntemp \tautemp \varrho} + \frac{\Ntemp\iglobal^2 \tauglobal}{\itemp^2 \tautemp \varrho} \right)^{-1} J(\alpha).
      \label{SMeq:Fisher_info_rate_AP_vs_channels_indep_channels}
\end{align}
As compared to the case of only one channel type that is temperature- and voltage-gated, the suppression factor $(\ldots)^{-1}$ thus has an additional factor $\frac{\ivolt^2 \Nvolt \tauvolt \rho}{\itemp^2 \Ntemp \tautemp \varrho}$, which quantifies the noise introduced by the voltage-gated relative to the temperature-gated channels.
This additional term is the smaller the smaller the current fluctuations through the voltage-gated channels compared to the temperature-gated channels, and is of order 1 if the two types of channels contribute a comparable amount to the current fluctuations.
Similarly, the information about temperature that is available if the whole voltage dynamics could be observed (instead of just the AP firing) is
\begin{align}
    \fisherinforatevoltage^{(T)} = \sigma \left( \der{\alpha}{T}\right)^2 = \frac{1}{\taurest^2} \frac{\omega^2}{\sigma} \left( \der{\delta \Vhalf}{T}\right)^2  = \fisherinforatechannel_{\Ntemp}^{(T)}  \left(1+\frac{\ivolt^2 \Nvolt \tauvolt \rho}{\itemp^2 \Ntemp \tautemp \varrho} + \frac{\Ntemp\iglobal^2 \tauglobal}{\itemp^2 \tautemp \varrho} \right)^{-1}, 
\end{align}
where we used Eqs.~\ref{SMeq:fisher_info_voltage_tot_der_alpha},~\ref{SMeq:Fisher_info_rate_entire_voltage_dynamics},~\ref{SMeq:Fisher_info_rate_AP_indep_channels} and~\ref{SMeq:Fisher_info_rate_AP_vs_channels_indep_channels}.
The additional term in the suppression factor thus reduces the information rate available in the entire voltage dynamics to the same extent as the information rate in the AP dynamics.
The ``information fidelity", corresponding to the ratio between the two, is thus simply given by $J(\alpha)$, as before:
\begin{align}
    \fisherinforate^{(T)} = \fisherinforatevoltage^{(T)} J(\alpha).
\end{align}


\begin{thebibliography}{10}

\bibitem{bullock_physiology_1952}
T.~H. Bullock and R.~B. Cowles, ``Physiology of an infrared receptor: The
  facial pit of pit vipers,'' {\em Science}, vol.~115, no.~2994, pp.~541--543,
  1952.

\bibitem{bullock_properties_1956}
T.~H. Bullock and F.~P.~J. Diecke, ``Properties of an infra-red receptor,''
  {\em The Journal of physiology}, vol.~134, no.~1, p.~47, 1956.

\bibitem{goris_infrared_1967}
R.~C. Goris and M.~Nomoto, ``Infrared reception in oriental crotaline snakes,''
  {\em Comparative Biochemistry And Physiology}, vol.~23, no.~3, 1967.

\bibitem{goris_infrared_2011}
R.~C. Goris, ``Infrared organs of snakes: An integral part of vision,'' {\em
  Journal of Herpetology}, vol.~45, no.~1, pp.~2--14, 2011.

\bibitem{sichert_snakes_2006}
A.~B. Sichert, P.~Friedel, and J.~L. Van~Hemmen, ``Snake’s perspective on
  heat: Reconstruction of input using an imperfect detection system,'' {\em
  Physical Review Letters}, vol.~97, no.~6, p.~068105, 2006.

\bibitem{bakken_imaging_2007}
G.~S. Bakken and A.~R. Krochmal, ``The imaging properties and sensitivity of
  the facial pits of pitvipers as determined by optical and heat-transfer
  analysis,'' {\em Journal of Experimental Biology}, vol.~210, no.~16,
  pp.~2801--2810, 2007.

\bibitem{ebert_behavioural_2006}
J.~Ebert and G.~Westhoff, ``Behavioural examination of the infrared sensitivity
  of rattlesnakes (crotalus atrox),'' {\em Journal of Comparative Physiology
  A}, vol.~192, pp.~941--947, 2006.

\bibitem{gracheva_molecular_2010}
E.~O. Gracheva, N.~T. Ingolia, Y.~M. Kelly, J.~F. Cordero-Morales,
  G.~Hollopeter, A.~T. Chesler, E.~E. Sánchez, J.~C. Perez, J.~S. Weissman,
  and D.~Julius, ``Molecular basis of infrared detection by snakes,'' {\em
  Nature}, vol.~464, no.~7291, pp.~1006--1011, 2010.

\bibitem{karashima_bimodal_2007}
Y.~Karashima, N.~Damann, J.~Prenen, K.~Talavera, A.~Segal, T.~Voets, and
  B.~Nilius, ``Bimodal action of menthol on the transient receptor potential
  channel {TRPA}1,'' {\em The Journal of Neuroscience}, vol.~27, no.~37,
  pp.~9874--9884, 2007.

\bibitem{voets_principle_2004}
T.~Voets, G.~Droogmans, U.~Wissenbach, A.~Janssens, V.~Flockerzi, and
  B.~Nilius, ``The principle of temperature-dependent gating in cold- and
  heat-sensitive {TRP} channels,'' {\em Nature}, vol.~430, no.~7001,
  pp.~748--754, 2004.

\bibitem{nilius_gating_2005}
B.~Nilius, K.~Talavera, G.~Owsianik, J.~Prenen, G.~Droogmans, and T.~Voets,
  ``Gating of {TRP} channels: A voltage connection?,'' {\em Journal of
  Physiology}, vol.~567, no.~1, pp.~35--44, 2005.

\bibitem{zheng_molecular_2013}
J.~Zheng, ``Molecular mechanism of {TRP} channels,'' {\em Comprehensive
  Physiology}, vol.~3, no.~1, pp.~221--242, 2013.

\bibitem{diaz-franulic_allosterism_2016}
I.~Diaz-Franulic, H.~Poblete, G.~Miño-Galaz, C.~González, and R.~Latorre,
  ``Allosterism and structure in thermally activated transient receptor
  potential channels,'' {\em Annual Review of Biophysics}, vol.~45,
  pp.~371--398, 2016.

\bibitem{voets_sensing_2005}
T.~Voets, K.~Talavera, G.~Owsianik, and B.~Nilius, ``Sensing with {TRP}
  channels,'' {\em Nature Chemical Biology}, vol.~1, no.~2, pp.~85--92, 2005.

\bibitem{cordero-morales_cytoplasmic_2011}
J.~F. Cordero-Morales, E.~O. Gracheva, and D.~Julius, ``Cytoplasmic ankyrin
  repeats of transient receptor potential a1 ({TRPA}1) dictate sensitivity to
  thermal and chemical stimuli,'' {\em Proceedings of the National Academy of
  Sciences}, vol.~108, no.~46, 2011.

\bibitem{baez-nieto_thermo-trp_2011}
D.~Baez-Nieto, J.~P. Castillo, C.~Dragicevic, O.~Alvarez, and R.~Latorre,
  ``Thermo-trp channels: biophysics of polymodal receptors,'' {\em Transient
  Receptor Potential Channels}, pp.~469--490, 2011.

\bibitem{ermentrout_mathematical_2010}
B.~Ermentrout and D.~H. Terman, {\em Mathematical Foundations of Neuroscience},
  vol.~35.
\newblock Springer New York, 2010.

\bibitem{hodgkin_quantitative_1952}
A.~L. Hodgkin and A.~F. Huxley, ``A quantitative description of membrane
  current and its application to conduction and excitation in nerve,'' {\em The
  Journal of Physiology}, vol.~117, no.~4, p.~500, 1952.

\bibitem{izhikevich2007dynamical}
E.~M. Izhikevich, {\em Dynamical systems in neuroscience}.
\newblock MIT press, 2007.

\bibitem{hathcock_reaction_2021}
D.~Hathcock and J.~P. Sethna, ``Reaction rates and the noisy saddle-node
  bifurcation: Renormalization group for barrier crossing,'' {\em Physical
  Review Research}, vol.~3, no.~1, 2021.

\bibitem{pi_critical_2021}
Z.~Pi and G.~Zocchi, ``Critical behavior in the artificial axon,'' {\em Journal
  of Physics Communications}, vol.~5, no.~12, p.~125013, 2021.

\bibitem{toyoizumi_fisher_2006}
T.~Toyoizumi, K.~Aihara, and S.-i. Amari, ``Fisher information for spike-based
  population decoding,'' {\em Physical Review Letters}, vol.~97, no.~9,
  p.~098102, 2006.

\bibitem{ramsey_introduction_2006}
I.~S. Ramsey, M.~Delling, and D.~E. Clapham, ``An introduction to {TRP}
  channels,'' {\em Annual Review of Physiology}, vol.~68, pp.~619--647, 2006.

\bibitem{rohacs_regulation_2007}
T.~Rohacs and B.~Nilius, ``Regulation of transient receptor potential ({TRP})
  channels by phosphoinositides,'' {\em Pflügers Archiv - European Journal of
  Physiology}, vol.~455, pp.~157--168, 2007.

\bibitem{taberner_trp_2015}
F.~J. Taberner, G.~Fernández-Ballester, A.~Fernández-Carvajal, and
  A.~Ferrer-Montiel, ``{TRP} channels interaction with lipids and its
  implications in disease,'' {\em Biochimica et Biophysica Acta -
  Biomembranes}, vol.~1848, no.~9, pp.~1818--1827, 2015.

\bibitem{hasan_ca2_2018}
R.~Hasan and X.~Zhang, ``Ca2+ regulation of {TRP} ion channels,'' {\em
  International Journal of Molecular Sciences}, vol.~19, no.~4, 2018.

\bibitem{vangeel_transient_2019}
L.~Vangeel and T.~Voets, ``Transient receptor potential channels and calcium
  signaling,'' {\em Cold Spring Harbor Perspectives in Biology}, vol.~11,
  no.~6, 2019.

\bibitem{sornette_critical_1992}
D.~Sornette, ``Critical phase transitions made self-organized: a dynamical
  system feedback mechanism for self-organized criticality,'' {\em Journal de
  Physique I}, vol.~2, no.~11, pp.~2065--2073, 1992.

\bibitem{sornette_mapping_1995}
D.~Sornette, A.~Johansen, and I.~Dornic, ``Mapping self-organized criticality
  onto criticality,'' {\em Journal de Physique I}, vol.~5, no.~3, pp.~325--335,
  1995.

\bibitem{di_santo_self-organized_2016}
S.~di~Santo, R.~Burioni, A.~Vezzani, and M.~A. Muñoz, ``Self-organized
  bistability associated with first-order phase transitions,'' {\em Physical
  Review Letters}, vol.~116, no.~24, p.~240601, 2016.

\bibitem{buendia_feedback_2020}
V.~Buendía, S.~di~Santo, J.~A. Bonachela, and M.~A. Muñoz, ``Feedback
  mechanisms for self-organization to the edge of a phase transition,'' {\em
  Frontiers in Physics}, vol.~8, 2020.

\bibitem{barkai_robustness_1997}
N.~Barkai and S.~Leibler, ``Robustness in simple biochemical networks,'' {\em
  Nature}, vol.~387, no.~6636, pp.~913--917, 1997.

\bibitem{zierenberg_homeostatic_2018}
J.~Zierenberg, J.~Wilting, and V.~Priesemann, ``Homeostatic plasticity and
  external input shape neural network dynamics,'' {\em Physical Review X},
  vol.~8, no.~3, p.~31018, 2018.

\bibitem{kinouchi_mechanisms_2020}
O.~Kinouchi, R.~Pazzini, and M.~Copelli, ``Mechanisms of self-organized
  quasicriticality in neuronal network models,'' {\em Frontiers in Physics},
  vol.~8, 2020.

\bibitem{zeraati_self-organization_2021}
R.~Zeraati, V.~Priesemann, and A.~Levina, ``Self-organization toward
  criticality by synaptic plasticity,'' {\em Frontiers in Physics}, vol.~9,
  no.~619661, 2021.

\bibitem{oleary_temperature-robust_2016}
T.~O'Leary and E.~Marder, ``Temperature-robust neural function from
  activity-dependent ion channel regulation,'' {\em Current Biology}, vol.~26,
  no.~21, pp.~2935--2941, 2016.

\bibitem{milewski_homeostatic_2017}
A.~R. Milewski, D.~Maoiléidigh, J.~D. Salvi, and A.~J. Hudspeth, ``Homeostatic
  enhancement of sensory transduction,'' {\em Proceedings of the National
  Academy of Sciences}, vol.~114, no.~33, pp.~E6794--E6803, 2017.

\bibitem{harraz_pip2_2020}
O.~F. Harraz, D.~Hill-Eubanks, and M.~T. Nelson, ``{PIP}2: A critical regulator
  of vascular ion channels hiding in plain sight,'' {\em Proceedings of the
  National Academy of Sciences}, vol.~117, no.~34, pp.~20378--20389, 2020.

\bibitem{moreau_balancing_2003}
L.~Moreau and E.~Sontag, ``Balancing at the border of instability,'' {\em
  Physical Review E}, vol.~68, no.~2, p.~020901, 2003.

\bibitem{levina_dynamical_2007}
A.~Levina, J.~M. Herrmann, and T.~Geisel, ``Dynamical synapses causing
  self-organized criticality in neural networks,'' {\em Nature Physics},
  vol.~3, no.~12, pp.~857--860, 2007.

\bibitem{bonachela_self-organization_2010}
J.~A. Bonachela, S.~d. Franciscis, J.~J. Torres, and M.~A. Muñoz,
  ``Self-organization without conservation: are neuronal avalanches generically
  critical?,'' {\em Journal of Statistical Mechanics: Theory and Experiment},
  vol.~2010, no.~2, p.~P02015, 2010.

\bibitem{magnasco_self-tuned_2009}
M.~O. Magnasco, O.~Piro, and G.~A. Cecchi, ``Self-tuned critical anti-hebbian
  networks,'' {\em Physical Review Letters}, vol.~102, no.~25, p.~258102, 2009.

\bibitem{landmann_self-organized_2021}
S.~Landmann, L.~Baumgarten, and S.~Bornholdt, ``Self-organized criticality in
  neural networks from activity-based rewiring,'' {\em Physical Review E},
  vol.~103, no.~3, p.~032304, 2021.

\bibitem{camalet_auditory_2000}
S.~Camalet, T.~Duke, F.~Jülicher, and J.~Prost, ``Auditory sensitivity
  provided by self-tuned critical oscillations of hair cells,'' {\em
  Proceedings of the National Academy of Sciences}, vol.~97, no.~7,
  pp.~3183--3188, 2000.

\bibitem{pascual_criticality_2005}
M.~Pascual and F.~Guichard, ``Criticality and disturbance in spatial ecological
  systems,'' {\em Trends in Ecology \& Evolution}, vol.~20, no.~2, pp.~88--95,
  2005.

\bibitem{Veatch2008}
S.~L. Veatch, P.~Cicuta, P.~Sengupta, A.~Honerkamp-Smith, D.~Holowka, and
  B.~Baird, ``Critical fluctuations in plasma membrane vesicles,'' {\em {ACS}
  Chemical Biology}, vol.~3, no.~5, pp.~287--293, 2008.

\bibitem{mora_are_2011}
T.~Mora and W.~Bialek, ``Are biological systems poised at criticality?,'' {\em
  Journal of Statistical Physics}, vol.~144, pp.~268--302, 2011.

\bibitem{machta_critical_2012}
B.~B. Machta, S.~L. Veatch, and J.~P. Sethna, ``Critical casimir forces in
  cellular membranes,'' {\em Physical Review Letters}, vol.~109, no.~13,
  p.~138101, 2012.

\bibitem{shew_functional_2013}
W.~L. Shew and D.~Plenz, ``The functional benefits of criticality in the
  cortex,'' {\em The Neuroscientist}, vol.~19, no.~1, pp.~88--100, 2013.

\bibitem{hidalgo_information-based_2014}
J.~Hidalgo, J.~Grilli, S.~Suweis, M.~A. Muñoz, J.~R. Banavar, and A.~Maritan,
  ``Information-based fitness and the emergence of criticality in living
  systems,'' {\em Proceedings of the National Academy of Sciences}, vol.~111,
  no.~28, pp.~10095--10100, 2014.

\bibitem{kimchi_ion_2018}
O.~Kimchi, S.~L. Veatch, and B.~B. Machta, ``Ion channels can be allosterically
  regulated by membrane domains near a de-mixing critical point,'' {\em The
  Journal of General Physiology}, vol.~150, no.~12, pp.~1769--1777, 2018.

\bibitem{munoz_colloquium_2018}
M.~A. Muñoz, ``Colloquium: Criticality and dynamical scaling in living
  systems,'' {\em Reviews of Modern Physics}, vol.~90, no.~3, p.~31001, 2018.

\bibitem{stanoev_organization_2020}
A.~Stanoev, A.~P. Nandan, and A.~Koseska, ``Organization at criticality enables
  processing of time-varying signals by receptor networks,'' {\em Molecular
  Systems Biology}, vol.~16, no.~2, p.~e8870, 2020.

\bibitem{autorino_critical_2022}
C.~Autorino and N.~I. Petridou, ``Critical phenomena in embryonic
  organization,'' {\em Current Opinion in Systems Biology}, p.~100433, 2022.

\bibitem{obyrne_how_2022}
J.~O’Byrne and K.~Jerbi, ``How critical is brain criticality?,'' {\em Trends
  in Neurosciences}, vol.~45, no.~11, pp.~820--837, 2022.

\bibitem{graf_thermodynamic_2022}
I.~R. Graf and B.~B. Machta, ``Thermodynamic stability and critical points in
  multicomponent mixtures with structured interactions,'' {\em Physical Review
  Research}, vol.~4, no.~3, p.~033144, 2022.

\bibitem{sourjik_receptor_2002}
V.~Sourjik and H.~C. Berg, ``Receptor sensitivity in bacterial chemotaxis,''
  {\em Proceedings of the National Academy of Sciences}, vol.~99, no.~1,
  pp.~123--127, 2002.

\bibitem{bray_receptor_1998}
D.~Bray, M.~D. Levin, and C.~J. Morton-Firth, ``Receptor clustering as a
  cellular mechanism to control sensitivity,'' {\em Nature}, vol.~393,
  no.~6680, pp.~85--88, 1998.

\bibitem{choe_model_1998}
Y.~Choe, M.~O. Magnasco, and A.~J. Hudspeth, ``A model for amplification of
  hair-bundle motion by cyclical binding of ca2+ to
  mechanoelectrical-transduction channels,'' {\em Proceedings of the National
  Academy of Sciences}, vol.~95, no.~26, pp.~15321--15326, 1998.

\bibitem{eguiluz_essential_2000}
V.~M. Eguíluz, M.~Ospeck, Y.~Choe, A.~J. Hudspeth, and M.~O. Magnasco,
  ``Essential nonlinearities in hearing,'' {\em Physical Review Letters},
  vol.~84, no.~22, pp.~5232--5235, 2000.

\bibitem{hudspeth_critique_2010}
A.~J. Hudspeth, F.~Jülicher, and P.~Martin, ``A critique of the critical
  cochlea: Hopf - a bifurcation - is better than none,'' {\em Journal of
  Neurophysiology}, vol.~104, no.~3, pp.~1219--1229, 2010.

\bibitem{reichenbach_physics_2014}
T.~Reichenbach and A.~J. Hudspeth, ``The physics of hearing: Fluid mechanics
  and the active process of the inner ear,'' {\em Reports on Progress in
  Physics}, vol.~77, no.~7, p.~076601, 2014.

\bibitem{vennettilli_multicellular_2020}
M.~Vennettilli, A.~Erez, and A.~Mugler, ``Multicellular sensing at a
  feedback-induced critical point,'' {\em Physical Review E}, vol.~102, no.~5,
  p.~52411, 2020.

\bibitem{bialek_coding_1990}
W.~Bialek and A.~Zee, ``Coding and computation with neural spike trains,'' {\em
  Journal of Statistical Physics}, vol.~59, pp.~103--115, 1990.

\bibitem{bialek_physical_1987}
W.~Bialek, ``Physical limits to sensation and perception,'' {\em Annual Review
  of Biophysics and Biophysical Chemistry}, vol.~16, no.~1, pp.~455--478, 1987.

\bibitem{goldwyn_what_2011}
J.~H. Goldwyn and E.~Shea-Brown, ``The what and where of adding channel noise
  to the hodgkin-huxley equations,'' {\em {PLoS} Computational Biology},
  vol.~7, no.~11, 2011.

\bibitem{kurten_thermoperception_1982}
L.~Kürten and U.~Schmidt, ``Thermoperception in the common vampire bat
  (desmodus rotundus),'' {\em Journal of Comparative Physiology}, vol.~146,
  pp.~223--228, 1982.

\bibitem{gracheva_ganglion-specific_2011}
E.~O. Gracheva, J.~F. Cordero-Morales, J.~A. González-Carcacía, N.~T.
  Ingolia, C.~Manno, C.~I. Aranguren, J.~S. Weissman, and D.~Julius,
  ``Ganglion-specific splicing of {TRPV1} underlies infrared sensation in
  vampire bats,'' {\em Nature}, vol.~476, no.~7358, pp.~88--91, 2011.

\end{thebibliography}
\end{document}